\documentclass[a4paper,11pt]{article}
\usepackage[margin=2.3cm]{geometry}
\usepackage{gensymb}
\usepackage{amssymb}
\usepackage{amsmath}
\usepackage{ffcode}
\usepackage{graphicx}
\usepackage{blindtext}
\usepackage{subcaption}
\usepackage[export]{adjustbox}
\usepackage{booktabs}
\usepackage{slashed}
\usepackage{physics}
\usepackage{multicol}
\usepackage{amsthm}
\usepackage{caption}
\usepackage{tikz}
\usepackage{wrapfig}
\usepackage[makeroom]{cancel}
\usepackage{mathtools}
\usepackage{esint}
\usepackage{fancyhdr}
\usepackage{orcidlink}
\usepackage[merge,sort&compress,numbers,colon]{natbib}

\makeatletter
\AddToHook{cmd/appendix/before}{\crefalias{section}{appendix}}

\AddToHook{cmd/appendix/before}{\crefalias{subsection}{appendix}}
\AddToHook{cmd/appendix/before}{\crefalias{subsubsection}{appendix}}
\makeatother

\usepackage[capitalise]{cleveref}

\makeatletter
\AddToHook{cmd/appendix/before}{\crefalias{section}{appendix}}
\makeatother

\AtBeginEnvironment{appendices}{\crefalias{section}{appendix}}

\usepackage{stmaryrd }
\usepackage{enumitem}
\usepackage{bm}
\usepackage [english]{babel}
\usepackage [autostyle, english = american]{csquotes}
\usepackage{slashed}
\usepackage{enumitem}
\usepackage{epigraph}
\MakeOuterQuote{"}

\usepackage{tikz} 
\usepackage{tkz-euclide}
\usetikzlibrary{backgrounds} 
\usetikzlibrary{decorations.pathmorphing}
\usetikzlibrary{arrows.meta}

\usepackage{tcolorbox}

\begin{document}

\begin{titlepage}
\vspace*{-1cm}
\phantom{hep-ph/????.?????}
\flushright
\hfil{CPPC-2026-06}

\vskip 1.5cm
\begin{center}
\mathversion{bold}
{\LARGE\bf
Unitarity Cuts, $t$-channel Divergences and the KLN Theorem for Unstable Particles
}\\[3mm]
\mathversion{normal}
\vskip 0.3cm
\end{center}
\vskip 0.5cm
\begin{center}
{\large Marko Beocanin$^a$ \orcidlink{0009-0009-0953-8513}},
{\large Michael A.~Schmidt$^a$ \orcidlink{0000-0002-8792-5537}}
\\
\vskip 0.7cm
{\footnotesize
$^a$ Sydney Consortium for Particle Physics and Cosmology, School of Physics, The University of New South Wales, Sydney, NSW 2052, Australia\\[0.3cm]

\vskip 0.5cm
\begin{minipage}[l]{.9\textwidth}
\begin{center}
\textit{E-mail:}
\tt{m.beocanin@unsw.edu.au}, \tt{m.schmidt@unsw.edu.au}
\end{center}
\end{minipage}
}
\end{center}
\vskip 1cm
\begin{abstract}
Many phenomenological calculations involving massless or unstable particles suffer from divergences as mediating particles go on-shell. One way to deal with these divergences is via the Kinoshita-Lee-Nauenberg (KLN) theorem, which guarantees that by summing over all physically-degenerate processes, the divergences cancel and inclusive observables remain finite. However, actually implementing this theorem in practice requires handling disconnected diagrams, ill-defined distributional objects, threshold behavior and subtle regulator dependence. In this work, we formulate practical prescriptions for dealing with some of these issues by studying the KLN cancellation in an illustrative model exhibiting a $t$-channel divergence. We demonstrate intricate cancellations across several regularization schemes, connect our results to the complex-analytic structure of the underlying amplitudes, and take steps towards constructing a finite, fixed-order, inclusive $t$-channel collider observable. This work highlights both the utility of the KLN theorem, and also the technical subtleties and open questions involved with applying it in practice.
\end{abstract}

\end{titlepage}
\pagenumbering{arabic}
\tableofcontents
\newpage

\section{Introduction}
The Standard Model and its many extensions contain massless and unstable particles. Unfortunately, by allowing for long-range interactions and decays, these particles violate one of the basic tenets of scattering theory: namely, that \textit{asymptotically-separated} states are well-approximated by \textit{free} states. In perturbation theory, this issue presents itself in the form of threshold singularities and infrared divergences associated with mediators going on-shell in pathological pinch or endpoint configurations \cite{Eden:1966dnq}. The ultimate consequence is that phenomenological calculations can become ill-defined or infinite.

To make matters worse, such divergences can arise even at tree-level. In $2 \rightarrow 2$ $t$-channel exchange, for instance, under certain collinear kinematics associated with unstable external particles~\cite{Grzadkowski:2021kgi}, mediators can go on-shell and the amplitude can become infinite. These $t$-channel divergences were first pointed out by Peierls in 1961 in pion-nucleon scattering~\cite{Peierls:1961zz}, and arise in a variety of other Standard Model processes relevant to colliders, such as neutrino-mediated muon collisions~\cite{Ginzburg:1995bc}, and elastic $Z$ or $W$ boson scattering with electrons \cite{Grzadkowski:2021kgi}, to name a few. Recently, they have also been of great interest in a finite-temperature cosmological setting: see, for example, in dark matter annihilation~\cite{Grzadkowski:2021kgi, Beneke:2014gla}, dark matter freeze-in production~\cite{Becker:2023vwd,Coy:2022unt}, and leptogenesis~\cite{Garbrecht:2013gd,Garbrecht:2013bia}.

To deal with divergences more broadly on a formal level, much work has gone into formulating a better-behaved set of asymptotic states and an IR-finite $S$-Matrix from the beginning; see, for instance,~\cite{Kulish:1970ut,Hirai:2020kzx, Carney:2018ygh, Giavarini:1987ts,  Contopanagos:1991yb,  Forde:2003jt,
Hannesdottir:2019umk, Gonzo:2019fai}. Although these approaches have been highly fruitful for studying soft divergences in theories like (massive electron) QED, they become significantly more complicated in the presence of collinear divergences, which arise in theories like QCD or QED with massless electrons. Moreover, even when IR-finiteness is formally guaranteed, the involved mathematical objects can be difficult to work with for actual practical calculations~\cite{Hannesdottir:2019umk}. In ordinary phenomenological calculations, then, the standard approach is often to work within the somewhat ill-defined machinery of the usual $S$-Matrix perturbation theory, but to invoke some sort of physical regularization mechanism that improves the infrared behavior.

Most commonly for $t$-channel divergences, this involves a resummation of self-energy insertions, which essentially amounts to replacing the mediating propagator with its Breit-Wigner version $[p^2-m^2]^{-1} \rightarrow [p^2-M^2+iM\Gamma]^{-1}$, where $M$ is a modified mass and $\Gamma$ a decay width. However, by mixing perturbative orders, this approach requires careful treatment to avoid double-counting \cite{Matak:2022qwc}, and to retain key theoretical properties like gauge invariance (see for eg. \cite{Kurihara:1994fz, Argyres:1995ym, Denner:1996gb, Carrington:2003ut}). It may also simply not work if the mediating particle is sufficiently stable with $\Gamma \approx 0$, although recent proposals  \cite{Grzadkowski:2021kgi, Iglicki:2022jjf} that consider the scattering as embedded in a surrounding thermal medium can ameliorate this issue if the stable mediator acquires a \textit{thermal} mass/width,\footnote{A special case is that of Goldstone bosons. Although they do not acquire a thermal mass, they acquire a thermal width. This has also been discussed in the context of thermal broadening of Goldstone modes for a model of an antiferromagnet~\cite{OBrien:2020bho,OBrien:2020hlp} and applied in a cosmological setting, for instance, in~\cite{Coy:2022unt}. It would be interesting to further explore the generality of thermal and in-medium effects. 
} which has been applied to regularize $\pi D^*$ scattering amplitudes in an expanding hadron gas~\cite{Braaten:2022qag,Braaten:2023ciy}.

Alternatively, one can also invoke finite-beam sizes and collider geometries \cite{Melnikov:1996iu, Melnikov:1996na, Dams:2002uy, Dams:2003gn} that provide natural regularizing soft/collinear cutoffs. Another interesting recent proposal \cite{Asai:2025cdb} involves the use of the Feynman $i\epsilon_F$ as a $t$-channel regulator, in which contributions proportional to $1/\epsilon_F$ are subtracted on the basis of double-counting. However, it is unclear how widely these approaches might generalize to higher perturbative orders, where virtual particles can still probe the entire (soft/collinear) phase space irrespective of the instrumented external kinematics, or where a finite regularizing $\epsilon_F$ can dramatically modify the analytic structure of the amplitude \cite{Hannesdottir:2022bmo}, respectively. 

Therefore, in this work, we feel motivated to study an approach that is fixed-order in perturbation theory, generalizable to arbitrary scattering processes, and completely first-principles in its theoretical input, requiring only unitarity. That is, the Kinoshita-Lee-Nauenberg (KLN) theorem~\cite{Kinoshita:1962ur, Lee:1964is} -- and in particular, the recent stronger result of~\cite{Frye:2018xjj}. In a word, this theorem guarantees that even though \textit{individual} scattering cross-sections/rates may be ill-defined or divergent, one can always recover an \textit{inclusive} cross-section/rate by summing over an appropriate set of physically-degenerate processes, in which all divergences cancel between them to yield an overall finite result. Accounting for degeneracies, it is this finite inclusive quantity that is actually measured in experiment. 

Importantly, however, this inclusive set often involves disconnected and forward-scattering interferences that localize the underlying amplitudes in unusual kinematic regimes~\cite{Lavelle:2005bt, Frye:2018xjj}. Thus, when one attempts to actually use the KLN theorem in practical calculations, one often runs into many technical complications with orders of limits, ill-defined distributional products, regulator dependence, and threshold behavior. Moreover, in the scenarios where cancellations are more involved than the physically-intuitive "real emissions $+ $ virtual corrections $=$ finite" (associated with the Bloch-Nordsieck theorem \cite{Bloch:1937pw}, for instance), it is often difficult to connect the results with observable quantities that might be measured in a physical collider setting \cite{Frye:2018xjj}. 

In this work, then, we have three main contributions. First, we provide a detailed illustration of the KLN cancellation in an example model exhibiting a $t$-channel divergence, developing a generalizable prescription for dealing with some of the technical complications that can arise in these rather involved calculations. Second, more broadly, we explore the connection between the KLN theorem and older, more well-known results about unitarity cuts of scattering amplitudes, which helps us demystify the complex-analytic origin of some of the subtleties of the calculation. Third, we construct a finite, fixed-order, inclusive $t$-channel cross-section and discuss its relation to collider observables, along the way exploring issues with positivity and physical interpretation. 

The main body of work will be structured as follows: in \cref{sec::background}, we review the relevant background on divergences, regularization and the KLN theorem, with a focus on connecting it to unitarity cuts of scattering amplitudes; in \cref{sec::illustrative}, we present our model and compute all the KLN cuts across several different regularization schemes, canceling our $t$-channel divergence in all cases; in \cref{sec::inclusiveobservable}, we take steps towards constructing an inclusive $t-$channel collider cross-section, and highlight open questions about its positivity; in \cref{sec::discussion}, we discuss our results, emphasize the key practical takeaways of the calculation, and make the link to the underlying analytic structure of the diagram being cut; and in \cref{sec::conclusion}, we conclude.

We have included many of the technical details of our calculations, along with extensive appendices, with the aim of being as pedagogical as possible in our account.

\section{Divergences, Regularization and Unitarity Cuts}
\label{sec::background}
\subsection{Infrared Divergences \& Landau Singularities}
In perturbative quantum field theory, scattering amplitudes are computed by integrating over products of propagators of the form $[p^2-m^2]^{-1}$, along with potential numerator factors. Naturally, then, one has to make sense of the poles arising at $p^2 = m^2$, associated with virtual particles going on-shell. The textbook story is that one shifts these poles slightly off the integration contour by augmenting the propagators as $[p^2-m^2]^{-1} \rightarrow [p^2-m^2+i\epsilon_F]^{-1}$, with $\epsilon_F > 0$ taken infinitesimally small. The positivity of $\epsilon_F$ here is crucial to enforce causal propagation in time.

Equivalently, this prescription can instead be thought of as a deformation of the integration contour itself into the complex plane, as opposed to the poles. This allows us to think of scattering amplitudes as functions of \textit{complexified} external momenta (or, more typically, complexified kinematic invariants), which makes available the tools of complex analysis \cite{Hannesdottir:2022bmo}. In fact, one modern perspective is to treat scattering amplitudes as multivalued functions, with all possible contour deformations around the poles encoded as different choices of \textit{Riemann sheet}, and the $i\epsilon_F$ prescription selecting out the \textit{physical sheet} \cite{Hannesdottir:2022bmo}.

On the other hand, when soft $E \rightarrow 0$ or collinear $\cos\theta \rightarrow \pm 1$ limits become kinematically accessible on-shell to scattering particles, the ${i\epsilon_F}$ prescription can fail, and the integral can become divergent.\footnote{Note that it may not always diverge, however, if factors arising from the numerators or integration measure sufficiently regulate the behavior of the integrand at the pole.} This is because such configurations allow the singularities to pinch the integration contour and/or arise at its endpoint/s, in which case a contour deformation around them becomes impossible \cite{Eden:1966dnq}. This is called a \textit{Landau singularity} or \textit{threshold} (see \cite{Fevola:2023fzn} for a modern discussion) when it arises only for specific values of the external invariants, and an \textit{infrared divergence} or \textit{permanent pinch} otherwise.\footnote{Feynman integrals often also exhibit \textit{ultraviolet divergences}, which are a consequence of integrating loop momenta to arbitrarily high scales. These are dealt with via renormalization (see for eg. \cite{Collins:1984xc}) and will not be the focus of this work, although, as we briefly comment on in \cref{sec::multiplediagrams}, the interplay of UV/IR divergences can introduce additional subtleties.} Physically, these configurations are often associated with the unique kinematics of massless particles, which can have arbitrarily soft energies, and unstable particles, which can decay into on-shell partices.

\subsection{Regularization}
To make sense of these ill-behaved configurations, and to hopefully extract something observable, we use \textit{regularization}. This involves computing the relevant quantity in a modified theory, where it exists and is finite, and then \textit{defining} the quantity back in the original theory by analytic continuation.

To be clear, the regularization we describe here is somewhat different to the "physical regularization" approaches described in the introduction, where the regulator has a physical motivation. In this perspective, the regulator is a purely mathematical procedure for extracting a finite quantity from a divergent integral, by parameterizing exactly how it diverges. It should be taken to zero at the end of the calculation for any physically-observable quantities.

The two regularization schemes we will consider here are:
\begin{itemize}
    \item \textbf{\textit{mass regularization}}, whereby one gives the relevant particles fictitious masses $m_0$, and sets their propagators (if there are any) to $[p^2 + {i\epsilon_F}]^{-1} \rightarrow [p^2 -{m_0^{2}} + {i\epsilon_F}]^{-1}$. This improves the infrared behavior by providing a strictly positive lower bound for the energy of on-shell particles $\abs{E_i} =\sqrt{\abs{\vec{p_i}}^2+m_0^2}\geq m_0 > 0$, which forbids singular pinch contributions from the soft $E \rightarrow 0$ limit altogether. Regulator masses can also sometimes be chosen such that on-shell collinear $\cos\theta \rightarrow \pm 1$ endpoint regions are forbidden by kinematics.

   Mass regularization is conceptually quite simple, but it is not always favorable because it may break gauge symmetry explicitly, or dramatically modify the analytic structure of the underlying amplitudes.
    \item \textbf{\textit{dimensional regularization}}, whereby one formally extends the dimensionality of spacetime from $d=4 \rightarrow 4-2\epsilon$, with $\epsilon$ lifted into the complex plane. While there are many technicalities around actually constructing the relevant $d$-dimensional objects (see \cite{Gnendiger:2017pys} for an overview), for our purposes, in a theory involving only scalars, it simply amounts to augmenting the integration measure to
    \begin{align}
    \label{eq::dimregmeasure}
    \int d^4q \rightarrow \mu^{2\epsilon} \int d^d q \equiv  \frac{2 \pi^{1-\epsilon}}{\Gamma(1-\epsilon)}\mu^{2\epsilon}\int_{-\infty}^\infty d{q_0} \int_0^\infty d \abs{\vec{q}} \abs{\vec{q}}^{2-2\epsilon} \int_{-1}^{1} dz (1-z^2)^{-\epsilon} ,
    \end{align}
    where $z \equiv \cos \theta$ and $\mu$ is an arbitrary mass scale introduced to preserve the mass-dimension.\footnote{Note that one has to be careful with the order of integration: the ${q_0}$ energy integral may not always exist if performed first without the regularization of the spacetime integrals, so one should always do it last, or ensure that an interchange can be safely performed if necessary \cite{Anselmi:2016fid, Gorda:2022yex}.} For $\epsilon$ sufficiently negative, the $\abs{\vec{q}}^{-2\epsilon}(1-z^2)^{-\epsilon}$ contribution suppresses the integrand in both soft and collinear regions, canceling any potential poles and improving the infrared behavior. Note that for more general theories, one must also make sure to account for additional subtleties involving the $d$-dimensional metric, fermion algebra, chirality, and so on, but these will not be the focus here. 
    
    Although conceptually more complicated, dimensional regularization is perhaps a more favorable choice since it better respects the symmetries of the original theory.
\end{itemize}
Let us emphasize here that the value of an initially divergent integral after analytic continuation of the regulator $m_0, \epsilon \rightarrow 0$ is \textit{definition} which can depend non-trivially on the precise region of regulator-space in $m_0, \epsilon$ with which the integral is evaluated. Importantly, as we shall see in this paper, even if the integral exists and is convergent across several such regions, the final analytically-continued result may still be different. But more on that in the main body of work.
\subsection{The Kinoshita-Lee-Nauenberg (KLN) Theorem}
\label{sec::kln}
Actually physical, observable quantities should be finite even in the limit $m_0, \epsilon \rightarrow 0$, and should not depend on the underlying regularization scheme used. So, what physical information can we extract from a theory whose amplitudes diverge? 

The basic idea is that even if \textit{individual} scattering amplitudes diverge, $S$-Matrix unitarity $S^\dagger S = 1$ guarantees that the overall \textit{sum} of probabilities must be well-defined and equal unity, in which case any divergences in the individual processes must cancel out. This idea was formalized by Kinoshita, Lee and Nauenberg in their famous KLN theorem \cite{Kinoshita:1962ur, Lee:1964is}, which showed that
\begin{align}
    \sum_{a, b} \sigma_{a \rightarrow b} < \infty
\end{align}
where $a, b$ index over \textit{all} initial \textit{and} \textit{all} final states respectively, and $\sigma_{a\rightarrow b}$ is the scattering cross-section (or equivalently the decay rate, where appropriate) for $a \rightarrow b$ scattering. 

This is commonly interpreted to mean that that the physically-meaningful quantity may not be a single cross-section, but rather some set of cross-sections summed together to something finite. A recent paper by Frye et al \cite{Frye:2018xjj} concerns itself with the question of what might be a minimal such set,\footnote{It should be mentioned that \cite{Frye:2018xjj} explores a variety of different ways to cancel divergences; in this paper, we will only focus on the one prescription outlined below.} since the mandate to sum over \textit{all} initial and \textit{all} final states is somewhat infeasible. 

To this end, Frye et al show that one need only sum over initial \textit{or} final states, not both, to achieve finiteness. In fact, remarkably, they demonstrate that all divergences (and in fact everything altogether\footnote{ Note that we will discuss one way to physically interpret the fact that the KLN sum of \cref{eq::klnschwartz} vanishes -- which may seem somewhat counterintuitive, given that we observe non-zero scattering -- in \cref{sec::physint}.}) must actually cancel -- up to some subtleties with renormalization and gauge invariance which we will discuss in \cref{sec::multiplediagrams} -- even just amongst the different \textit{cuts} of a \textit{single} forward-scattering Feynman diagram.

Let us elaborate on what this means. While we leave a full review of the rich field of cuts to, say, \cite{Britto:2024mna}, a cut $C$ from this perspective is simply a convenient book-keeping device that selects out a subset of internal states $i_C$ from some Feynman diagram $F$ by drawing a line through the diagram, and puts them on-shell as external states with positive energy via the replacement
\begin{align}
    \frac{i}{p^2-m^2 + {i\epsilon_F}} \rightarrow 2\pi \delta^+(p^2-m^2)
\end{align}
with $\delta^+(p^2 -m^2) \equiv \delta(p^2-m^2)\theta(p_0)$. Any cut external lines are left unchanged. This splits the diagram $F$ into the left-hand side $F_L$ and the right-hand (shaded) side $F_R$ of the cut. The left-hand side $F_L$ is computed normally, while the right-hand side $F_R$ is conjugated. 

\begin{figure}[tb!] 
\begin{multline*}
\left[\includegraphics[width=0.32\linewidth, valign = c]{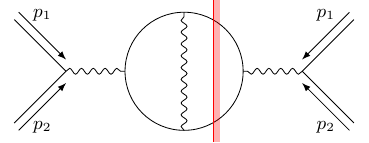}\right] = \int d\Pi_{3,4}\left[\includegraphics[width=0.23\linewidth, valign = c]{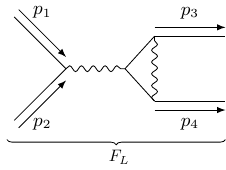}\hspace{1.5ex}\right] \left[\includegraphics[width=0.2\linewidth, valign = c]{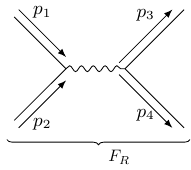}\right]^* 
\\\text{where }d\Pi_{3,4}\equiv \left[\prod_{i=3,4} \frac{d^d p_i}{(2\pi)^d} (2\pi)\delta(p_i^2-m_i^2)\theta(p_i)\right] \times (2\pi)^d \delta^d(p_1+p_2-p_3-p_4)
\end{multline*}
\caption{The cut on the left-hand side can be written as the interference $S$-matrix element between diagrams $F_L$ and $F_R^*$, integrated over the on-shell cut momenta $p_3, p_4$. As discussed in \cref{sec::kln}, both the initial and final state momenta are fixed as $p_1, p_2$, so it can be interpreted as a cross-section integral. Since the right-hand side of the cut diagram is conjugated, we draw the outgoing $p_1, p_2$ as \textit{incoming} for $F_R$, even though they are \textit{outgoing} in the corresponding uncut diagram.}
\label{fig:FL-FR-cut} 

\end{figure}

When the underlying Feynman diagram consists of the same initial and final state $a$, then one can fix the initial and final momenta of corresponding pairs of external particles to be equal, ie. $p_i = p_f$. With this, the cut diagram can (up to an overall normalization) be thought of as an interference contribution to a cross-section, since loop integrals involving cut internal lines are of the form $\int\frac{d^d p}{(2\pi)^d}\, 2\pi \delta^+(p^2-m^2)$, and can thus be interpreted as cross-section phase space integrals. Under this interpretation, the associated interference $S$ matrix element is given by
\begin{align}
    \abs{S}_{\text{int}}^2 \supset 
    S_{F_L, a \rightarrow i_C}\times [S_{F_R, a \rightarrow i_C}]^*
\end{align}
between diagram $F_L$ for $a \rightarrow i_C$ scattering, and (the conjugate of) diagram $F_R$ for $a \rightarrow i_C$ scattering.  An example highlighting the notation and momentum conventions is shown in \cref{fig:FL-FR-cut}.

With this notion of cuts, then the particular incarnation of the KLN theorem mentioned above can be stated precisely as
\begin{align}
    \label{eq::klnschwartz}
    \sum_{\text{cuts }C}\sigma^C_{F, a \rightarrow a} = 0,
\end{align}
 where $a$ is any external state of particles, and $\sigma_{F, a \rightarrow a}^C$ is the interference contribution to the total $\sigma_{a \rightarrow a}$ cross-section from a single cut Feynman diagram $F$. This is the link between the KLN theorem and cuts: cuts of diagrams with the same initial and final state momenta compute interference cross-sections, and all these cuts must sum to zero.

\subsection{Disconnected Diagrams}
\label{sec::disconnecteddiagrams}
An important and somewhat unintuitive consequence of \cref{eq::klnschwartz} is that the inclusive sums associated with the KLN theorem can include disconnected and forward-scattering contributions \cite{Lavelle:2005bt,Frye:2018xjj}; indeed, this is the case in our model, with three cuts shown in \cref{fig:Feynman diagrams} containing such contributions, for example. This can be understood in terms of the Lehmann-Symanzik-Zimmermann (LSZ) reduction formula~\cite{Lehmann:1954rq} for relating $S$-matrix elements to time-ordered products, which indeed contains disconnected pieces that are often ignored in practice (see for eg. \cite{Collins:2019ozc} for a detailed derivation which keeps track of such disconnected pieces). As the computation of cross-sections involve $\abs{S}^2$, these disconnected pieces can still contribute via their interferences.

\begin{figure}[tbp!]\centering 
\begin{subfigure}{0.2\linewidth}
\includegraphics[width=\linewidth]{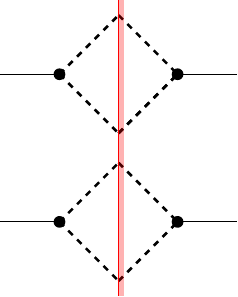}
\caption{}
\end{subfigure}\hspace{7ex}
\begin{subfigure}{0.2\linewidth}
\includegraphics[width=\linewidth]{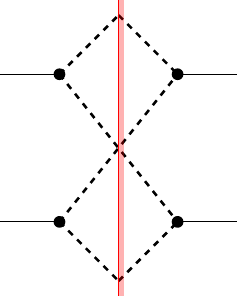}
\caption{}
\end{subfigure}
\caption{The interference shown in (a) can be written as a  cut of an underlying disconnected diagram, and does not contribute to the KLN cancellation. However, the interference shown in (b) is the cut of an underlying connected diagram, and can contribute to the KLN cancellation, even though the diagrams on either side of the cut both contain two disconnected components. Note that dots are added to the three-point interaction vertices to distinguish them from the points where dashed lines cross due to the non-planarity of (b).}
\label{fig:disconnectedinterferences}
\end{figure}

At a glance, this may seem to contradict the well-known result (see, for eg. \cite{Weinberg:1995mt}) that disconnected interferences factorize and do not contribute anything to the physical amplitude, due to the appearance of additional momentum-conserving $\delta$-functions which scale at an infinitely higher rate than their connected counterparts. But there is no contradiction: one must simply be careful to only include those interferences which scale at the same rate as connected ones. Namely, these are the interferences that can be represented as cuts of a single, connected diagram, regardless of how many connected components each interfering diagram contains individually \cite{Lavelle:2005bt, Frye:2018xjj}. Interferences which cannot be written this way factorize and can be dropped. A simple example is shown in \cref{fig:disconnectedinterferences}.
\subsection{The Cutting Rules of 't Hooft and Veltman}
\label{sec::cuttingrules}
With the awareness that disconnected and forward-scattering interferences can contribute to the KLN sum, we can in fact understand the particular result of \cref{eq::klnschwartz} as following from an older theorem by 't Hooft and Veltman, the so-called cutting rules \cite{tHooft:1973wag, Veltman:1994wz, Cutkosky:1960sp}. Namely, for any Feynman diagram $i\mathcal{M}(p_i, p_f)$, with $p_i$ indexing over initial momenta and $p_f$ indexing over final momenta, one can find the imaginary part of $\mathcal{M}(p_i, p_f)$ by
\begin{align}
\label{eq::cuttingrules}
    i\mathcal{M}(p_i,p_f) + \left[i\mathcal{M}(p_i,p_f)\right]^* = -\sum_{\text{cuts } C'} \left[i\mathcal{M}(p_i, p_f)\right]_{C'}.
\end{align}
Here, $C'$ indexes all possible cuts \textit{except} those two cuts which go through all initial state lines only and all final state lines only. This result is essentially the perturbative incarnation of the well-known Optical Theorem (see for eg. \cite{Schwartz:2014sze}), although there are some subtle differences we will comment on in \cref{sec::unstableparticles} regarding which states appear in the cuts. 

When one considers $a \rightarrow a$ diagrams with $p_i = p_f$, \cref{eq::cuttingrules} is precisely \cref{eq::klnschwartz}! To see that this is the case, observe that $i\mathcal{M}(p_i, p_i) + [i\mathcal{M}(p_i, p_i)]^*$ on the left-hand side of \cref{eq::cuttingrules} can be interpreted as the two titular forward-scattering cuts of \cite{Frye:2018xjj}, where only final or only initial states are cut. One can then simply take everything to one side, and divide by the flux factor to obtain the cross section. 

This tells us that there is a deep connection between the KLN theorem and the analytic structure of the underlying diagram being cut, since all the non-forward cut cross-sections appearing in the KLN sum of \cref{eq::klnschwartz} are in fact different contributions to the imaginary part of the underlying diagram $\mathcal{M}$, restricted to the $p_i =p_f$ hypersurface. These contributions can be associated with the monodromies arising from encircling different combinations of singularities in the complex plane (see eg. \cite{Britto:2024mna, Bourjaily:2020wvq, Abreu:2017ptx}), which involves an analytic continuation off the physical sheet.

\subsection{Multiple Diagrams}
\label{sec::multiplediagrams}
Now, one subtlety we alluded to earlier is that, in order to identify the different cuts $\sigma_C$ appearing inside \cref{eq::klnschwartz} with physical cross-sections, there are certain circumstances where one may need to sum over cuts of \textit{multiple} diagrams in order to achieve a cancellation. Two such circumstances are:
\begin{itemize}
    \item in gauge theories, one may need to consider a gauge-invariant \textit{combination} of diagrams when working in a non-unitary choice of gauge (such as the covariant $R_\xi$ gauges commonly used in practice) \cite{Frye:2018xjj}. This is because, in these gauges, the propagators have numerators which include contributions from unphysical polarizations that cannot be identified with the sum over physical spin-states (see for eg. \cite{Schwartz:2014sze}); thus, cutting a line no longer corresponds to identifying it with a physical, external state. Of course, the Ward identities guarantee that these unphysical contributions all cancel out, but this only holds for the full amplitude summed across all different topologies at the same order, and not for individual diagrams. In non-abelian theories, this may also require contributions involving ghosts.
    \item in UV-divergent theories, one also needs to include counterterm diagrams so that the cross-sections are properly renormalized. In principle, as \cite{Anselmi:2016fid} points out, this simply amounts to summing together the cutting equation for the original diagram with the cutting equations for all the counterterm diagrams used to subtract off the overall UV divergence and any UV subdivergences. However, one must be careful to properly keep track of UV and IR divergences independently; for instance, the vanishing of scaleless integrals in dimensional regularization arises from a cancellation \textit{between} UV and IR divergences, which different renormalization schemes treat differently (see for eg. \cite{Collins:1984xc, Schwartz:2014sze}).\footnote{ Some of our arguments about selecting a fixed region in regulator-space for the entire calculation may also require additional qualification when the same regulator is used for both UV and IR divergences. For instance, one often takes $d > 4$ for the IR divergent pieces of an integral, and $d < 4$ for the UV ones; in the presence of both, there is no single region in $\epsilon$ that can regulate the entire integral at once. We intend to explore this with practical examples in future work.}
\end{itemize}
In our model, we consider a UV-finite diagram involving only scalars, so these issues do not apply. 

\subsection{Unstable Particles at Fixed Order}
\label{sec::unstableparticles}
One final subtlety important to clarify here is that, in the full, non-perturbative theory, unstable particles are of course no longer part of the outgoing asymptotic spectrum, since after asymptotic times, they will have all decayed. In principle, then, one does not even have an $S$-Matrix involving unstable states as outgoing states, and as such, it can be shown that cuts through unstable propagators (such as those considered in this work) resum to zero \cite{Veltman:1963th, Hannesdottir:2022bmo}. One consequence of this is that cuts involving outgoing unstable states do not contribute to the Optical Theorem, which is the non-perturbative analogue of \cref{eq::klnschwartz}.

Alternatively, at \textit{fixed-order}, that is, without resummation, the contribution of such cuts is not necessarily vanishing, and is indeed crucial for the "perturbative" KLN theorem as stated in \cref{eq::klnschwartz} to hold. After all, \cref{eq::klnschwartz} follows from the Largest Time Equation (see for eg. \cite{tHooft:1973wag, Veltman:1994wz, Anselmi:2016fid}), which is ultimately nothing more than an algebraic relation on free propagators, and has no knowledge about the spectrum of the full, non-perturbative theory. 

The fact that unstable particles are not true asymptotic states may therefore seem to discredit the validity of any calculation that treats them as such. However, one must recall that physical detectors do not really probe asymptotic times, and instead operate on some finite experimental timescale. So long as this timescale is sufficiently short compared to the lifetime of the unstable particle, it can still be a useful approximation to nonetheless consider it as a quasi-asymptotic state; unstable muons identified by tracks in the detector are a routine part of collider physics, for instance. 

\section{KLN Cancellations in an Illustrative Model}
\label{sec::illustrative}
\subsection{The Model \& Related Diagrams}
\label{sec::themodel}
\begin{figure}[t]
\centering 

\begin{subfigure}{0.35\linewidth} 
\includegraphics[width=\linewidth]{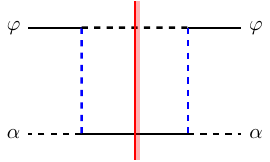}%
\caption{}
\end{subfigure}\hspace{3ex} 
\begin{subfigure}{0.35\linewidth}
\includegraphics[width=\linewidth]{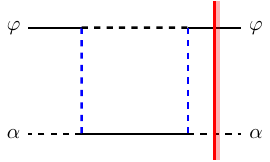}%
\caption{}
\end{subfigure}

\vspace{3ex}

\begin{subfigure}{0.35\linewidth}
\includegraphics[width=\linewidth]{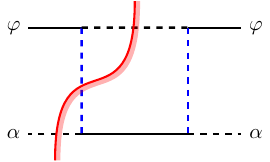}%
\caption{}
\end{subfigure}\hspace{3ex}
\begin{subfigure}{0.35\linewidth}
\includegraphics[width=\linewidth]{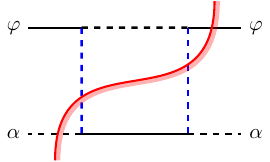}%
\caption{}
\end{subfigure}
\caption{KLN cuts of the underlying box diagram. The corresponding squared matrix element is obtained by multiplying the left-hand side with the conjugate of the right-hand (shaded) side and adding the complex conjugate of the graph for graphs which are not symmetric, namely diagrams (b), (c) and (d). All other cuts not shown are forbidden by kinematics.}
    \label{fig:Feynman diagrams}
\end{figure}

The illustrative model we are going to consider has a Lagrangian given by
\begin{align}
    \mathcal{L} = \frac{1}{2} \partial_\mu \alpha \partial^\mu \alpha + \frac{1}{2}\partial_\mu \varphi \partial^\mu \varphi - \frac{1}{2}m^2 \varphi^2 + \frac{g}{2!} \alpha^2 \varphi. 
\end{align}
Due to the instability of $\varphi$ with its three-body interaction, this theory exhibits a collinear $t$-channel divergence at $s=2m^2$, with the diagram shown in \cref{fig:Feynman diagrams} (a). For $s > 2m^2$, even though the divergence is no longer collinear, the amplitude still contains an ill-defined infinite squared $\delta$-function contribution that is not regulated by the $i\epsilon_F$ prescription. Moreover, since the $\alpha$ particle is completely stable, it does not acquire a decay width in vacuum, and so a resummation of $\alpha$ self-energies cannot resolve the divergence.\footnote{There is one caveat to this. Even though the stability of $\alpha$ prevents it from acquiring a width, there is no obvious symmetry in this model that prohibits it from acquiring a real mass under self-energy corrections. However, so long as this mass remains sufficiently below the mass $m$ of $\varphi$, then $\varphi$ will continue to be unstable, and the divergence will remain an issue. Since the scale of this mass is ultimately fixed by the coupling $g$, we will assume that $g$ is small enough that these mass-corrections to $\alpha$ do not play an essential role. If this is not the case, then a fixed-order, tree-level perturbative calculation is no longer necessarily reliable to begin with.} 

The divergence appears in the squared $t$-channel amplitude, but also in the interferences between the $t$-channel and all other $\varphi \alpha \rightarrow \varphi \alpha$ scattering processes, even if those other processes are finite on their own. At tree-level, the only other relevant interfering process at the same order of perturbation theory is $s$-channel scattering, and we will explore that in \cref{sec::stinterference}. For now, let us focus on the $t$-channel process alone. 

To cancel the divergence, we follow the KLN prescription of \cite{Frye:2018xjj}, treating the amplitude-squared $t$-channel cross-section as just one cut of an underlying box diagram, which should cancel against all the other cuts shown in \cref{fig:Feynman diagrams}. There are essentially three subsets of the physical region $s > m^2$ to consider:
\begin{itemize}
    \item $s \in (m^2, 2m^2)$, where the $t$-channel amplitude is finite and well-defined on its own;
    \item $s = 2m^2$, where the collinear divergence arises; and
    \item $s>2m^2$, where the $t$-channel amplitude can go on-shell at non-collinear angles. 
\end{itemize}
As the theory is singular for $s \geq 2m^2$, we are motivated to consider how it behaves under regularization. Thus, we consider three types of regulator:
\begin{itemize}
    \item \textbf{"Large" mass regulator (LM)}, with masses $m_0$ given to the blue dashed lines shown in \cref{fig:Feynman diagrams}. In this case, we take $m_0 > m$ and then analytically continue to $m_0 \rightarrow 0$ after integration. While it certainly seems unphysical to give these particles such a (relatively) large fictitious mass to only then continue it back to zero, it is necessary to do so if one wishes to actually regulate the singularity with a mass regulator, due to a pole of the form $s = 2m^2-m_0^2$ appearing in the amplitudes.\footnote{For $s \in (m^2, 2m^2]$ and $m_0 >m$, this pole can never be reached.} Nonetheless, by rendering the incoming $\varphi$ stable, this choice of regulator introduces physical and mathematical consequences that we will discuss in \cref{sec::largemassexclude}.
    \item \textbf{"Small" mass regulator (SM)}, with masses $0<m_0 < m$ given to the same blue dashed lines shown in \cref{fig:Feynman diagrams}. Note that although this regulates the pole at $s=2m^2$, it  produces a new pole in the physical region at $s=2m^2-m_0^2$, and so it fails as a regulator; we will only consider its off-pole contributions. However, it is still interesting to compare how the results for the small and large regulators differ away from their respective poles.
    \item \textbf{Dimensional regularization (DR)}, with $d = 4-2\epsilon$ and $\epsilon < -1$ taken for convergence. Note that, perhaps analogously to the large mass regulator, we need to go far away from $d=4$ and all the way up to $d > 6$ to actually regulate the integral, although this does not modify the stability of the incoming $\varphi$. 
\end{itemize}
We will also compute the cuts off-the-pole at $s \neq 2m^2$ \textbf{without any regulator (NR)}, to compare how the analytically-continued regulated results compare to the unregulated (but still finite\footnote{To be precise, we will be calculating each cut summed with its complex conjugate (save for the $t$-channel cut, which does not come with a $+c.c.$) altogether, since this will allow us to make several simplifications. For instance, the unregulated forward-scattering cut of \cref{fig:Feynman diagrams} (b) is not finite on its own; a divergence arises in the imaginary part, but this cancels against the complex conjugate term. For $s > 2m^2$, we will also only be calculating the finite contribution, up to an additional divergent squared $\delta$-function contribution for the $t$-channel and $3$-body diagrams.}) results. The fact that an unregulated result exists off the pole is one of the interesting features of this model, since it allows for such a comparison.

Those more interested in the key conceptual takeaways and discussion should skip to \cref{sec::discussion}, while those more interested in the technical details of the calculation should forge ahead. Let us also mention that several kinematical identities are listed in \cref{app::kinematics}, and a review on the relevant tricks and identities involving the $\delta$-function is outlined in \cref{app::deltafunctions}. 

Finally, note that in the calculations that follow, many of the initial steps are quite similar across the different regulators, and so we have discussed them together for brevity. To be completely clear: when a regulator is actually used to regularize an integral, all other regulators are set to zero; at no point do we use multiple regulators together. 

\mathversion{bold}
\subsection{The \texorpdfstring{$t$}{t}-channel Diagram}
\mathversion{normal}
\label{sec::tchannel}

\begin{figure}[t]\centering 

\includegraphics[width=0.4\linewidth]{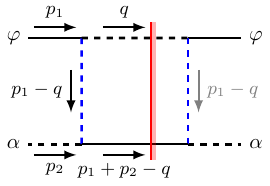}
\caption{The $t$-channel cross-section drawn as a cut diagram, where everything to the right of the cut is to be complex-conjugated. The dashed lines in blue are given the mass $m_0$ in the LM and SM regulator schemes. The momentum labels in black correspond to the unconjugated side of the cut, while the momentum label in gray corresponds to the conjugated side.}
\label{fig:t-channel}
\end{figure}

Our $t$-channel cross-section can be defined in terms of the cut loop integral, incorporating the vertex coupling and cross-section flux factor, as
\begin{align}
    \sigma_t = \frac{g^4}{2(s-m^2)} I_t , 
\end{align}
where
\begin{multline}
\label{eq::tchannelcut}
    I_t(s; d, m_0) = \mu^{2\epsilon}\int \frac{d^dq}{(2\pi)^d} \frac{1}{(p_1-q)^2-m_0^2 + {i\epsilon_F}} \frac{1}{(p_1-q)^2-m_0^2-{i\epsilon_F}} \\ \times 2\pi\delta^+(q^2) 2\pi\delta^+((p_1+p_2-q)^2-m^2) 
\end{multline}
and where we do not sum with the complex conjugate (since this is the unique cut along the symmetric middle of the diagram). The cut and our choice of momenta is indicated in \cref{fig:t-channel}.

Writing out explicitly the $d$-dimensional measure of \cref{eq::dimregmeasure} with $\epsilon = 0$ in the prefactors out the front (including $\mu^{2\epsilon}=1$), since we anticipate no $1/\epsilon$ poles in the final result, and $\abs{\vec{q}} = \omega_q$ since the particle $\alpha$ with 4-momentum $q$ here is massless, and then evaluating $\delta^+(q^2) \equiv \frac{1}{2\omega_q} \delta^+({q_0}-\omega_q)$ against the $d\abs{\vec{q}}$ integral, we get 
\begin{align}
    I_t = \frac{1}{4\pi} \int_0^\infty d{q_0} {q_0}^{1-2\epsilon}\int_{-1}^1dz(1-z^2)^{-\epsilon} \frac{\delta^+((p_1+p_2-q)^2-m^2)}{[(p_1-q)^2-m_0^2 ]^2+ \epsilon_F^2}.
\end{align}
The denominator can be simplified to
\begin{align}
    (p_1-q)^2-m_0^2 = (m^2-m_0^2)-2{q_0}(E_1-\abs{\vec{p_1}}z),
\end{align}
while the $\delta$-function can be simplified to 
\begin{equation} 
\begin{aligned}
    \delta^+((p_1+p_2-q)^2-m^2) &= \delta(s -2\sqrt{s}{q_0}-m^2) \theta(E_1+E_2-q_0)
    \\ &
    = \frac{1}{2\sqrt{s}} \delta\left({q_0}-\frac{s-m^2}{2\sqrt{s}}\right) \theta(\sqrt{s}-{q_0}),
\end{aligned}
\end{equation}
where we have chosen the centre-of-mass  frame in which $p_1 + p_2 = (\sqrt{s}, \vec{0})$. Substituting these expressions in and using the relevant kinematical identities, we get
\begin{equation}
\label{eq::tchannelint}
\begin{aligned}
    I_t &= \frac{1}{8\pi} \frac{1}{\sqrt{s}} \int_0^{\sqrt{s}} d{q_0} {q_0}^{1-2\epsilon} \delta\left({q_0}-\frac{s-m^2}{2\sqrt{s}}\right) \int_{-1}^1dz\frac{(1-z^2)^{-\epsilon}}{[(m^2-m_0^2) - 2{q_0}(E_1-\abs{\vec{p_1}}z)]^2+\epsilon_F^2}
    \\ &= \frac{s(s-m^2)}{4\pi} \int_{-1}^1 dz \frac{(1-z^2)^{-\epsilon}}{[m^4+2(m^2-m_0^2)s-s^2+z(s-m^2)^2]^2+\epsilon_F^2}.
\end{aligned}
\end{equation}
 This expression allows us to discuss the different regulators, one at a time. For each case, we take all other regulators to zero.  

Now, below the pole in $s\in(m^2,2m^2)$ (or $s \in (m^2, 2m^2-m_0^2)$ for the SM case), the integral converges nicely and we can simply set the regulator (and Feynman $i\epsilon_F$) to zero inside the integral in order to recover the limit $m_0, \epsilon \rightarrow 0$, which should otherwise be done \textit{after} the integration. More precisely, this is because of the dominated convergence theorem. The integrand $f(x, \epsilon, m_0)$ is finite and well-defined for all $m_0 \geq 0, \epsilon \leq 0$. Moreover, the limit $\lim_{m_0 \rightarrow 0^+} f(x, 0, m_0)$ converges pointwise to a single $f(x, 0,0)$ expression across the entire region of integration, and the limit $\lim_{\epsilon \rightarrow 0^-} f(x, \epsilon, 0)$ converges "almost everywhere" to $f(x,0,0)$ (that is, everywhere except at the measure-zero endpoints $z = \pm 1$, which can be excluded). Thus, the required assumptions for interchanging limit and integral hold.

Doing so, we arrive at
\begin{align}
\sigma_t^{\text{LM, SM, DM, NR}} =  -\frac{1}{16\pi} \frac{g^4}{m^4} \frac{1}{s-2m^2}  \quad s \in (m^2, 2m^2) 
\end{align}
where we have set all the regulators to zero, which also takes $2m^2 - m_0^2 \rightarrow 2m^2$ for the SM case.

At the pole, however, the integral is no longer convergent due to the endpoint collinear singularity at $z=-1$, so this interchange no longer works. Moreover, since the singularity is at the endpoint, the $i\epsilon_F$-prescription cannot be used to deform around it. Thus, we need to retain our regulators, and write
\begin{align}
    I_t(s=2m^2)=\frac{1}{2\pi} \int_{-1}^1 dz \frac{(1-z^2)^{-\epsilon}}{[m^2(1+z)-4m_0^2]^2},
\end{align}
which can be evaluated straightforwardly for both mass and dimensional regularization.

Doing so and sending $m_0, \epsilon \rightarrow 0$ after integration gives
\begin{align}
\label{eq::polevaluest}
\begin{split}
    \sigma_t^{\text{LM}} &= -\frac{1}{16\pi} \frac{g^4}{m^4} \frac{1}{m_0^2}- \frac{1}{8\pi} \frac{g^4}{m^6} 
    \\ \sigma_t^{\text{DR}} &= -\frac{1}{4 \pi } \frac{g^4}{m^6}
\end{split}
\qquad\qquad s=2m^2 .
\end{align}
Interestingly, observe that even though the integral was ill-defined for $\epsilon = 0$ in dimensional regularization, the result of the analytic continuation back to $\epsilon \rightarrow 0$ is completely finite. In the mass-regulator case, a pole of the form $1/m_0^2$ still remains.

Above the pole, the situation is somewhat more complicated. In LM, the denominator simply cannot go on-shell, so the result for $m^2<s < 2m^2$ extends trivially to $s > 2m^2$. Meanwhile, in all the other regularization schemes, the denominator \textit{can} go on-shell, and since the singular point is no longer at $z = \pm 1$ for $s>2m^2$, it is in fact un-regulated even by DR! Moreover, even a finite choice of $\epsilon_F$ throughout the integration fails to regulate the full result upon analytic continuation $\epsilon_F \rightarrow 0$; to see this, observe that one can write
\begin{multline}
    \frac{1}{(p_1-q)^2-m_0^2 + i\epsilon_F} \frac{1}{(p_1-q)^2+m_0^2-i\epsilon_F} = \frac{[(p_1-q)^2-m_0^2]^2-\epsilon_F^2}{([(p_1-q)^2-m_0^2]^2+\epsilon_F^2)^2}\\+2\pi^2\left(\frac{\epsilon_F}{\pi} \frac{1}{[(p_1-q)^2-m_0^2]^2+\epsilon_F^2}\right)^2.
\end{multline}
 The first term essentially selects out the finite piece as $\epsilon_F \rightarrow 0$, but the second term can be recognized as (twice) the problematic square of a $\delta$-function, $2\pi^2 [\delta^\epsilon((p_1-q)^2-m_0^2)]^2$, as per the discussion in \cref{app::iepsrep}.\footnote{Note that the factor of $2$ appearing in front of the squared $\delta$-function comes from subtracting off an additional $\epsilon_F^2$ in the numerator of the first term. Without this subtraction, the first term would not be finite after integration as $\epsilon_F \rightarrow 0$. Essentially the same identity, and factor of 2, appears in Eq. (13) of \cite{Matak:2022qwc}.} Therefore, except in the LM scheme, the result is an unregulated divergence for $s>2m^2$, when this $\delta$-function acquires support. Note that at $s =2m^2$, its support is in the collinear region suppressed by DR, which is why it does not appear in \cref{eq::polevaluest}. 

Nonetheless, evaluating the finite contribution (which turns out to be the same as below the pole), we can still still write
\begin{align}
\label{eq::fulltchannelexppole}
\sigma_t^{\text{LM, SM, DR, NR}} =  -\frac{1}{16\pi} \frac{g^4}{m^4} \frac{1}{s-2m^2} + \Xi^{\text{LM, SM, DR, NR}}  \quad s \ne 2m^2,
\end{align}
where 
\begin{multline}
\label{eq::squaredeltat}
    \Xi^{\text{LM, SM, DR, NR}} \equiv \frac{g^4}{2(s-m^2)} \times2\pi^2 \int \frac{d^dq}{(2\pi)^d}
    [\delta((p_1-q)^2-m_0^2)]^2 \\ \times 2 \pi \delta^+(q^2)2\pi \delta^+((p_1+p_2-q)^2-m^2)
\end{multline}
encodes the divergent square of the $\delta$-function, which only has support for for $s \geq 2m^2$ (except in LM, where it always vanishes). Here, the superscript tells us which regulator to select for the calculation; all other regulators should be set to zero.

This simply tells us that the $t$-channel contribution is not well-defined for $s \geq 2m^2$ on its own. Indeed, in \cref{sec::3body}, we will show that the $3$-body cut produces precisely the correct $-\Xi$ term to cancel this one, such that their inclusive cross-section is finite.

\subsection{The Box Diagram}
\label{sec::boxdiagram}

\begin{figure}[t]\centering 

\includegraphics[width=0.4\linewidth]{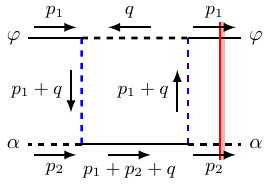}
\caption{The forward-scattering cut of the box diagram, to be summed with its complex conjugate. Everything to the right of the cut is to be complex-conjugated. The dashed lines in blue are given the mass $m_0$ in the LM and SM regulator schemes.}
\label{fig:box diagram}
\end{figure}
Our forward-scattering interference contribution with the box diagram can be similarly defined as
\begin{align}
    \sigma_{\text{box}} =\frac{g^4}{2(s-m^2)} I_{\text{box}}
\end{align}
with 
\begin{equation} 
\begin{aligned}
    I_{\text{box}}(s; d, m_0) \equiv \mu^{2\epsilon}\int \frac{d^d q}{(2\pi)^d} \frac{1}{q^2+{i\epsilon_F}} \frac{1}{(p_1 + q)^2-m_0^2 + {i\epsilon_F}} &\frac{1}{(p_1+p_2+q)^2-m^2+{i\epsilon_F}} 
    \\  & \times
    \frac{1}{(p_1 + q)^2-m_0^2 + {i\epsilon_F}} + c.c.
\end{aligned}
\end{equation}
with the cut and choice of momenta shown in \cref{fig:box diagram}. In \cref{app::FeynparamFULL}, we go through the Feynman parameterization, including the case where the final state momenta differ from the initial state momenta, ie. $p_1 \neq p_3$; 
doing so, one arrives at 
\begin{align}
\label{eq::t0integral}
    I_{\text{box}} = \frac{i}{16\pi^2}\int_0^1 dx \int_0^{1-x}dy \frac{{y}}{(\Delta-{i\epsilon_F})^{2+\epsilon}} + c.c.
\end{align}
with
\begin{align}
\label{eq::boxdenom}
    \Delta\equiv sx(-1+x+y)-m^2(-1+x+y+xy) +m_0^2y,
\end{align}
where we have again set $\epsilon =0$ in all prefactors.

\subsubsection{Mass Regulators and No Regulator}
\label{sec::massboxnoreg}
To study the large/small/no regulator cases, which run similarly in their overall structure, we set $\epsilon = 0$. This allows us to eliminate the square in the denominator using the distributional identity outlined in \cref{app::deltaderiv}, namely
\begin{align}
\label{eq::loopderivdelta}
    \lim_{\epsilon_F \rightarrow 0^+}\frac{1}{(\Delta-{i\epsilon_F})^2}-\frac{1}{(\Delta+{i\epsilon_F})^2} = -2\pi i \partial_\Delta \delta(\Delta) ,
\end{align}
where the limit should be understood as being taken \textit{after} integration has been performed. Then, we get
\begin{equation} 
\begin{aligned}
    I_{\text{box}} &= \frac{i}{16 \pi^2} (-2\pi i) \int_0^1 dx \int_0^{1-x} dy\; y \,\partial_\Delta \delta(\Delta) 
    \\ \label{eq::derivdeltamassreg}
    &= \frac{1}{8 \pi} \int_0^1 dx \int_0^{1-x} dy \;y \left(\pdv{\Delta}{y}\right)^{-1} \pdv{}{y} \delta(\Delta) ,
\end{aligned}
\end{equation}
where we have used the chain rule to rewrite it only in terms of a derivative over $y$.
Explicitly calculating this, we get
\begin{align}
    I_{\text{box}} = \frac{1}{8\pi} \int_0^1 dx \int_0^{1-x} dy \frac{y \partial_y\delta(y-y_0)}{(sx-m^2(1+x) + m_0^2)^2} \eta_\pm ,
\end{align}
where
\begin{align}
    \eta_\pm \equiv \text{sgn}(sx-m^2(1+x)+m_0^2  )
\qquad\mathrm{and}\qquad
    y_0 \equiv \frac{(1-x)(sx - m^2)}{sx - m^2(1+x) +m_0^2} .
\end{align}
Integrating by parts and being careful to keep the boundary contribution, we arrive at
\begin{multline} 
    I_{\text{box}} = -\frac{1}{8\pi}\int_0^1 dx \left[\frac{1}{[sx-m^2(1+x)+m_0^2]^2} \eta_{\pm}\mathcal{I}_{y_0}\right]
    \\  
    +\frac{1}{8\pi}\int_0^1 dx \left[ \frac{1}{[sx-m^2(1+x)+m_0^2]^2} \eta_\pm [y\delta(y-y_0)]^{1-x}_0\right] ,
\end{multline}
where $\mathcal{I}_{y_0}$ is unity for $y_0 \in [0, 1-x]$, and zero elsewhere. Finally, note that one can simplify the boundary contribution
\begin{align}
    [y\,\delta(y-y_0)]^{1-x}_0 = (1-x) \left[\abs{\frac{s-2m^2+m_0^2}{m^2-m_0^2}} \delta(x-1)+ \abs{\frac{m^4-m_0^2s}{m^4-m^2 m_0^2}}\delta(x-m_0^2/m^2)\right] .
\end{align}

It is in evaluating these remaining terms that the distinction between large, small and no mass regulators becomes important for the box diagram.
\paragraph{Large Mass Regulator}
\label{sec::boxmassreg}
For the large mass regulator, $\eta_\pm = 1$. One can also check that the intersection of $\mathcal{I}_{y_0}$ with $x \in [0, 1]$ is the set
\begin{align}
    \frac{m^2}{s} \leq x \leq 1.
\end{align}
Now, for the $\delta$-functions in the boundary term. For $m_0 > m$, the $\delta(x-m_0^2/m^2)$  is excluded completely from the region of integration, so it can be dropped. Meanwhile, the first $\delta$-function can also be dropped due to the $(1-x)$ prefactor out the front.\footnote{In principle, if the denominator of the integrand also carried some factors of $1-x$, this argument would not hold, since they could cancel against the $1-x$ in the numerator. However, except for $s=2m^2$ in NR which anyways does not exist, this is not the case here.} So, the boundary term contributes nothing in this case.

Therefore, we can simply calculate that
\begin{equation} 
\begin{aligned}
    I_{\text{box}} &= -\frac{1}{8\pi} \int_{m^2/s}^1 dx \frac{1}{[sx-m^2(1+x)+m_0^2]^2} 
    \\ &=\frac{1}{8\pi} \frac{s-m^2}{(s-2m^2+m_0^2)(m^4-m_0^2s)} ,
\end{aligned}
\end{equation}
which gives for $m_0 \rightarrow 0$
\begin{align}
    \sigma^{\text{LM}}_{\text{box}} = \begin{dcases}
    +\frac{1}{16 \pi}\frac{g^4}{m^4} \frac{1}{s-2m^2} & s \neq 2m^2
    \\ +\frac{1}{16\pi} \frac{g^4}{m^4} \frac{1}{m_0^2} + \frac{1}{8\pi} \frac{g^4}{m^6} & s =2m^2
    \end{dcases}.
\end{align}
\paragraph{Small Mass Regulator}
For the small mass regulator, we exclude the point $s =2m^2-m_0^2$. Then, we have $\eta_\pm = -1$.  One can also check that the intersection of $\mathcal{I}_{y_0}$ with $x \in [0, 1]$ is given by
\begin{align}
    m_0^2/m^2 \leq x \leq m^2/s \quad \text{ and } \quad x=1.
\end{align}
Note that we can immediately drop the $x=1$ contribution, since the integrand is non-singular there, and it is a region of measure zero. 

Meanwhile, for the $\delta$-functions in the boundary term, we can similarly drop the $\delta(1-x)$ since it cancels against the $1-x$ in the numerator. However, the $\delta(x-m_0^2/m^2)$ term does contribute, since $m_0^2/m^2 \in (0, 1)$. Thus, our expression is given by
\begin{multline}
    I_{\text{box}} = \frac{1}{8\pi} \int^{m^2/s}_{m_0^2/m^2} dx \frac{1}{[sx-m^2(1+x)+m_0^2]^2}  \\
    - \frac{1}{8\pi} \frac{m^4 -sm_0^2}{m^4-m^2 m_0^2} \int_0^1 dx \frac{1-x}{[sx-m^2(1+x)+m_0^2]^2} \delta(x-m_0^2/m^2),
\end{multline}
which evaluates straightforwardly to zero, giving a total cross-section
\begin{align}
    \sigma^{\text{SM}}_{\text{box}} = 0.
\end{align}
Importantly, observe that in their overlapping region of convergence, we have $\sigma^{\text{SM}}_{\text{box}} \neq \sigma^{\text{LM}}_{\text{box}}$, even though we have sent $m_0 \rightarrow 0$ in both cases, and even though both results are finite. This illustrates concretely the non-commutativity of the continuation $m_0 \rightarrow 0$ and the integration. Evidently, all of this is rooted in the non-analyticity of the boundary $\delta$-function, whose support only "switched on" in the small mass case. We will discuss the implications of this more in \cref{sec::discussion}.  
\paragraph{No Regulator}
\label{sec::boxnoreg}
For the no regulator case, we consider physical $s > m^2$, with the restriction that $s\neq 2m^2$, so that our resulting integral is well-defined. Following similar steps, we arrive at
\begin{align}
    I_{\text{box}} = \frac{1}{8\pi} \int_0^{m^2/s} dx \frac{1}{[sx-m^2(1+x)]^2} - \frac{1}{8\pi} \int_0^1 dx \frac{1-x}{[sx-m^2(1+x)]^2}\delta(x).
\end{align}
The boundary term now seems rather dubious; we have a $\delta(x)$ localized at the $x=0$ boundary, which in general is an ill-defined, representation-dependent object. However, following \cref{app::deltabdary}, we can understand this in terms of a limiting sequence of \textit{even} nascent $\delta$-functions constructed from the ${i\epsilon_F}$, in which case the boundary localization requires us to include an additional factor of $1/2$. Thus, we have
\begin{align}
    I_{\text{box}} = \frac{1}{8\pi} \left[\frac{1}{m^4}-\frac{1}{2} \frac{1}{m^4}\right] ,
\end{align}
which gives
\begin{align}
    \sigma^{\text{NR}}_{\text{box}} = \frac{1}{32\pi}\frac{g^4}{m^4} \frac{1}{s-m^2}.
\end{align}
Observe, then, that taking care with the boundary $\delta$-functions can drastically modify the structure of the final result. 

\subsubsection{Dimensional Regularization}
\label{sec::loopdimreg}
In dimensional regularization, the loop calculation plays out a little differently, since we cannot use the identity \cref{eq::loopderivdelta}.\footnote{There is an analogous identity \cite{Britto:2025wzt} which can be used directly 
\begin{align}
    \lim_{\epsilon_F \rightarrow 0^+}\frac{1}{(\Delta -{i\epsilon_F})^\lambda} - \frac{1}{(\Delta + {i\epsilon_F})^\lambda} = -\frac{\theta(-\Delta)}{(-\Delta)^\lambda} 2i \sin{\pi \lambda} \quad \lambda \notin \mathbb{Z}_+, 
\end{align}
where the limit should be understood as being taken \textit{after} integration has been performed. However, our method is essentially equivalent to proving this identity.} Let us focus on what happens off the pole $s \neq 2m^2$ with $s>m^2$, since the pole behavior follows similarly. Defining $s = \kappa m^2$ and pulling out overall constants, the loop integral of \cref{eq::t0integral} with $m_0=0$ and $\epsilon < -1$ takes the form
\begin{align}
\label{eq::Iexpdimreg}
    I_{\text{box}} = \frac{1}{16\pi^2} \frac{1}{m^4}[iK + c.c.]
\end{align}
with
\begin{align}
    K\equiv \int_{0}^1dx \int_0^{1-x}dy\frac{y}{[\kappa x(-1+x+y) - (-1+x+y+xy)-{i\epsilon_F}]^{2+\epsilon}}.
\end{align}
The basic technique (see \cite{Jones:2025jzc} for a more general exposition) is to split up the integral into two regions, $R_+$ and $R_-$, which correspond to the sign of the denominator. In $R_+$, where the denominator is positive, the ${i\epsilon_F}$ can be dropped altogether, while in $R_-$, where the denominator is negative, one can make the manipulation
\begin{equation} 
\begin{aligned}
    \frac{1}{[\Delta_0-{i\epsilon_F}]^{2+\epsilon}} &\stackrel{\epsilon_F \rightarrow 0^+}{=} \frac{1}{(-\Delta_0)^{2+\epsilon}} \times (-1-{i\epsilon_F})^{-2-\epsilon}
    \\ &\stackrel{\epsilon_F \rightarrow 0^+}{=}\frac{1}{(-\Delta_0)^{2+\epsilon}} e^{i\pi \epsilon } ,
\end{aligned}
\end{equation}
where in the first line we redefined ${i\epsilon_F} \rightarrow (-\Delta_0){i\epsilon_F}$ since $\Delta_0 < 0$ in $R_-$, and in the second line we used the fact that $(-1-{i\epsilon_F})^{-2-\epsilon} = \exp[(-2-\epsilon)\ln (-1-{i\epsilon_F})]$, with $\ln(-1-{i\epsilon_F}) \stackrel{\epsilon_F \rightarrow 0^+} = -i\pi$.

There is of course also the question of what happens in the measure-zero region where the denominator vanishes. However, measure-zero regions can be dropped when the integrand is integrable over those regions, which is precisely what taking $\epsilon <-1$ allows us to do. So, we need only focus on $R_\pm$. 

In fact, since $R_+$ is the integral of a real function over a real domain, it cannot contribute anything imaginary to the relevant $iK + c.c.$ factor, and so the only important contribution is the integral over $R_-$. Thus, dropping the irrelevant contributions, we have
\begin{align}
    K=e^{i\pi\epsilon}\int_{R_-}dx\,dy\;\frac{y}{[\kappa x(1-x-y) + (-1+x+y+xy)]^{2+\epsilon}}, 
\end{align}
where a simple computation shows that
\begin{align}
    \int_{R_-} dx\, dy \equiv \int_0^{1/\kappa} dx\int_{y_0}^{1-x}dy + \int_{1/\kappa}^{1}dx\int_0^{1-x}dy
\end{align}
with the integration boundary $y_0$ defined by
\begin{align}
    y_0 \equiv \frac{-1+x+x\kappa -x^2\kappa}{-1-x+x\kappa}, 
\end{align}
where the denominator vanishes. Integrating explicitly over $y$ and rearranging slightly, this becomes
\begin{multline}
    K = \frac{e^{i\pi\epsilon}}{(1+\epsilon)}\frac{1}{\epsilon}\left\{ 
    \int_{1/\kappa}^1dx\left[ \frac{ (x\kappa -1)^{-\epsilon}(1-x)^{-\epsilon}  }{(1+x-x\kappa)^2}\right]-\int_0^{1}dx \left[ \frac{ x^{-\epsilon}(1-x)^{-\epsilon} }{(1+x-x\kappa)^2}\right]\right\}
    \\ -\frac{e^{i\pi\epsilon}}{(1+\epsilon)}\left\{\int_0^1 dx\left[ \frac{ x^{-\epsilon-1}(1-x)^{-\epsilon} }{(1+x-x\kappa)}\right]\right\}.
\end{multline}
This can be integrated directly to yield the final result, but we find it illustrative to manipulate our expression slightly more so that direct comparison can be made to the other regularization schemes. First, since we eventually intend to take $\epsilon \rightarrow 0$, observe that
\begin{align}
     e^{i\pi\epsilon} = 1+i\pi \epsilon + \mathcal{O}(\epsilon^2)
\end{align}
and moreover since we only care about the imaginary part which contributes to $iK + c.c.$, we can write
\begin{multline}
    K = i\pi \left\{\int_{1/\kappa}^1 \frac{ (x\kappa -1)^{-\epsilon}(1-x)^{-\epsilon}   }{(1+x-x\kappa)^2} - \int_0^{1}dx  \left[\frac{  x^{-\epsilon}(1-x)^{-\epsilon}  }{(1+x-x\kappa)^2}\right]\right\}
    \\ -i\pi \epsilon \int_0^{1}dx\left[ \frac{  x^{-\epsilon-1}(1-x)^{-\epsilon} }{(1+x-x\kappa)}\right]  +\text{real} + \mathcal{O}(\epsilon).
\end{multline}
The terms on the first line converge nicely as $\epsilon \rightarrow 0^-$, so as with our discussion on the $t$-channel diagram, the dominated convergence theorem allows us to set $\epsilon = 0$ inside the integral to recover the $\epsilon \rightarrow 0^-$ limit. Importantly, however, we have not dropped the term on the second line, even though naively it appears to be $\mathcal{O}(\epsilon)$. This is because the $x^{-\epsilon -1}$ term inside the integral renders it divergent when one sends $\epsilon \rightarrow 0^-$ -- the convergence is no longer dominated by an integrable function, so to speak -- and so we cannot set $\epsilon = 0$ before performing the integration.

However, observe that this term actually forms a representation of the $\delta$-function 
\begin{align}\label{eq:deltaDim}
    \delta_{\text{dim}}(x) = \lim_{\epsilon \rightarrow 0^-} - \frac{1}{2} \epsilon \abs{x}^{-\epsilon -1}, 
\end{align}
which we discuss in more detail in \cref{app::deltaDim}. What this means is that we can understand this term as a boundary contribution, totally analogous to the boundary contribution we found using the other regulator methods! 

Thus, in the limit of $\epsilon \rightarrow 0$, and restoring the appropriate factors given by \cref{eq::Iexpdimreg} for $I_\text{box}$, we get (after combining the integration regions of common terms)
\begin{align}
    I_{\text{box}} = \frac{1}{8\pi} \left[\int_0^{m^2/s} dx\, \frac{1}{[sx-m^2(1+x)]^2} -\frac{2}{m^2} \int_0^1dx\, \frac{ \delta_{\text{dim}}(x) }{[m^2(1+x)-sx]}\right],
\end{align}
which has a very similar structure to the $d=4$ case.
Evaluating this, and being careful to instate an additional factor of $1/2$ as per \cref{app::deltaDim}, since we have a $\delta_{\text{dim}}(x)$ localized at the boundary of integration (and we constructed it as a sequence of even functions), the integral vanishes 
and thus
\begin{align}
    \sigma^{\text{DR}}_{\text{box}} = 0.
\end{align}
A very similar calculation can be done at $s=2m^2$, which gives $0$ also.

Thus, the lesson from the box calculation is that one must be careful with the boundaries of integration: both to actually include boundary terms (either of the form $\epsilon \times \text{divergent}$ in dimensional regularization, or those arising from integration by parts of $\delta'$ as in the mass/no-regulator cases); and to keep track of any potential factors of $1/2$ coming from boundary-localization.

\subsection{The 3-body Diagram}
\label{sec::3body}

\begin{figure}[t]\centering 

\includegraphics[width=0.4\linewidth]{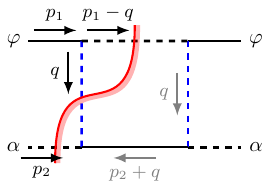}
\caption{The $3$-body cut of the box diagram, to be summed with its complex conjugate. Everything to the right of the cut is to be complex-conjugated. The dashed lines in blue are given the mass $m_0$ in the LM and SM regulator schemes. The momentum labels in black correspond to the unconjugated side of the cut, while the momentum labels in gray correspond to the conjugated side.}
\label{fig:3body}
\end{figure}

Once again, we take
\begin{align}
    \sigma_3 = \frac{g^4}{2(s-m^2)}I_3 ,
\end{align}
where
\begin{align}
    I_3(s; d,m_0) =\mu^{2\epsilon} \int \frac{d^dq}{(2\pi)^{d}} 2\pi\delta^+((p_1-q)^2)\frac{1}{(p_2+q)^2-m^2 -{i\epsilon_F}} \frac{2\pi\delta^+(q^2-m_0^2)}{q^2-m_0^2+{i\epsilon_F}}  + c.c.
\end{align}
with the cut and choice of momenta shown in \cref{fig:3body}.
Note that in the large mass case, this diagram immediately vanishes by momentum conservation, so we will only consider $m_0$ in the small mass region.

To understand the rather unintuitive $\delta(x)/(x+{i\epsilon_F})$ object appearing in this expression, let us simplify the $+c.c.$; using the $\epsilon_F$-representation of \cref{app::iepsrep} for the first propagator, one has
\begin{multline}
    \frac{1}{(p_2+q)^2-m^2 -{i\epsilon_F}} \frac{\delta(q^2-m_0^2)}{q^2-m_0^2+{i\epsilon_F}}  + c.c. = \left[\frac{(p_2+q)^2-m^2 }{[(p_2+q)^2-m^2]^2+\epsilon_F^2}\right] \left[\frac{\delta(q^2-m_0^2)}{q^2-m_0^2+i\epsilon_F}+c.c.\right]\\-2\pi^2 \delta((p_2+q)^2-m^2)[\delta(q^2-m_0^2)]^2.
\end{multline}
The remaining $+c.c.$ in the first term
can be understood in terms of the distributional identity outlined in \cref{app::deltaderiv}
\begin{align}
    \lim_{\epsilon_F \rightarrow 0^+}\frac{\delta(q^2-m_0^2)}{q^2-m_0^2 + {i\epsilon_F}} + c.c. = -\delta'(q^2-m_0^2)   = -\delta'(q_0^2 - \omega_q^2) 
\end{align}
 with $\omega_q \equiv \sqrt{\abs{\vec{q}}^2 +m_0^2}$, where as always, the limit $\epsilon_F \rightarrow 0^+$ should be understood as being performed \textit{after} the integration.
 
 Meanwhile, the second term consists of (twice) the problematic square of a $\delta$-function, which is indeed exactly the same as what we found for our $t$-channel amplitude in \cref{sec::tchannel}. Indeed, changing variables in this second term from $q \rightarrow p_1-q$, one can straightforwardly show that the contribution to $\sigma_3$ cancels precisely against the divergent $\Xi$ as defined for our $t$-channel amplitude in \cref{eq::squaredeltat}. In this way, our \textit{inclusive} cross-section is once-again well defined across different regularization schemes.  

Thus, dropping the divergent $\Xi$ contribution for now, we are left with 
\begin{align}
\label{eq::pvterm}
    I_3 = -(2\pi)^2\int \frac{d^dq}{(2 \pi)^d} \delta^+((p_1-q)^2)\mathcal{P}\left[\frac{1}{(p_2+q)^2-m^2}\right] \delta'(q^2-m_0^2). 
\end{align}
where we define our principal value as
\begin{align}
\mathcal{P}\left[\frac{1}{(p_2+q)^2-m^2}\right] \equiv  \left[\frac{(p_2+q)^2-m^2 }{[(p_2+q)^2-m^2]^2+\epsilon_F^2}\right]
\end{align}
and will omit it from future expressions for visual clarity. Then, rearranging, we get
\begin{align}
    I_3 = -\frac{1}{2\pi}\int_{0}^{(m^2+s)/(2\sqrt{s})} d{q_0}\int_0^\infty d\abs{\vec{q}}\abs{\vec{q}}^{2-2\epsilon} \int_{-1}^1 dz(1-z^2)^{-\epsilon} \frac{\delta((p_1-q)^2)}{(p_2+q)^2-m^2} \delta'({q_0^2}-\omega_q^2), 
\end{align}
where we used \cref{eq::dimregmeasure} with $z = \vec q \cdot \vec p_1 /( \abs{\vec q} \abs{\vec p_1})$, and have explicitly written out our integration measure with $\epsilon  =0$ in the prefactors out the front. The bounds on the ${q_0}$ integral come from enforcing energy-positivity of the cut lines $q_0 \geq 0, E_1 -q_0 \geq 0$ with the $\theta$-functions. 

The next step is to evaluate the angular integral with our $\delta((p_1-q)^2)$ integral. A simple calculation shows that
\begin{align}
    \delta((p_1-q)^2) &= \frac{1}{\abs{\vec{q}}}\frac{\sqrt{s}}{s-m^2}\delta(z-z_0)
\end{align}
with
\begin{align}
\label{eq::z0def}
    z_0 \equiv \frac{m^2 \sqrt{s}-m^2 {q_0}-s {q_0}+\sqrt{s} {q_0^2}-\sqrt{s} \abs{\vec{q}}^2}{\abs{\vec{q}} \left(m^2-s\right)}.
\end{align}
With this choice of $z_0$, we can simplify
\begin{align}
    (p_2+q)^2-m^2 = \frac{2m^2}{\sqrt{s}} ({q_0}s/m^2-\sqrt{s}),
\end{align}
in which case putting it all together gives
\begin{align}
    I_3 = -\frac{1}{4\pi} \frac{1}{m^2}\frac{s}{s-m^2}\int_{0}^{(m^2+s)/(2\sqrt{s})} d{q_0} \int_0^\infty d\abs{\vec{q}}\abs{\vec{q}}^{1-2\epsilon} \int_{-1}^1 dz(1-z^2)^{-\epsilon} \frac{\delta(z-z_0)}{{q_0}s/m^2-\sqrt{s}} \delta'({q_0^2}-\omega_q^2) .
\end{align}
Now, in order for $\delta(z-z_0)$ to be satisfied, we need $z_0 \in (-1, 1)$. With appropriate regularization from either the implicit principal value c.f.~\cref{eq::pvterm} or the SM, LM and DR regulators, any potential measure-zero regions (such as those associated with $z_0 = \pm 1$, for instance) can be ignored, since the integrand is non-singular there. This produces new bounds on our ${q_0}, \abs{\vec{q}}$ integrals, given by $R_1 \cup R_2$, with 
\begin{align}
    R_1 &\equiv \int_0^{m^2/\sqrt{s}}d{q_0} \int_{m^2/\sqrt{s}-{q_0}}^{\sqrt{s}-{q_0}}d\abs{\vec{q}} , 
    & R_2 &\equiv \int_{m^2/\sqrt{s}}^{(m^2+s)/(2\sqrt{s})}d{q_0} \int_{{q_0}-m^2/\sqrt{s}}^{\sqrt{s}-{q_0}} d\abs{\vec{q}}.
\end{align}
Let us take a moment to emphasize that these bounds treat $q_0$ and $\omega_q$ as independent variables, since we have not yet fixed them to be equal with the $\delta$-function.\footnote{Of course, any regions outside the support of $q_0 = \omega_q$ will eventually end up contributing nothing to the final result, but this way of notating the calculation makes the boundary contributions highly explicit. One might alternatively use indicator functions to enforce the appropriate regions of integration, for example, and be careful to act upon them when performing integration by parts.} This is vital. Had we simply set $q_0 =\omega_q$ when determining boundaries of integration, we would miss an important boundary contribution arising from integrating the $\delta'(q_0-\omega_q)$ by parts. 
With these bounds, then, our integral becomes
\begin{align}
    I_3 = -\frac{1}{4\pi} \frac{1}{m^2} \frac{s}{s-m^2}\int_{R_1 \cup R_2}d{q_0} d\abs{\vec{q}} \abs{\vec{q}}^{1-2\epsilon} (1-z_0^2)^{-\epsilon} \frac{\delta'({q_0^2}-\omega_q^2)}{{q_0}s/m^2-\sqrt{s}}.
\end{align}
Next, in order to simplify the derivative of the delta function, we can write
\begin{equation} 
\begin{aligned}
    \delta'({q_0^2}-\omega_q^2) &= \left(\pdv{{(q_0^2}-\omega_q^2)}{\omega_q}\right)^{-1} \partial_{\omega_q} \delta({q_0^2}-\omega_q^2)
    \\ &= -\frac{1}{2\omega_q } \partial_{\omega_q}\frac{1}{2\abs{{q_0}}}\left[\delta({q_0}-\omega_q) + \delta({q_0}+\omega_q)\right].
\end{aligned}
\end{equation}
We can drop the negative root, because for the mass regulators $q_0 \geq 0$ and $\omega_q >0$ inside this region, and for dimensional regularization and the no-regulator case the discussion in \cref{app::delta2theta2} applies when $\omega_q \rightarrow 0$. Thus we get 
\begin{align}\label{eq:I3}
    I_3 = \frac{1}{16\pi}\frac{1}{m^2} \frac{s}{s-m^2}\int_{R_1 \cup R_2} d{q_0} d\abs{\vec{q}} \frac{\abs{\vec{q}}^{1-2\epsilon}}{\omega_q} (1-z_0^2)^{-\epsilon} \frac{\partial_{\omega_q} \delta({q_0}-\omega_q)}{{q_0}({q_0}s/m^2 -\sqrt{s})}. 
\end{align}
Here, the details of the calculation change, depending on which regulator one takes.

\subsubsection{Mass Regulators}
For a large mass regulator, this diagram vanishes by momentum conservation, so we focus only on the small regulator case with $s \neq 2m^2-m_0^2$.

Setting $\epsilon = 0$, we  change variables in our integration to $\omega_q = \sqrt{\abs{\vec{q}}^2+m_0^2}$, in which case, selecting out the positive, physical contribution,  $R_1 \cup R_2$ combines into a single region, $R_{12}$, given by
\begin{align}
     R_{12} &\equiv \int_{0}^{(m^2+s)/(2\sqrt{s})}d{q_0} \int_{\omega_1({q_0})}^{\omega_2({q_0})} d\omega_q,
\end{align}
where we define
\begin{align}
    \omega_1({q_0}) &= \sqrt{(m^2/\sqrt{s}-{q_0})^2+m_0^2}, 
    & 
    \omega_2({q_0}) &= \sqrt{(\sqrt{s}-{q_0})^2+m_0^2}.
\end{align}
With this, our integral becomes
\begin{align}
    I_3 = \frac{1}{16\pi} \frac{1}{m^2} \frac{s}{s-m^2} \int_0^{(m^2+s)/(2\sqrt{s})} d{q_0} \int_{\omega_1({q_0})}^{\omega_2({q_0})}d\omega_q \frac{\partial_{\omega_q} \delta({q_0}-\omega_q)}{{q_0}({q_0}s/m^2-\sqrt{s})} .
\end{align}
Under an integration by parts, the only term which survives is the boundary term, since there is no other $\omega_q$-dependence in the integrand. Thus,
\begin{align}
    I_3 = \frac{1}{16\pi} \frac{1}{m^2} \frac{s}{s-m^2} \left[\int_0^{(m^2+s)/(2\sqrt{s})} d{q_0} \frac{\delta({q_0}-\omega_2({q_0}))-\delta({q_0}-\omega_1({q_0}))}{{q_0}({q_0}s/m^2-\sqrt{s})}\right].
\end{align}
Dropping contributions outside this region of integration, the $\delta$-functions simplify immediately to
\begin{align}
    \delta({q_0} - \omega_1({q_0})) &= \left[\frac{1}{2} + \frac{m_0^2 s}{2m^4}\right]\frac{2m^4}{m^4 + m_0^2s}\delta\left({q_0}-\left[\frac{m^2}{2\sqrt{s}} + \frac{m_0^2}{m^2} \frac{\sqrt{s}}{2}\right]\right), 
    \\ \delta({q_0} - \omega_2({q_0})) &= \left[\frac{1}{2} + \frac{m_0^2}{2s}\right]\delta\left({q_0} - \left[\frac{\sqrt{s}}{2} + \frac{m_0^2}{2 \sqrt{s}}\right]\right).
\end{align}
Plugging this back into the integral and simplifying, we get
\begin{align}
    I = \frac{1}{8\pi} \frac{s-m^2}{(s-2m^2+m_0^2)(m^4-m_0^2s)}
\end{align}
which, after analytic continuation to $m_0 \rightarrow 0$, gives
\begin{align}\label{eq::sigma3SM}
    \sigma_3^{\text{SM}} = \frac{1}{16\pi} \frac{g^4}{m^4} \frac{1}{s-2m^2} - \Xi^{\text{SM}},  
\end{align}
where we have restored the divergent contribution involving a squared $\delta$-function as defined in \cref{eq::squaredeltat}.

\subsubsection{No Regulator}
In the case of no regulator, we have $m_0 = 0, \epsilon = 0$. With this, the integral \eqref{eq:I3} becomes
\begin{align} 
    I_3 = \frac{1}{16\pi}\frac{1}{m^2} \frac{s}{s-m^2}\int_{R_1 \cup R_2} d{q_0} d\omega_q \frac{\partial_{\omega_q} \delta({q_0}-\omega_q)}{{q_0}({q_0}s/m^2 -\sqrt{s})}. 
\end{align}
Once again, under an integration by parts, the only surviving term is the boundary term given by
\begin{align}
\label{eq::noregexp3bdy}
    I_3 = \frac{1}{16\pi}\frac{1}{m^2}\frac{s}{s-m^2} \left[\int_0^{m^2/\sqrt{s}} d{q_0} \frac{\delta(2{q_0} - \sqrt{s}) - \delta(2{q_0}-m^2/\sqrt{s})}{{q_0}({q_0}s/m^2-\sqrt{s})}\right],
\end{align}
where we have dropped the $\delta$-functions which have no support inside the integration region. This can be evaluated straightforwardly as
\begin{align}
    I_3 &=  \frac{1}{8\pi} \frac{1}{m^4} \frac{s-m^2}{s-2m^2},
\end{align}
in which case, after restoring the divergent contribution defined in \cref{eq::squaredeltat}, we get
\begin{align}
    \sigma_{3}^{\text{NR}} = \frac{g^4}{16\pi} \frac{1}{m^4} \frac{1}{(s-2m^2)}- \Xi^{\text{NR}}.
\end{align}
There are no additional factors of half, because the $\delta$-functions lie strictly on the interior of the region of ${q_0}$ integration.

\subsubsection{Dimensional Regularization}
In dimensional regularization, we set $m_0 =0$ in \cref{eq:I3}, in which case $\abs{\vec{q}} = \omega_q$, such that 
\begin{align}
    I_3 = \frac{1}{16\pi} \frac{1}{m^2} \frac{s}{s-m^2} \int_{R_1 \cup R_2} d{q_0} d\omega_q \omega_q^{-2\epsilon}(1-z_0^2)^{-\epsilon}  \frac{ \partial_{\omega_q} \delta({q_0}-\omega_q)}{{q_0}({q_0}s/m^2-\sqrt{s})}.
\end{align}
Once again, we need to do integration by parts; doing so, we get
\begin{align}
    I_3 = \frac{1}{16\pi} \frac{1}{m^2} \frac{s}{s-m^2} \left[K_3^{\text{bdary}}-K_3^{\text{bulk}}\right],
\end{align}
where we used that the denominator does not depend on $\omega_q$ to obtain
\begin{align}
K_3^{\text{bdary}} &\equiv \int_{R_1 \cup R_2}d{q_0} d\omega_q \frac{1}{{q_0}({q_0}s/m^2-\sqrt{s})} \partial_{\omega_q}\left[\omega_q^{-2\epsilon} (1-z_0^2)^{-\epsilon} \delta({q_0}- \omega_q) \right] ,
    \\K_3^{\text{bulk}} &\equiv \int_{R_1 \cup R_2}d{q_0} d\omega_q \frac{1}{{q_0}({q_0}s/m^2-\sqrt{s})} \partial_{\omega_q}\left[\omega_q^{-2\epsilon} (1-z_0^2)^{-\epsilon}\right] \delta(\omega_q-{q_0}) .
\end{align}
Let us start with the boundary term. As with the no-regulator case, we can drop the contribution from region $R_2$ and thus find
\begin{align}
    K_3^\text{bdary}  =\int_0^{m^2/\sqrt{s}} d{q_0} \left[\frac{\delta(2{q_0} - \sqrt{s}) - \delta(2{q_0}-m^2/\sqrt{s})}{{q_0}({q_0}s/m^2-\sqrt{s})}\right] \left[\omega_q^{-2\epsilon}(1-z_0^2)^{-\epsilon}\right]\lvert_{\omega_q={q_0}}.
\end{align}
Now, the $\epsilon$-dependent piece simplifies as
\begin{align}
   \left[\omega_q^{-2\epsilon}(1-z_0^2)^{-\epsilon}\right] \lvert_{\omega_q={q_0}} = \left[(2{q_0} - \sqrt{s})(m^2/\sqrt{s}-2{q_0})\right]^{-\epsilon} \left[\frac{m^2 s}{(s-m^2)^2}\right]^{-\epsilon} , 
\end{align}
which means that the integral takes the form\footnote{At the pole $s= 2m^2$, the integral is instead of the form $\int dx x^{-\epsilon -1} \delta(x)$, since the $[q_0s/m^2-\sqrt{s}]^{-1}$ denominator also contributes to the $[\sqrt{s}-2q_0]^{-\epsilon}$ term. However, with $\epsilon < -1$, the following discussion still holds.} $\int dx\, x^{-\epsilon}\, \delta(x)$ for each of the $\delta$-functions. This object appears rather problematic, demonstrating a severe non-analyticity in $\epsilon$: for $\epsilon < 0$, it vanishes; for $\epsilon = 0$, it is finite and non-vanishing; and for $\epsilon > 0$, it is infinite.

However, the question of what to do with integrals of this form is really a question of what order we take $\epsilon \rightarrow 0^-$ and ${i\epsilon_F} \rightarrow 0$, since there are essentially factors of $\epsilon_F$ contained inside the $\delta$-function as per \cref{app::iepsrep}. We have a definite answer in accordance to the prescription outlined in \cref{sec::practicaltakeaways}, in which we first perform the integration, then take $i\epsilon_F \rightarrow 0$, and only then finally analytically continue $\epsilon \rightarrow 0^-$. Thus, since $\epsilon$ is strictly negative while performing the integral and evaluating the $\delta-$function (which corresponds to sending $i\epsilon_F \rightarrow 0$), we should set such integrals to zero and take
\begin{align}
    K_3^\text{bdary} = 0.
\end{align}
Meanwhile, for $K_3^{\text{bulk}}$ we evaluate the $\omega_q$ integral using the $\delta$-function
\begin{align}
    K_3^{\text{bulk}} = \int_0^{(m^2+s)/(2\sqrt{s})} d{q_0}\frac{\partial_{\omega_q}\left[\omega_q^{-2\epsilon} (1-z_0^2)^{-\epsilon}\right]\lvert_{\omega_q = {q_0}}}{{q_0}({q_0}s/m^2-\sqrt{s})} \mathcal{I}_{R_1 \cup R_2},
\end{align}
where the indicator function $\mathcal{I}_{R_1 \cup R_2}$ enforces $\delta({q_0}-\omega_q)$ is satisfied inside the region $R_1 \cup R_2$. Explicitly, then, we have 
\begin{multline}
    K_3^{\text{bulk}} 
    =-\frac{2\epsilon}{s} \left(\frac{m^2 \sqrt{s}}{(s-m^2)^2}\right)^{-\epsilon} 
    \times\\ 
    \int_{m^2 /(2\sqrt{s})}^{\sqrt{s}/2} \frac{dq_0}{m}\frac{1+s^2/m^4 -2\sqrt{s}q_0(s/m^2+1)/m^2}{(q_0s/m^3-\sqrt{s}/m)} \left[(\sqrt{s}/m-2{q_0}/m)(2{q_0}/m-m/\sqrt{s})\right]^{-1-\epsilon} .
\end{multline}
As we saw with the loop in dimensional regularization in \cref{sec::loopdimreg}, we cannot simply set this to zero despite it being naively $\mathcal{O}(\epsilon)$, because the remaining integral is divergent for $\epsilon \rightarrow 0$. So, rewriting in terms of dimensionless quantities $s = \kappa m^2, {q_0} = mx$ and expanding the prefactor to $\mathcal{O}(\epsilon)$, we get
\begin{align}\label{eq:K3bulk}
    K_3^{\text{bulk}} = \frac{-2\epsilon}{s} \int_{1/(2\sqrt{\kappa})} ^{\sqrt{\kappa}/2} dx f_\kappa(x)\left[\left(\sqrt{\kappa }-2 x\right) \left(2 x-\frac{1}{\sqrt{\kappa }}\right)\right]^{-1-\epsilon }
\end{align}
with
\begin{align}
    f_\kappa(x) \equiv \frac{1 +\kappa^2-2 \sqrt{\kappa} x(\kappa+1) }{\kappa  x-\sqrt{\kappa }}.
\end{align}
Let us first consider the off-pole calculation for $s > m^2$ and $s \neq 2m^2$. To simplify the computation, we can again make use of the representation defined in \cref{app::deltaDim},
$    \delta_\text{dim}(x) = \lim_{\epsilon \rightarrow 0}   -\frac{1}{2} \epsilon \abs{x}^{-1-\epsilon}
$,
in which case we can write
\begin{align}
    K_3^{\text{bulk}} = \frac{4}{s}\int_{1/(2\sqrt{\kappa})}^{\sqrt{\kappa}/2} dx f_\kappa(x) \delta_\text{dim}\left[(\sqrt{\kappa}-2x)(2x-\frac{1}{\sqrt{\kappa}})\right].
\end{align}
This can be simplified with the usual identity
\begin{align}
\delta_\text{dim}\left[(\sqrt{\kappa}-2x)(2x-\frac{1}{\sqrt{\kappa}})\right] = \frac{\sqrt{\kappa}}{2(\kappa-1)}\left[\delta_{\text{dim}}\left(x-\frac{\sqrt{\kappa}}{2}\right) + \delta_{\text{dim}}\left(x-\frac{1}{2\sqrt{\kappa}}\right) \right].
\end{align}
Now, since both $\delta$-functions are localized at the boundaries of integration, we again need to add an additional factor of half following \cref{app::deltaDim}, since $\delta_{\text{dim}}$ is constructed as a limiting sequence of even functions. Thus, we have
\begin{align}
    K_3^{\text{bulk}} &= \frac{1}{s} \frac{\sqrt{\kappa}}{(\kappa -1)} \left[
    f_\kappa\left(\frac{\sqrt{\kappa}}{2}\right)
    +f_\kappa\left(\frac{1}{2\sqrt{\kappa}}\right)
    \right]
    = -\frac{2}{m^2 s} \frac{(s-m^2)^2}{s-2m^2},
\end{align}
in which case
\begin{align}
    I_3 = \frac{1}{8\pi} \frac{1}{m^4} \frac{s-m^2}{s-2m^2} .
\end{align}
Meanwhile, at the pole, let us substitute in $\kappa = 2$ ($s=2m^2$) in \cref{eq:K3bulk} before evaluating the integral, in which case we get
\begin{align}
    K_3^{\text{bulk}} = -\frac{\epsilon}{m^2} \int_{1/(2\sqrt{2})}^{1/\sqrt{2}} dx\, \left(-5 + 6\sqrt{2}x\right) \left[\sqrt{2}-2x\right]^{-2-\epsilon}\left[2x-\frac{1}{\sqrt{2}}\right]^{-1-\epsilon} , 
\end{align}
whereby we can no longer use the $\delta_{\text{dim}}$ representation, because the denominator of $f_\kappa(x)$ also contributes to the $\epsilon$-dependent exponential piece at $\kappa = \sqrt{2}$. Nonetheless, this integral is easy enough to evaluate directly, giving
\begin{align}
    K_3^{\text{bulk}}&=-\frac{\sqrt{\pi } 2^{3 \epsilon +1} \epsilon  (\epsilon +2) \Gamma (-\epsilon -1)}{m^2 \Gamma \left(\frac{1}{2}-\epsilon \right)}
    \;\;\stackrel{\epsilon \rightarrow 0}{=}\;\; 
    - \frac{4}{m^2} .
\end{align}
Thus, combining everything, we arrive for $ s> m^2$ at
\begin{align}\label{eq::sigma3DR}
    \sigma_3^{\text{DR}} = \begin{dcases}
        \frac{1}{16 \pi} \frac{g^4}{m^4} \frac{1}{s-2m^2} - \Xi^{\text{DR}}& s \neq 2m^2
        \\ \frac{1}{4\pi} \frac{g^4}{m^6} & s = 2m^2
    \end{dcases} 
\end{align}
where once again we have restored the divergent contribution involving the squared $\delta$-function as defined in \cref{eq::squaredeltat}.

We would like to point out the interesting role of bulk/boundary here. In a certain sense, for all the different regulators, one cannot get around (representations of) $\delta$-functions at the boundary; either they arise in the literal boundary contribution from an integration by parts of $\delta'({q_0^2}-\omega_q^2)$, or they arise as a boundary contribution coming from $\delta_{\text{dim}}$ in dimensional regularization. In all cases, a key contribution is somehow contained at the boundaries of integration.

\subsection{The 4-Body Diagram}
\label{sec::4body}

\begin{figure}[t]\centering 

\includegraphics[width=0.4\linewidth]{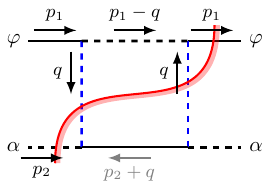}
\caption{The $4$-body cut of the box diagram, to be summed with its complex conjugate. Everything to the right of the cut is to be complex-conjugated. The dashed lines in blue are given the mass $m_0$ in the LM and SM regulator schemes. The momentum labels in black correspond to the unconjugated side of the cut, while the momentum label in gray corresponds to the conjugated side.}
\label{fig:4body}
\end{figure}

This diagram is the strangest diagram we will consider, because, kinematically, it is only allowed when the outgoing particles with momenta $q_1, q_2$ have identically zero energy. Moreover, as we shall argue, it vanishes for any $\epsilon, m_0 > 0$, but gives a non-vanishing contribution at $\epsilon, m_0 = 0$. Once again, this is where continuing $m_0, \epsilon \rightarrow 0$ \textit{after} the integration (and also after sending ${i\epsilon_F} \rightarrow 0$) matters; the limits do not commute.

Since both cases with large and small mass regulators clearly cause this diagram to be kinematically forbidden, we will only focus on dimensional regularization and the no regulator cases, which are more subtle.

As usual, we have
\begin{align}
    \sigma_4 \equiv \frac{g^4}{2(s-m^2)} I_4 , 
\end{align}
where
\begin{align}
    I_4 &\equiv \frac{1}{4\pi^2}\int d^dq\frac{1}{(p_1-q)^2 + {i\epsilon_F}} \frac{1}{(p_2 +q)^2 -m^2 - {i\epsilon_F}} \theta({{q_0}}) \delta(q^2)\theta(-{{q_0}})\delta(q^2) + c.c.
\end{align}
with the cut and choice of momentum shown in \cref{fig:4body}. In particular, observe that the $\theta(q_0)\theta(-q_0)$ comes from enforcing energy-positivity along the cut, since momentum-conservation at each vertex mandates that the two cut internal lines with momentum $q$ point in opposite directions, such that one comes with a $+q_0$ energy flow along the cut, and the other with a $-q_0$. Also, note that, as usual, we have set $d=4$ in the prefactors out the front. 

Evidently, this looks rather absurd. We have a square of a $\delta$-function integrated over a measure-zero region of phase space, $q_0 = \omega_q= 0$, permitted by the $\theta({q_0}) \theta(-{q_0})$. However, as we argue in \cref{app::delta2theta2}, in this particular setting it can be understood by going back to the position-space origins of the cutting rules; the result is that 
\begin{align}
[\delta(q^2)]^2\theta({q_0})\theta(-{q_0}) = \frac{\pi}{8} \delta^{4}(q) \quad \text{ in $d=4$}
\end{align}
and
\begin{align}
[\delta(q^2)]^2\theta({q_0})\theta(-{q_0}) = 0 \quad \text{ in $d>4$} , 
\end{align}
wherein the integrand is sufficiently suppressed by the measure factors. 

Thus, we find $\sigma_4^{\rm DR}=0$. As we outline in \cref{app::delta2theta2}, the vanishing of this diagram in $d>4$ is intimately related to the order of $\epsilon \rightarrow 0^-$ and $\epsilon_F \rightarrow 0^+$ limits; had we taken $\epsilon\rightarrow 0^-$ first, the diagram would not vanish, and we would not achieve a cancellation. 

Finally, choosing $f(q) = \frac{1}{4\pi^2} [(p_1-q)^2+i\epsilon_F]^{-1}[(p_2+q)^2-m^2-i\epsilon_F]^{-1} + c.c.$ and applying our identity, the only non-vanishing cross-section is $\sigma_4^{\text{NR}}$, with 
\begin{align}
    \sigma^{\text{NR}}_4 = -\frac{1}{32\pi} \frac{g^4}{m^4} \frac{1}{s-m^2}.
\end{align}
Let us emphasize here that, as we outline in \cref{app::delta2theta2}, this is a factor of $1/4$ off what one would get without accounting for the implicit boundary localization. 

\subsection{Canceling the Divergent \texorpdfstring{$s,t$}{s,t} Interference}
\label{sec::stinterference}
\begin{figure}[t]
\centering 

\begin{subfigure}{0.35\linewidth} 
\includegraphics[width=\linewidth]{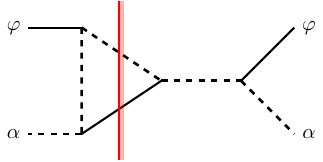}%
\caption{}
\end{subfigure}\hspace{3ex} 
\begin{subfigure}{0.35\linewidth}
\includegraphics[width=\linewidth]{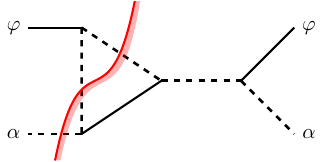}%
\caption{}
\end{subfigure}

\vspace{3ex}

\begin{subfigure}{0.35\linewidth}
\includegraphics[width=\linewidth]{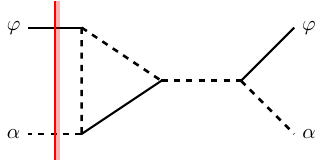}%
\caption{}
\end{subfigure}\hspace{3ex}
\begin{subfigure}{0.35\linewidth}
\includegraphics[width=\linewidth]{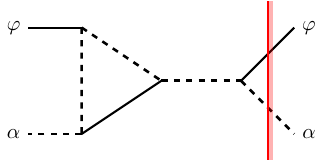}%
\caption{}
\end{subfigure}
\caption{KLN cuts of the underlying triangle diagram. The complex conjugates of (a) and (b) do not appear as cuts (and thus should not be included), while (c) and (d) are complex conjugates of one another. All other cuts not shown are forbidden by kinematics. Note that in addition to the diagrams shown above, there is a horizontally mirrored diagram for each diagram shown above.}
    \label{fig:mixed diagrams}
\end{figure}

As we mentioned in \cref{sec::themodel}, the $t$-channel divergence can also arise in its interferences with other scattering processes, even if those processes alone are finite. At tree-level, the only other relevant $\varphi \alpha \rightarrow \varphi \alpha$ process at the same order to consider is the $s$-channel scattering, which is finite on its own. In this section, then, we will illustrate that the divergent interference between the $s$ and $t$ channel diagrams shown in \cref{fig:mixed diagrams} (a) can also be canceled in accordance with the KLN theorem. 

Once again, we consider this interference as just one cut of an underlying triangle diagram, whose divergence must cancel against all the other cuts shown in \cref{fig:mixed diagrams}.  Since this is a simpler calculation with many similar steps to the previous section, we will be rather brief in this discussion and will only work in dimensional regularization, with $d = 4-2\epsilon$ and $\epsilon <0$. 

Finally, note that for each of the diagrams shown in \cref{fig:mixed diagrams}, there is an additional diagram which is obtained by mirroring the underlying diagram horizontally, which corresponds to complex conjugation.\footnote{We did not have to worry about this for the diagrams associated with the squared $t$-channel amplitude, because the underlying box diagram was symmetric under horizontal mirroring.} For instance, the diagram in \cref{fig:mixed diagrams} (a) corresponds to $\mathcal{M}_t \mathcal{M}_s^*$, while the mirrored diagram would correspond to $\mathcal{M}_t^* \mathcal{M}_s$; both of these contribute to $\abs{\mathcal{M}_s + \mathcal{M}_t}^2$. We will include the total real contribution $\mathcal{M}_t \mathcal{M}_s^* + \mathcal{M}_t^* \mathcal{M}_s$  in the final result in \cref{tab::mixedsummary}, but not in the following calculations.
\subsubsection{The \texorpdfstring{$s$, $t$}{s,t} Interference}
\label{sec::stdiagraminterference}
\begin{figure}[t]\centering 

\includegraphics[width=0.4\linewidth]{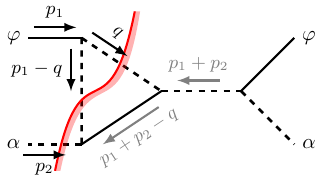}
\caption{The cut of the triangle diagram representing interference between the $s$ and $t$ channel processes. Everything to the right of the cut is to be complex-conjugated. The momentum labels in black correspond to the unconjugated side of the cut, while the momentum labels in gray correspond to the conjugated side. }
\label{fig:stcut}
\end{figure}
For this diagram, we can write
\begin{align}
    \sigma_{st} = \frac{g^4}{2(s-m^2)} \frac{1}{s}I_{st},
\end{align}
where we define 
\begin{align}
    I_{st} \equiv \mu^{2\epsilon} \int \frac{d^dq}{(2\pi)^d} \frac{1}{(p_1-q)^2+i\epsilon_F}  2\pi \delta^+(q^2) 2\pi \delta^+((p_1+p_2-q)^2-m^2)
\end{align}
with a factor of $1/s$ pulled out from the $[(p_1 + p_2)^2 - i\epsilon_F]^{-1}$ propagator. Momentum labels are shown in \cref{fig:stcut}. Following very similar steps to \cref{sec::tchannel}, this simplifies to
\begin{equation} 
\begin{aligned}
    I_{st}&= \frac{(4\pi)^{\epsilon}\mu^{2\epsilon}}{\Gamma(1-\epsilon)}\frac{1}{8\pi} \frac{1}{\sqrt{s}} \int_0^{\sqrt{s}} dq_0 q_0^{1-2\epsilon} \delta(q_0 -\frac{s-m^2}{2\sqrt{s}}) \int_{-1}^1 dz \frac{(1-z^2)^{-\epsilon}}{(m^2-2q_0(E_1-\abs{\vec{p_1}}z)) + i \epsilon_F}
    \\&=\left(\frac{16 \pi s \mu^2}{(s-m^2)^2} \right)^{\epsilon} \frac{1}{\Gamma(1-\epsilon)}\frac{s-m^2}{8\pi}  \int_{-1}^1 dz \frac{\left(1-z^2\right)^{-\epsilon }}{m^4 (z+1)-2 m^2 s (z-1)+s^2 (z-1) + i\epsilon_F},
\end{aligned}
\end{equation}
where we now keep track of the $\epsilon$-dependent prefactors out the front, as per \cref{eq::dimregmeasure}.

This integral can be evaluated at $s\in (m^2, 2m^2)$, $s=2m^2$ and $s >2m^2$ separately, and then expanded around $\epsilon = 0$; doing so for $s > m^2$ gives
\begin{align}
    \sigma_{st} = \begin{dcases}
        -\frac{1}{16 \pi} \frac{g^4}{s \left(s-m^2\right)^2} \log\left(\abs{\frac{(2m^2-s)s}{m^4}}\right) + \text{imag. part for $s>2m^2$} & s \neq 2m^2
        \\ -\frac{1}{32\pi}\frac{g^4}{m^6} \left[\frac{1}{\epsilon } + \log \left(\frac{8\pi e^{-\gamma_E}\mu^2}{m^2}\right) \right] & s =2m^2
    \end{dcases},
\end{align}
where we ignore the imaginary part arising from the $i\epsilon_F$ contour deformation for $s > 2m^2$, because it will cancel out in the final result summed with its complex conjugate.
\subsubsection{The Triangle Diagram}
\label{sec::triangle}
\begin{figure}[t]\centering 

\includegraphics[width=0.4\linewidth]{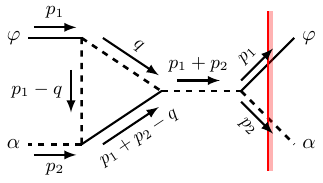}
\caption{The forward-scattering cut of the triangle diagram, to be summed with its complex conjugate.}
\label{fig:fwdstcut}
\end{figure}
Now we consider our triangle diagram, given by
\begin{align}
    \sigma_\Delta = \frac{g^4}{2(s-m^2)} \frac{1}{s} I_\Delta,
\end{align}
where
\begin{align}
    I_\Delta \equiv \mu^{2\epsilon}\int \frac{d^d q}{(2\pi)^d} \frac{1}{q^2+i\epsilon_F} \frac{1}{(p_1-q)^2+i\epsilon_F} \frac{1}{(p_1+p_2-q)^2-m^2 + i\epsilon_F} + c.c.
\end{align}
is the relevant triangle loop integral, with a factor of $1/s$ pulled out from the $[(p_1+p_2)^2+i\epsilon_F]^{-1}$ propagator. Momentum labels are shown in \cref{fig:fwdstcut}. Now, we can rewrite this in terms of the Symanzik representation outlined in \cref{app::graphtheory} as
\begin{align}
    I_\Delta = -\frac{i\mu^{2\epsilon} \pi^{d/2}}{(2\pi)^d} \int_{x,y,z \geq 0} \delta(1-x-y-z) \frac{1}{(\mathcal{F}-i\epsilon_F)^{1+\epsilon}} + c.c.,
\end{align}
where our second Symanzik polynomial is defined by
\begin{align}
    \mathcal{F} \equiv -xzs-yzm^2+ xm^2
\end{align}
and the first Symanzik polynomial $\mathcal{U}$ evaluates to $1$ after evaluating the $\delta$-function.
Explicitly choosing $x=1-y-z$, this integral becomes
\begin{align}
    I_\Delta = -\left(\frac{4\pi\mu^2}{m^2}\right)^{\epsilon}\frac{1}{16\pi^2}\frac{1}{m^2} K_\Delta ,
\end{align}
with
\begin{align}
    K_\Delta \equiv  i\int_0^1 dz \int_0^{1-z} dy \frac{1}{[\kappa z (y+z-1)- (y z+y+z-1)-i\epsilon_F]^{1+\epsilon}} + c.c.,
\end{align}
where we have defined $\kappa \equiv s/m^2$ to make the integral dimensionless. Since only regions where the denominator is negative contribute to the sum with $+c.c.$, we can isolate the relevant region and write
\begin{multline}
    K_\Delta \equiv \left[i(-1-i\epsilon_F)^{-1-\epsilon}+c.c.\right]
    \\ \times\left[\left(\int_0^{1/\kappa} dz\int_{y_0}^{1-z}dy+\int_{1/\kappa}^1dz \int_0^{1-z}  dy \right) \left(\frac{1}{[y (-\kappa  z+z+1)-(z-1) (\kappa  z-1)]^{1+\epsilon}}\right) \right],
\end{multline}
where we have factored out the $i\epsilon_F$ out the denominator in this region, and defined
\begin{align}
    y_0 \equiv \frac{(1-z) (\kappa  z-1)}{(\kappa -1) z-1}.
\end{align}
Let us consider the case $\kappa > 1$ and $\kappa \neq 2$ first. Then, to $\mathcal{O}(\epsilon)$, the prefactor can be expanded to 
\begin{align}
    \lim_{\epsilon_F \rightarrow 0^+}[i(-1-i\epsilon_F)^{-1-\epsilon} + c.c] = -ie^{i\pi \epsilon} + c.c. \stackrel{\epsilon \rightarrow 0}{=}2\pi \epsilon + \mathcal{O}(\epsilon^2),
\end{align}
in which case, integrating over $y$, our expression to $\mathcal{O}(\epsilon^0)$ becomes
\begin{multline}
    K_\Delta = -2\pi\int_0^{1/\kappa} dz \frac{1}{1+z-z\kappa}z^{-\epsilon}(1-z)^{-\epsilon}
    \\ +2\pi \int_{1/\kappa}^1dz \left[\frac{1}{1+z-z\kappa}(1-z)^{-\epsilon}(z \kappa-1)^{-\epsilon}-\frac{1}{1+z-z\kappa}(1-z)^{-\epsilon} z^{-\epsilon }\right].
\end{multline}
Since the $\epsilon \rightarrow 0^-$ limit of these integrands converges pointwise to an integrable function $[1+z-z\kappa]^{-1}$ (except at certain measure-zero boundaries where it vanishes, which can be ignored), the limit $\epsilon \rightarrow 0^-$ and the integration can be exchanged, and our expression becomes
\begin{align}
    K_\Delta &= - 2\pi \int_0^{1/\kappa} \frac{1}{1+z-z\kappa}
    = -2\pi \frac{\log \kappa}{\kappa -1}.
\end{align}
At $s=2m^2$ ($\kappa=2$) the integral is easy enough to do directly, retaining the full dimensional dependence and expanding in $\epsilon$ at the end. Collecting results and restoring prefactors, one arrives for $s >m^2$ at
\begin{align}
    \sigma_{\Delta} = \begin{dcases}
        \frac{1}{16\pi} \frac{g^4}{s(s-m^2)^2} \log \left(\frac{s}{m^2}\right)& s \neq 2m^2
        \\ \frac{1}{32\pi} \frac{g^4}{m^6} \log 2 & s =2m^2
    \end{dcases}.
\end{align}
Observe that the finite result at $s=2m^2$ is consistent with the extension of the $s \neq 2m^2$ result. 
\subsubsection{The Triple Emission}
\label{sec::tripleemission}
\begin{figure}[t]\centering 

\includegraphics[width=0.4\linewidth]{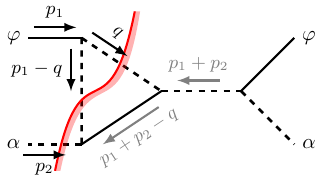}
\caption{The triple-emission cut of the triangle diagram. Everything to the right of the cut is to be complex-conjugated. The momentum labels in black correspond to the unconjugated side of the cut, while the momentum labels in gray correspond to the conjugated side.}
\label{fig:tripcut}
\end{figure}
Here, we define
\begin{align}
    \sigma_T = \frac{g^4}{2(s-m^2)}\frac{1}{s} I_T,
\end{align}
where
\begin{align}
    I_T \equiv \mu^{2\epsilon} \int \frac{d^d q}{(2\pi)^d} \frac{1}{(p_1+p_2-q)^2-m^2-i\epsilon_F} 2\pi \delta^+(q^2)2\pi\delta^+((p_1-q)^2) 
\end{align}
is the relevant cut integral, with a factor of $1/s$ pulled out from the $[(p_1+p_2)^2-i\epsilon_F]^{-1}$ propagator. Momentum labels are shown in \cref{fig:tripcut}. Plugging in the measure of \cref{eq::dimregmeasure} and evaluating the $\delta$-functions in a similar way as outlined \cref{sec::3body}, we arrive at
\begin{align}
    I_T = \frac{\left(4\pi\mu^{2}\right)^{\epsilon}}{4\pi\Gamma(1-\epsilon)}\frac{\sqrt{s}}{s-m^2} \int_0^{E_1} dq_0q_0^{-2\epsilon} \frac{1}{s-2\sqrt{s}q_0-m^2-i\epsilon_F}\int_{-1}^1 dz  (1-z^2)^{-\epsilon}\delta(z-z_0)
\end{align}
with
\begin{align}
    z_0 \equiv \frac{m^2 \left(\sqrt{s}-q_0\right)-q_0 s}{q_0 \left(m^2-s\right)}.
\end{align}
Evaluating the $\delta(z-z_0)$ (and being careful to adjust the bounds of the $q_0$ integral accordingly), this becomes
\begin{align}
    I_T=\left(\frac{4\pi(s-m^2)^2}{\sqrt{s}} \frac{\mu^2}{m^2}\right)^{\epsilon}\frac{1}{4\pi\Gamma(1-\epsilon)}\frac{\sqrt{s}}{s-m^2} \int_{\frac{m^2}{2\sqrt{s}}}^{\frac{\sqrt{s}}{2}} dq_0 \frac{(\sqrt{s}-2q_0)^{-\epsilon}(2q_0\sqrt{s}-m^2)^{-\epsilon}}{s-2\sqrt{s}q_0-m^2-i\epsilon_F},
\end{align}
which evaluates to
\begin{align}
    \sigma_{T} = \begin{dcases}
        \frac{1}{16\pi} \frac{g^4}{s(s-m^2)^2} \log \left(\abs{\frac{2m^2-s}{m^2}}\right) + \text{imag. part for $s>2m^2$}& s\neq 2m^2 
        \\ \frac{1}{32\pi} \frac{g^4}{m^6} \left[\frac{1}{\epsilon}+\log\left(\frac{4\pi \mu^2 e^{-\gamma_E}}{m^2}\right) \right] & s =2m^2
    \end{dcases},
\end{align}
where once again we ignore the imaginary part arising from the $i\epsilon_F$ contour deformation for $s > 2m^2$, because it will cancel out in the final result summed with its complex conjugate.

\subsection{Summary of Results}

\begin{table}[tbp!]
    \centering\def\arraystretch{1.5}
    \begin{tabular}{c c c c c }\toprule
         & LM  & SM  & DR  & NR \\\midrule
        $\bar{\sigma}_t$ & $-[s-2m^2]^{-1}$  & $-[s-2m^2]^{-1}$  & $-[s-2m^2]^{-1}$  & $-[s-2m^2]^{-1}$ \\
         $\bar{\sigma}_\text{box}$ & $+[s-2m^2]^{-1}$   & $0$  & $0$  & $ +\frac{1}{2}[s-m^2]^{-1}$\\
         $\bar{\sigma}_3$ & $0$  & $+[s-2m^2]^{-1}$  & $+[s-2m^2]^{-1}$  & $+[s-2m^2]^{-1}$ \\
         $\bar{\sigma}_4$ &  $0$ & $0$  & $0$  & $ -\frac{1}{2}[s-m^2]^{-1}$ \\ \bottomrule
    \end{tabular}
    \caption{Summary of the finite parts of squared $t-$channel cancellations off the pole for $s > m^2$ and $s \neq 2m^2$ with a large mass regulator $m_0 > m$ (LM), small mass regulator $0<m_0<m$ (SM), dimensional regularization (DR) and no regulator (NR). In each column, all results sum to zero, as expected by the KLN theorem. Results are written in terms of $\bar{\sigma}$ defined by $\text{finite}( \sigma)  \equiv \frac{1}{16\pi} \frac{g^4}{m^4} \bar{\sigma}$. Note that $\sigma_t$ and $\sigma_3$ have an additional squared $\delta$-function divergent contribution for $s > 2m^2$ which also cancels. See Eqs.~\eqref{eq::fulltchannelexppole}, \eqref{eq::squaredeltat}, \eqref{eq::sigma3SM}, and \eqref{eq::sigma3DR}.} 
    \label{tab::summaryoffpole}
\end{table}

\begin{table}[tbp!]
    \centering\def\arraystretch{1.5}
    \begin{tabular}{c c c }\toprule
         & LM & DR   \\\midrule
        $\bar{\sigma}_t$ &$-(1/m_0^2 + 2/m^2)$  &  $-4/m^2$  \\
         $\bar{\sigma}_{\text{box}}$& $+(1/m_0^2+2/m^2)$ &  $0$ \\
         $\bar{\sigma}_3$& $0$  &  $+4/m^2$ \\
         $\bar{\sigma}_4$ & $0$ & $0$  \\ \bottomrule
    \end{tabular}
    \caption{Summary of squared $t-$channel cancellations at the pole $s=2m^2$ with a large mass regulator $m_0 > m$ (LM) and dimensional regularization (DR). In each column, all results sum to zero, as expected by the KLN theorem. Results are written in terms of $\bar{\sigma}$ defined by $\sigma \equiv \frac{1}{16\pi} \frac{g^4}{m^4} \bar{\sigma}$.}
    \label{tab::summaryonpole}
\end{table}

\begin{table}[tbp!]
    \centering\def\arraystretch{1.5}
    \begin{tabular}{c c c }\toprule
         & $s \neq 2m^2$ & $s=2m^2$\\\midrule
        $\hat{\sigma}_{st}$ &$-\log\mid2-s/m^2\mid - \log \left(s/m^2\right)$  &  $-1/\epsilon - \log(2 \bar{\mu})$  \\
         $\hat{\sigma}_{\Delta}$& $\log\left(s/m^2\right)$ &  $\log 2$ \\
         $\hat{\sigma}_T$& $\log\mid2-s/m^2\mid$  &  $1/\epsilon + \log \bar{\mu}$ \\\bottomrule
    \end{tabular}
    \caption{Summary of mixed $s, t$-channel cancellations off the pole for $s > m^2$ and $s \neq 2m^2$ in the first column, and on the pole for $s=2m^2$ in the second column. In both columns, all results sum to zero, as expected by the KLN theorem. Results are written in terms of $\hat{\sigma}$ defined by $\sigma \equiv \frac{1}{8\pi}
    \frac{g^4}{s(s-m^2)^2} \hat{\sigma}$, and $\bar{\mu} \equiv 4\pi \mu^2 e^{-\gamma_E}/m^2$. All calculations are done in dimensional regularization.}    \label{tab::mixedsummary}
\end{table}

For the $t$-channel divergence in the squared amplitude, a summary of our cancellations off the pole for $s > m^2$ and $s \neq 2m^2$ is shown in \cref{tab::summaryoffpole}, and on the pole at $s=2m^2$ in \cref{tab::summaryonpole}.  
For the mixed $s- t$ divergence, a summary of our cancellations is shown both on and off the pole in  \cref{tab::mixedsummary}, which includes also the horizontally mirrored diagrams associated with the conjugate process.  
We were able to achieve the cancellation in all cases.

\section{Constructing an Observable}
\label{sec::inclusiveobservable}
In this section, we will make use of the cancellations we have just demonstrated to schematically outline how one might construct an inclusive quantity more in line with a real collider observable. We will explore some of the subtleties of what is and is not physically-degenerate when unstable particles are involved, and discuss issues of finiteness, regulator independence and non-negativity.

\subsection{All Cross-Sections Sum to Zero}
\label{sec::physint}
It may initially appear somewhat contradictory that all non-trivial cross-sections sum to \textit{zero} in the KLN prescription of \cref{eq::klnschwartz}, given that we evidently observe non-zero scatterings in experiments. However, one way to understand this is to recall that cross-sections should only be summed together inclusively in the regions of phase space where they correspond to degenerate or experimentally-indistinguishable processes. Outside these regions, one should treat non-degenerate processes independently. 

The textbook example \cite{Schwartz:2014sze} is the indistinguishability of $e^+e^- \rightarrow \mu^+\mu ^-$ and $e^+e^- \rightarrow \mu^+\mu ^-(+\gamma)$: when $\gamma$ is sufficiently soft/collinear (relative to a given detector resolution), it cannot be differentiated from the outgoing $\mu^+ \mu^-$ pair, and thus both processes contribute to the cross-section that is actually measured. Meanwhile, outside this region, say, with some non-collinear, high-energy $\gamma$, the processes are distinguishable and should be accounted for separately without summation.

In the example above, the identification of degenerate processes is reasonably straightforward. This is not always the case, as we shall see in a moment. More generally, however, for each process indexed by a cut $C$ of some underlying diagram, the phase space of external kinematics can always be divided into two regions: a region $ R_d^C$ which contains any divergences and singularities, and a region $ R_f^C$ 
where all contributions are finite.\footnote{For instance, one could always take $ R^C_d$ to be the entire phase space and $ R_f^C$ empty, in which case $\sigma_{\text{inc}} = 0 < \infty$.} Then, the KLN theorem of \cref{eq::klnschwartz} can be written
\begin{align}
    \label{eq::splitphasespace} \sum_{\text{cuts $C$}} [\sigma^C_{F, a \rightarrow a}]_{ R_d^C} = -\sum_{\text{cuts $C$}}[\sigma^C_{F, a \rightarrow a}]_{ R_f^C} < \infty ,
\end{align}
where $[\sigma]_R$ denotes a cross-section integrated only over the region $R$. If the $ R_d^C$ can be chosen in such a way that they correspond to processes which are degenerate or indistinguishable with respect to some experimental setup, then the sum on the left-hand side can be identified with the measured, inclusive cross-section, and the sum on the right-hand side is by construction some finite (but not necessarily vanishing) number. In this way, the KLN theorem still allows for non-zero scattering whilst also ensuring the finiteness of inclusive observables.

\subsection{Physical Criteria and On-Shell Regions}
\label{sec::criteria}
Let us now consider what criteria govern this partition into degenerate and non-degenerate regions for $t$-channel scattering. Certainly, there are three basic mathematical requirements for an inclusive cross-section $\sigma_{\text{inc}}(s)$ in the physical region $s \geq m^2$:
\begin{itemize}
    \item \textbf{Finiteness}: $\sigma_{\text{inc}}(s) < \infty$;
    \item 
    \textbf{Regulator Independence}: $\sigma_{\text{inc}}^{\text{NR}} =\sigma_{\text{inc}}^{\text{SM}}  = \sigma_{\text{inc}}^{\text{DR}}$ for all the regions where the various regulators are defined. Note that we have excluded the large-mass regulator $\sigma_{\text{inc}}^{\text{LM}}$ from this requirement, for reasons which will be discussed in \cref{sec::largemassexclude};
    \item \textbf{Non-negativity}: $\sigma_{\text{inc}}(s) \geq 0$.
\end{itemize}
Importantly, however, we can only demand that these properties hold at the level of the inclusive cross-section, summed 
over all contributing topologies up to the relevant order in perturbation theory. 
\textit{Individual} diagrams or subdominant orders of perturbation theory need not satisfy this on their own, since they are not physical in their own right.

For the $t$-channel process, divergences can only occur when the propagator denominator of \cref{eq::tchannelint} goes on-shell at
\begin{align}
z^*(\kappa) \equiv 1-\frac{2}{(\kappa -1)^2},
\end{align}
where $\kappa \equiv s/m^2$. Any compact subset $R_f^t$ of $z \in [-1,1]$ not containing this point involves a positive, non-singular integrand, that yields a $[\sigma]_{R_f^t}$ satisfying all the desired properties listed above. This is why for $\kappa <2$, the $t$-channel amplitude is well-behaved on its own, since $z^* < -1$. 

Thus, the region of degeneracy $R_d^t$ must be some window containing $z^*(\kappa)$. To make contact with collider observables, notice that the precision with which we know $z^*(\kappa)$ is fixed entirely by the precision with which we know $\kappa$. Thus, supposing that $\kappa$ can be measured with a precision of $\pm \Lambda$, where $0 < \Lambda \ll 1$ is related to some small energy resolution (since $\kappa \equiv s/m^2=E_{\text{COM}}^2/m^2$), let us define
 \begin{align}
 \label{eq::onshellregion}
  R^t_d \equiv [z^*(\kappa -\Lambda), z^*(\kappa + \Lambda)] \cap [-1,1],
 \end{align}
 where we have added the intersection with $[-1,1]$ to ensure that $z \equiv \cos \theta$ remains within the appropriate range.

Before going on, let us stress once again that it is the \textit{energetic} resolution on $\kappa$ which determines the size of $ R^t_d$, rather than the perhaps more intuitive \textit{angular} resolution on $z$; $ R^t_d$ may intersect with one or multiple angular bins of a given experiment, depending on the relative resolutions. This is because $ R^t_d$ is defined as the on-shell region where processes become degenerate, and has nothing to do with how precisely we can resolve angles, which may or may not probe into off-shell regions. Since this region $z^*(\kappa)$ is defined purely as a function of $\kappa$, our ability to specify it is fixed entirely by the uncertainty inherited from that $\kappa$.\footnote{To illustrate this concretely, suppose that we only had one angular bin that covered the full phase space, and could not resolve the angle of any events whatsoever. Were we to use this angular resolution to define $ R^t_d$, then the prediction would be a measured $t$-channel contribution of zero, since $\sigma_{\text{inc}}$ would consist of $\sum \sigma^C = 0$. Meanwhile, as we will show, our prescription here perfectly accommodates for non-zero scattering even in this scenario, supposing our energy detection still works reasonably precisely.}

\subsection{Unstable Particles and Degeneracies}
\label{sec::unstabledegeneracies}
\begin{figure}[tbp!]\centering 
\begin{subfigure}{0.27\linewidth}
\includegraphics[width=\linewidth]{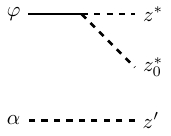}
\caption{}
\end{subfigure}\hspace{7ex}
\begin{subfigure}{0.35\linewidth}
\includegraphics[width=\linewidth]{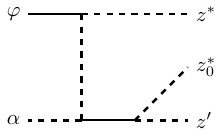}
\caption{}
\end{subfigure}
\caption{Two possible $\varphi$ decay channels degenerate with on-shell $t$-channel scattering. The labels $z^*, z_0^*, z'$ on the right hand side of either diagram label the angular cosine of each of the outgoing $\alpha$ particles. All internal lines of (b) are on-shell in the region of degeneracy.}
\label{fig:decaychannels}
\end{figure}
When $z \in R_d^t$, \textit{all lines of the $t$-channel diagram go on-shell}, and the mediating $\alpha$ can be thought of as the physical out-state of a real $\varphi \rightarrow \alpha \alpha$ decay, that then goes on to scatter again with another incoming $\varphi$. This gives us a clue into the possible degeneracies: since $\varphi$ is unstable, one cannot know \textit{a priori} at precisely which point in the potential series of sub-scatterings an on-shell $\varphi$ decay will occur, if at all, over the timescales of a given experiment.\footnote{Recall as per \cref{sec::unstableparticles}, that since $\varphi$ is unstable, it is not actually in the asymptotic spectrum of outgoing states for the interacting theory, and, in principle, any $S$-matrix element involving an outgoing $\varphi$ should vanish automatically. However, physical experiments operate on a finite timescale, and thus do not really probe the asymptotic spectrum.} Indeed, there are other possibilities which need to be accounted for:
\begin{itemize}
    \item as shown in \cref{fig:decaychannels} (a), the outgoing $\alpha$ states of the $\varphi \rightarrow \alpha \alpha$ sub-decay could hit the detector \textit{without} interacting with the other incoming $\alpha$, which would then forward-scatter until it too hit the detector;
    
    \item or, as shown in \cref{fig:decaychannels} (b), the on-shell $t$-channel scattering could occur, but the outgoing $\varphi$ could then decay \textit{again} before detection, in which case two additional $\alpha$ particles would be detected as out-states instead;  
\end{itemize}
along with iterations of these at higher and higher perturbative order. At $\mathcal{O}(g^4)$, however, which is the order of the $t$-channel process, it is precisely the interference between the two possibilities listed above that comprises the on-shell part of $\sigma_3$!

To be clear, it is only in the region where all internal lines are on-shell that this degeneracy holds, since one cannot identify virtual particles of one process with real particles of another. It can be shown straightforwardly that this occurs precisely at the point
\begin{align}
    z_0^*(\kappa) = \frac{1}{\kappa -1},
\end{align}
where $z_0^*$ is shown in \cref{fig:decaychannels}. Therefore, once again recalling our resolution $\pm \Lambda$ on $\kappa$, our claim is that in the region defined by
\begin{align}
\label{eq::onshell3bodyregion}R^3_d \equiv [z_0^*(\kappa +\Lambda),z_0^*( \kappa -\Lambda)] \cap [-1,1],
\end{align}
the $3$-body process is physically-degenerate with the $t$-channel process restricted to the region $ R^t_d$, 
and their cross-sections need to be summed inclusively together.\footnote{Since for the $3$-body decay, $z_0, z$ fix one another in general via momentum conservation as
\begin{align}
\label{eq::z0zdef}
    z_0 =1-\frac{2 (z+1)}{\kappa ^2(1-z)+z+1},
\end{align}
 one can trivially verify that $z_0^* = z_0(z^*)$, which confirms the physical intuition that the on-shell regions of both processes can be identified with one another.} Outside this region, summation is no longer appropriate.

Now, one subtlety with our physical interpretation is that \textit{degenerate} here is somewhat different to \textit{experimentally indistinguishable}. Writing $z_0^* \equiv \cos \theta_0^*, z^*\equiv \cos \theta^*$, it follows that
\begin{align}
\label{eq::noncollinear}
    \cos 2\theta_0^* =-\cos \theta^*,
\end{align}
and since the angle $\theta^*$ in the t-channel scattering process sweeps out all possible angles, this tells us that the particle in the three-body decay with angle $\theta_0^*$ is not in general restricted to be collinear with any other particles in the problem. Moreover, its energy $E_1 -E_2 = \frac{m^2}{\sqrt{s}}$ is not necessarily soft relative to our detector resolution. In other words, for general $s > 2m^2$, one should in principle be able to distinguish if such a decay process has occurred relative to the $t$-channel scattering by simply measuring an additional detection at $z_0^*$. 

However, the point is not that one cannot tell the difference between these two processes after the fact; rather, it is that in an ensemble of particles prepared in a scattering experiment, \textit{both} possibilities will be occurring at once, due to the intrinsically probabilistic nature of particle decay. It is the degenerate combination of both of these processes, which will be measured in a detector at the angle $z^*$, some of the time also including additional detections at $z_0^*$. The key subtlety is that the degeneracy is \textit{between} internal and external lines, depending on where exactly the $\varphi$ decays, rather than just among external ones. The importance of inclusively summing together with on-shell decays has also been observed in \cite{Bahl:2021rts}, for instance. This is in contrast to the complex mass scheme, where self-energy corrections to external legs are included by shifting the poles of the physical particles~\cite{Denner:1999gp,Denner:2005fg} and thus effectively resumming the self-energy corrections. 

\subsection{The Inclusive Cross-Section}
\label{sec::fullincresult}
With this physical interpretation, let us now explicitly state the result for the $t$-channel and $3$-body interference integrated over the region of degeneracy; we leave the details of the calculation to \cref{app::contributionstoint}. Note that outside this region, the $t$-channel process is manifestly finite and positive by definition, and requires no special consideration, so we do not state it here. 

Due to the constraint that $\cos \theta \in [-1,1]$, there are three relevant regions to consider: $\kappa < 2-\Lambda$, where the window around the on-shell point is excluded from the region of integration altogether, such that the contribution automatically vanishes (and will be omitted in the following for brevity); $\kappa \in [2-\Lambda, 2+\Lambda)$ where only the lower bound is included within $[-1,1]$; and $ \kappa \geq 2+\Lambda$, where both bounds appear inside $[-1,1]$. 

Individually, then, the $t$-channel and $3$-body contributions are given by
\begin{align} [\sigma_t]_{ R^t_d}&= \frac{g^4}{\pi m^6}\begin{dcases}
         -\frac{(\kappa +\Lambda -2) (\kappa +\Lambda )}{16 (\kappa -2) \Lambda  (2 \kappa +\Lambda -2)}& 2-\Lambda \leq \kappa < 2+\Lambda
        \\ -\frac{(\kappa -1) \kappa }{4 \Lambda  (2 \kappa -\Lambda -2) (2 \kappa +\Lambda -2)}& \kappa \geq 2+\Lambda 
    \end{dcases}
   \\ [\sigma_3]_{ R^3_d}&= \frac{g^4}{\pi m^6}\begin{dcases}
        \frac{(\kappa +\Lambda -2) (\kappa +2 \Lambda )}{32 (\kappa -2) \Lambda  (\kappa +\Lambda -1)}& 2-\Lambda \leq \kappa < 2+\Lambda
        \\ \frac{\kappa ^2-\kappa -\Lambda ^2}{16 \Lambda  (\kappa -\Lambda -1) (\kappa +\Lambda -1)}& \kappa \geq 2+\Lambda 
    \end{dcases},
\end{align}
where we have omitted the squared $\delta$ function contribution which cancels between the two processes c.f. Eqs.~\eqref{eq::fulltchannelexppole}, \eqref{eq::squaredeltat}, \eqref{eq::sigma3SM}, and \eqref{eq::sigma3DR}. In addition to the squared $\delta$-function contribution, there are divergences in the collinear region $\kappa = 2$, and also as the resolution $\Lambda \rightarrow 0$. These contributions can be summed together and expanded in $\Lambda$ as 
\begin{align}
    \sigma_{\text{inc, t $+$ 3}}^{(g^4)} = \frac{g^4}{\pi m^6}  \begin{dcases}
       \frac{\kappa -2}{64 (\kappa -1)^2}+\frac{(4-\kappa ) \Lambda }{128 (\kappa -1)^3}+\mathcal{O}\left(\Lambda ^2\right)& 2-\Lambda \leq \kappa < 2+\Lambda
        \\ \frac{(4-\kappa) \Lambda }{64 (\kappa -1)^3}+\mathcal{O}\left(\Lambda ^2\right)& \kappa \geq 2+\Lambda 
    \end{dcases}.
\end{align}

\subsection{A Negative Cross-Section?}
\label{sec::nonneg}

Returning to the criteria stated in \cref{sec::criteria}: this cross-section is finite for all $\kappa$,  even when $\Lambda \rightarrow 0$;\footnote{However, note that one still needs non-zero $\Lambda$ for this quantity to make sense, since $t$-channel contributions integrated over adjacent angular windows scale like $\sim 1 /\Lambda$ as they encroach upon the angular point $z^*(\kappa)$ when $\Lambda \rightarrow 0$.} and with the LM regulator excluded for reasons outlined in \cref{sec::largemassexclude}, regulator independence follows straightforwardly from the fact that in all cases, the calculation essentially amounted to evaluating the same $\delta$-functions, either in the $\delta_\text{dim}$ or $\delta^\epsilon$ representation, as per \cref{sec::3body} and \cref{app::deltafunctions} . 

Non-negativity, however, is \textit{not} satisfied for all $\kappa$. Even though we have extended the region of non-negativity from $\kappa < 2$ for the $t$-channel process alone to $\kappa \lessapprox 4$ for our inclusive quantity, when one retains the full $\Lambda$-dependence, this cross-section goes negative for $\kappa > 
\frac{1}{2}(5+\sqrt{9+4\Lambda^2})$, and thereafter asymptotically approaches zero from below.

At a glance, however, this is not immediately a problem: $ \sigma_{\text{inc, t $+$ 3}}^{(g^4)}$ is only one contribution to $\varphi \alpha \rightarrow \varphi \alpha$ scattering, computed at a single perturbative order. The physical cross-section for which one demands positivity consists of all topologies and all degeneracies summed over all the relevant orders of perturbation theory. However, as we explicitly compute in \cref{app::contributionstoint}, all of the obvious additional topologies one would usually include either cancel amongst themselves or are subdominant with respect to the scaling in $\kappa$, and do not resolve the issue of non-negativity.

Although it is beyond the scope of this work to fully resolve the question of non-negativity and construct a physical cross-section for all $\kappa \gtrapprox 4$, we suspect that it could be related to one or both of the following issues: the non-inclusion of additional, physically-degenerate processes; or the breakdown of a fixed-order perturbative calculation altogether. Let us now briefly comment on both of these with future research in mind.

Should this issue be related to unaccounted-for degeneracies, recall that one of the primary insights of \cite{Frye:2018xjj} was that there are often multiple ways to cancel divergences. In particular, although the approach we took in this work involved cuts which kept the initial state fixed while varying the final states, Frye et al  found that there were scenarios in which it was perhaps more natural to consider also modifying the initial states themselves. Although the physical interpretation of initial state sums is somewhat more obscure in a collider setting, the idea has been discussed in the literature (see for eg. \cite{Contopanagos:1991yb, Taylor:1993bz, Contopanagos:1996qx, Lavelle:2005bt, Frye:2018xjj}), and is perhaps more in line with the original proposal of Lee and Nauenberg \cite{Lee:1964is}. Using initial state cuts, one may even be able to retain the fixed-order spirit of our work here.  

Alternatively, on the question of the reliability of a fixed-order calculation altogether, recent works \cite{Bouzoud:2024bom, Becker:2025yvb} studying ALPs in a thermal setting resolved the comparable issue of negative rates using a resummation of thermal propagators. As we have already mentioned for our model, the standard approach involving a simple resummation of the mediating $\alpha$ propagator does not cure the $t$-channel divergence due to its vanishing width on-shell, but may still affect the positivity properties of the cuts amongst which the divergence cancels, particularly if one also resums the unstable $\varphi$ propagators. Additionally, since in principle our physical interpretation requires accounting for all possible on-shell $\varphi \rightarrow \alpha \alpha$ sub-decays, it is possible that some sort of ladder resummation of $\varphi \alpha \rightarrow \varphi \alpha$ box diagrams may be relevant for capturing these degeneracies; see, for instance, \cite{FuentesZamoro:2025exp}, for some brief comments on the connection between resummed ladder diagrams and $t$-channel divergences.          

Finally, we also remark as a passing curiosity that as $\Lambda \rightarrow 0$, the value $\kappa = 4$ at which $\sigma_{\text{inc, t $+$ 3}}^{(g^4)}$ changes sign corresponds precisely to the kinematic threshold for the crossed on-shell $\alpha \alpha \rightarrow \varphi \varphi$ scattering, although the significance of this fact is presently unclear. 

\section{Discussion}
\label{sec::discussion}
\subsection{Practical Takeaways}
\label{sec::practicaltakeaways}
With practical calculations in mind, we now emphasize the key takeaways for both computing unitarity cuts in general, and for constructing finite inclusive observables from those cuts.
\subsubsection{Computing Unitarity Cuts}
\mathversion{bold}\paragraph{The correct order of operations is: first loop integration; then $i\epsilon_F \rightarrow 0$; then finally sending the regulator $\epsilon$ in dimensional regularization or $m_0$ in mass regularization to zero,  $m_0, \epsilon \rightarrow 0$.} \mathversion{normal} Importantly, these three operations do \textit{not} in general commute.

That loop integration should come first is straightforward: the limit $i\epsilon_F \rightarrow 0$ is a \textit{distributional} one which only exists after integration (see \cref{app::iepsrep} for more details); and the regulated theory is in principle a separate, independent theory from the unregulated one,\footnote{After all, one may very well be interested in computing loops in $d$-dimensions or with additional masses \textit{without} eventually performing the analytic continuation $m_0, \epsilon \rightarrow 0$ at the end.} supposing that the relevant integrals even exist in the unregulated theory in the first place. This latter point is also why $i\epsilon_F \rightarrow 0$ must be performed \textit{before} $m_0, \epsilon \rightarrow 0$. For each theory defined by some non-zero $m_0, \epsilon$, the KLN theorem holds independently, since these regulators are unitary.\footnote{See \cite{Gnendiger:2017pys} for a review on both mathematical and practical considerations relating to dimensional regularization and its variants, and see  \cite{Anselmi:2016fid} for a discussion of dimensional regularization in the context of the cutting rules.} But, in order to even define the cuts and propagators appearing inside the KLN theorem, one must already have taken $i\epsilon_F \rightarrow 0$; for any small but finite choice, there will be $\mathcal{O}(\epsilon_F)$ deviations, and the $i\epsilon_F$ will no longer implement the contour deformation used to select the physical sheet \cite{Hannesdottir:2022bmo}. 

Now, let us comment what this order of operations practically means for calculations.
\begin{itemize}
    \item One must specify the precise range over which one's regulator lies, and keep it consistent across all calculations - even if terms appear superficially finite upon sending the regulator to zero. As we saw with our mass regulator, for example, the cancellation played out completely differently with large $m_0 > m$ and small $m_0 < m$ mass regulator cases. One can only interchange limits when mathematical results such as the dominated or monotone convergence theorems allow one to do so. 
    \item In dimensional regularization, one has relations of the form 
    \begin{align}
    \label{eq::delxeps}
        \int dx \delta(x) x^{-\epsilon} = 0
    \end{align}
    since, in accordance with \cref{app::deltafunctions} and the order of limits outlined above, the left-hand side of this expression should really be understood as
    \begin{align}
    \label{eq::olimsdisc}
        \int dx \delta(x) x^{-\epsilon} \equiv \lim_{\epsilon \rightarrow 0^-} \lim_{\epsilon_F \rightarrow 0^+} \int dx \left[\frac{\epsilon_F}{\pi} \frac{1}{x^2 + \epsilon_F ^2} \right]\ x^{-\epsilon}. 
    \end{align}
    This prescription allows us to argue why certain boundary contributions in dimensional regularization vanish, even though they do not vanish in the $\epsilon = 0$ no-regulator computation. Note that \cref{eq::delxeps} would not hold had we swapped the $\epsilon \rightarrow 0^-$ and $\epsilon_F \rightarrow 0^+$ limits in \cref{eq::olimsdisc}.

    Let us also mention briefly that a similar claim about the vanishing of these integrals was also made by \cite{Abreu:2014cla} in the context of on-shell massless three-point vertices, while alternatively, in \cite{Bourjaily:2020wvq}, it was argued that setting such integrals to zero may yield a result for certain cuts that is inconsistent with the computation involving monodromies around branch points. Although those calculations are beyond the scope of this work, we remark that it would be interesting to see if the learnings from this model might affect either the equivalence of the cut calculation to the monodromy calculation, or the details of how those calculations play out.

    \item In dimensional regularization, expressions which naively appear to be $\mathcal{O}(\epsilon)$ cannot always immediately be dropped, particularly if they involve an integral that diverges as $\epsilon \rightarrow 0$. One useful identity is that
    \begin{align}
    \label{eq::dimregdelta}
        \delta_{\text{dim}}(x) = \lim_{\epsilon \rightarrow 0^-} -\frac{1}{2} \epsilon \abs{x}^{-\epsilon -1}
    \end{align}
     forms a "dimensionally regulated" representation of the $\delta$-function, which picks up a boundary contribution at $x=0$ despite appearing naively $\mathcal{O}(\epsilon)$. Note also that the limit should be understood as being performed \textit{after} the integration in a way totally analogous to the discussion in \cref{app::iepsrep}. See \cref{app::deltaDim} for more technical details about this representation. 
\end{itemize}
\paragraph{\textbf{Boundary contributions cannot be neglected.}}
There are many ways which boundary contributions can appear in the calculations of cuts. As we saw above in \cref{eq::dimregdelta}, in dimensional regularization they can arise from products of the form $\epsilon \times \text{divergent}$; while in $d=4$, they can arise from kinematical constraints like the zero-energy condition of our four-body decay; or also from the integration-by-parts of derivatives of $\delta$-functions obtained using identities like
\begin{align}
    \frac{1}{(\Delta - i\epsilon_F)^n} - \frac{1}{(\Delta+i\epsilon_F)^n} = 2i \frac{(-1)^{n+1}}{(n-1)!} \partial^{n-1}_{\Delta} \delta(\Delta)
\end{align}
for computing imaginary parts in Feynman parameter space, and
\begin{align}
    \frac{\delta(q^2-m^2)}{q^2-m^2 + {i\epsilon_F}} + c.c. = -\delta'(q^2-m^2)
\end{align}
for computing cuts involving on-shell propagators. In all cases, the boundary contributions are vital for achieving the correct cancellation.

Importantly, one must also take care with the precise boundaries of integration, since they result in factors which are otherwise easy to miss.
As explained in \cref{app::deltafunctions}, since we construct our $\delta$-functions as limiting sequences of even functions, a $\delta$-function localized at the boundary of integration\footnote{Note that not every $\delta$-function arising from the boundary term of an integration by parts sits on the boundary of a subsequent integral. For example, this is the difference between the small-mass and no regulator cases of our box calculation in \cref{sec::massboxnoreg}: for the former, we have $\int_0^1 dx \delta(x-m_0^2/m^2) = 1$, while for the latter we have $\int_0^1 dx \delta(x) = 1/2$, even though both of these $\delta$-functions essentially came from the "same" integration by parts step.} requires an additional factor of half as per
\begin{align}
    \int_0^\infty dx \delta(x) = \frac{1}{2}.
\end{align}

When there is more than one $\delta-$function localized at a boundary, the factors can become somewhat more subtle (and do not necessarily amount to a $1/2^N$). See \cref{app::deltafunctions} for an example of how to deal with the $\delta[q^2]^2\theta(q_0)\theta(-q)$ arising from our four-body cut, which results in an additional factor of $1/4$; and see Appendix G of \cite{Bourjaily:2020wvq} for an example where a factor of $1/N!$ arises from products of angular $\delta$-functions. 

\paragraph{\textbf{Diagrams involving zero-energy on-shell processes may contribute to the cancellation.}} This was the lesson of our four-body decay. For the emitted on-shell $\alpha$ particles with momenta $q_1, q_2$, only a single point in the phase space at $q_1 = q_2 = 0$ was kinematically allowed. Although this diagram vanishes in both dimensional regularization and mass regularization when one takes the order of limits correctly, it is necessary for achieving a KLN cancellation in the no-regulator case.

\subsubsection{Constructing Inclusive Observables}
\paragraph{\textbf{Processes should only be summed together inclusively in the regions of phase space where they are degenerate or experimentally-indistinguishable.}} As we argued in \cref{sec::physint}, the physical, measured quantity is not the sum of total cross-sections (which will always vanish in accordance with the KLN theorem), but is rather the sum of cross-sections integrated \textit{only} over the regions of external kinematics where processes are degenerate or experimentally-indistinguishable. These regions can be identified as a window around the locus of external kinematics for which internal lines go on-shell, whose size is ultimately fixed by the angular or energetic resolution of the involved experimental parameters.

\paragraph{\textbf{For finiteness, processes in which unstable particles interact without decaying on-shell must be considered in tandem with those in which they do decay on-shell, even if the outgoing states of each can be distinguished in a detector.}} In particular, we saw that even though for general $s > 2m^2$, one of the decay products associated with the on-shell $3$-body cut of \cref{sec::3body} was neither soft nor collinear relative to the other scattering particles, its contribution was crucial to cancel the on-shell $t$-channel divergence of \cref{sec::tchannel}. 

Our interpretation was that the on-shell $t$-channel contribution only enumerated \textit{one} of the possible on-shell decay channels available to the unstable $\varphi$; the other channels shown in \cref{fig:decaychannels} incorporated in the on-shell $3$-body cut would also physically occur with some probability related to their cross-section. Since $\varphi$ is not a well-defined asymptotic state, in order to still make use of $S$-Matrix perturbation theory, one must account for all of these degenerate on-shell decay channels together, since no channel can be excluded in advance (even though they may be in principle distinguishable \textit{after} the scattering has occurred). 

Importantly, however, this interpretation only holds in the on-shell region where real particles are identified with other real particles; it does not make sense to identify a real particle with a virtual one, for instance.

\paragraph{\textbf{Among the cuts of a single diagram, the KLN theorem guarantees finiteness, but not non-negativity.}} Although we were able to construct a finite inclusive cross-section from the cuts of a single diagram integrated over the relevant on-shell region, this quantity went negative for $s \gtrapprox 4m^2$. While this is not intrinsically problematic since individual diagrams are not observable, it transpired that the contributions of all the other obvious topologies and degenerate processes at the same order were insufficient to render the final result positive for all physical $s > m^2$. We speculate that this may be related to either the necessity to include other degeneracies such as initial-state sums, or alternatively the breakdown of a fixed-order perturbative approach altogether. See \cref{sec::nonneg} for more details.  

\subsection{Analytic Structure \& Regularization}
\label{sec::analyticstructure}
Significant technical subtleties involving boundary contributions and regulator dependence arose in our calculation. In this section, we relate these subtleties to the analytic structure of the underlying diagram being cut.
\subsubsection{Subtleties with the \texorpdfstring{$p_i = p_f$}{p_i=p_f} Restriction}

\paragraph{\textbf{The identification $p_i = p_f$ in the KLN theorem can localize the underlying integral at a singular hypersurface.}}
For $2 \rightarrow 2$ scatterings like ours, fixing $p_i = p_f$ for the box diagram (which was necessary to interpret the cut diagrams as scattering cross-sections c.f. \cref{sec::cuttingrules}) amounts to restricting the underlying amplitude to the $t=0$ subspace of external invariants. This restriction may not be so trivial, however; in fact, for our model, in the NR case it turns out that $t=0$ corresponds to a singular hypersurface where the (imaginary part of the) amplitude jumps discontinuously. Throughout the calculation, the appearance of $\delta$-functions localized at boundaries of integration was a reflection of this fact.

The simplest way to see that there is such a singularity at $t=0$ in our model is to calculate the $t \rightarrow 0^-$ limit, which corresponds to the amplitude in the interior of the  $t= (p_1-p_3)^2 \leq 0$ physical region, and to then compare it to the direct $t=0$ boundary computation of \cref{sec::boxdiagram}. In \cref{app::trightarrow0minus}, we show that in the no regulator case for $s \in (m^2, 2m^2)$, one has
\begin{align}
    \lim_{t\rightarrow 0^-} \sigma^\text{NR}_{\text{box}}(s,t) = 0 , 
\end{align}
while in \cref{sec::boxdiagram} we computed
\begin{align}
    \sigma^\text{NR}_{\text{box}}(s, 0) = \frac{g^4}{16\pi} \frac{1}{m^4} \frac{1}{(s-2m^2)} \neq0. 
\end{align}
The difference between these two values indicates that some threshold has "switched on" at $t=0$, undetectable to the $t < 0$ region. Of course, this is precisely the 4-body diagram of \cref{sec::4body}, which has no kinematic support for $t < 0$, but phase-space support at $t=0$ corresponding to the zero-energy emissions of two $\alpha$ particles.

\paragraph{\textbf{In this model, mass regulators shift this singular hypersurface away to $t = 4m_0^2 >0$, away from the $t=0$ restriction associated with identifying $p_i=p_f$.}} In \cref{app::trightarrow0minus}, we show this explicitly by investigating the kinematic regions in $s, t$ in which the relevant integrals contributing to $\sigma_{\text{box}}^{\text{SM, LM}}$ acquire support. What this means, however, is that the mass regulator renders the amplitude better-behaved at the point $t=0$ probed by the KLN theorem; one has, for instance, that for $s \in (m^2,2m^2)$,
\begin{equation}
\begin{aligned}
\lim_{t\rightarrow 0}\sigma_{\text{box}}^{\text{SM}}(s,t) &= \sigma_{\text{box}}^{\text{SM}}(s,0) , 
    \\ \lim_{t \rightarrow 0}\sigma_{\text{box}}^{\text{LM}}(s,t) &= \sigma_{\text{box}}^{\text{LM}}(s,0), 
\end{aligned}
\end{equation}
where the limit $t \rightarrow 0$ can now be taken from either direction in a sufficiently small neighborhood, in contrast to the sudden threshold exhibited by the NR result. The fact that no $\delta$-functions localized at a boundary of integration appeared in the mass regulator KLN calculations is a reflection of this improved behavior.

In much the same way as before, this threshold is also associated with the 4-body diagram gaining support. For general $p_1 \neq p_3$, momentum conservation for the 4-body process is given by
\begin{align}
    p_1 = p_3 + q_1 + q_2, 
\end{align}
which can be rearranged and squared to give
\begin{align}
    t = 2\left(m_0^2 + \sqrt{m_0^2 + \abs{\vec{q}_1}^2}\sqrt{m_0^2 + \abs{\vec{q}_2}^2} -\abs{\vec{q}_1} \abs{\vec{q}_2} \cos \theta_{12} \right),  
\end{align}
where $\theta_{12}$ is the angle between $\vec{q_1}$ and $\vec{q_2}$. However, for physical $\abs{\vec{q}_1}, \abs{\vec{q}_2} \geq 0$ and $-1 \leq \cos\theta_{12} \leq 1$, the minimum value acquired by the right-hand side is $4m_0^2$, which tells us that this process is only kinematically allowed for $t \geq 4m_0^2$, which is precisely the threshold we identified in our loop calculation in \cref{app::trightarrow0minus}. 

\subsubsection{Regulators and Regulator Dependence}
\paragraph{\textbf{KLN partial sums can be regulator dependent.}} 
Even though the total KLN sum always vanishes in accordance with \cref{eq::klnschwartz} irrespective of the particular choice of (unitary) regulator, the partial sums -- that is, particular proper subsets of cross-sections where divergences might already cancel -- clearly \textit{can} depend on the particular regularization scheme. Our results summarized in \cref{tab::summaryoffpole} and \cref{tab::summaryonpole} demonstrate this quite conclusively. 

Having formulated the KLN theorem as a statement about cuts in \cref{sec::cuttingrules}, however, we have no right to be surprised. Cuts in general may not be analytic in kinematic invariants and masses \cite{Britto:2024mna}, nor in the spacetime dimensionality $4-2\epsilon$. This non-analyticity ultimately stems from the presence of the $\delta$-functions, whose supports can discontinuously "switch on" as one varies masses/kinematics, and which can select out function values that jump discontinuously for different values of $\epsilon$, the prototypical example being $\delta(x)x^{-\epsilon}$. So, there is no reason to expect that individual cuts or partial sums computed in one region of regulator space should agree with those computed in another; the only regulator independent quantity \textit{in general} is the "$0$" on the right-hand side of \cref{eq::klnschwartz}. 

Alternatively, there are of course many situations where the partial sums \textit{are} regulator independent (at least, for mass and dimensional regularization); this is the case with $e^+ e^- \rightarrow \mu^+ \mu^-(+\gamma)$ mentioned above, for instance \cite{Schwartz:2014sze}. As we saw in \cref{sec::analyticstructure}, our model is somewhat unique in that it exhibits a singular hypersurface at $t=0$, corresponding to the 4-body zero-emission "switching on", which is precisely the region where the KLN theorem localizes the cuts. Since the different regulators treat this threshold differently, the cancellations play out rather differently across the different schemes.

The key takeaway, however, is that \textit{the finiteness of some partial sum of cuts in the limit of setting a regulator to zero does not in general guarantee its regulator independence}. When an apparently physical quantity appears to depend on the choice of regulator, it indicates that either the quantity is unphysical, or the  regulator is a poor choice. As we demonstrated, the inclusive cross section defined in \cref{sec::fullincresult} is regulator independent, so long as one excludes the large mass regulator for reasons we will now outline.

\paragraph{\textbf{In this model, the large mass regulator dramatically modifies the analytic structure of the underlying box diagram, rendering it a poor choice for computing observables.}} \label{sec::largemassexclude}To see this, let us study how the analytic structure of the box integral being cut changes as a function of the regulator mass. Recall that the denominator of the integrand in Feynman parameter space was given by $(\Delta - i\epsilon_F)^{2+\epsilon}$, where 
\begin{align}
     \Delta\equiv \kappa x(-1+x+y)-(-1+x+y+xy) +\beta y,
\end{align}
and where we have set $s = \kappa m^2, m_0^2 = \beta m^2$ and $m^2 =1$, which amounts to assuming that all $m^2$-dependence has been factored out c.f. \cref{eq::boxdenom}.

Now, when $\Delta > 0$, the $i\epsilon_F$ can be neglected altogether because there are no poles to traverse around; it is only when $\Delta \leq 0$ that singular or multivalued behavior emerges. Thus, in order to avoid any singular behavior at all, we would require $\Delta > 0$ for the entire region of Feynman parameter space $F \equiv \left\{(x,y) : x, y \geq 0 \cap x+y \leq 1\right\}$ swept out by the integration contour. The region in $\kappa$ (and external invariants more generally) for which this is the case, if it exists, is called the \textit{Euclidean region}. Translated into a constraint on (real) $\kappa$, then, this amounts to insisting that
\begin{align}
    \kappa < \kappa^*(x,y) \equiv \frac{x y+x-\beta  y+y-1}{x^2+x y-x} \quad \forall (x,y) \in F,
\end{align}
which ultimately means that
\begin{align}
     \kappa < \min{\kappa^*(x,y)}\vert_{{(x,y) \in F}}.
\end{align}

This is where the distinction between regulators becomes important. In the large mass-regulator case $\beta \geq 1$, $\min{\kappa^*(x,y)}\vert_{{(x,y) \in F}} = 1$, while for $\beta <1$, $\kappa^*(x,y)$ can go arbitrarily negative and has no finite minimum. 

This means that when $\beta \ge 1$, any choice of $\kappa < 1$ corresponds to a non-singular box-diagram; in other words, $\kappa < 1$ is the Euclidean region. But when $\beta < 1$, there is \textit{no} choice of $\kappa$ for which the integral does not develop any potential singularities. This corresponds to a branch cut in $\kappa$ across the entire real axis, since approaching from above or below (which would correspond to different $i\epsilon_F$ prescriptions) could generically give different results.  

Of course, our actual calculation takes place with $\kappa > 1$, so why would the behaviour for $\kappa < 1$ be at all relevant? The answer here is given by the Schwarz Reflection Principle; namely, the existence of the Euclidean region allows us to connect the upper and lower half planes for $\kappa > 1$ along a path of analytic continuation going through $\kappa < 1$. In turn, for $\kappa>1$, where there \textit{is} a branch cut, one can then straightforwardly relate
\begin{align}
\label{eq::imdisc}
    \Im (f(\kappa)) = \frac{1}{2i} \text{Disc}_\kappa f(\kappa),
\end{align}
 where $\Im(f(\kappa))$ is the imaginary part of the amplitude obtained by approaching from \textit{above} the real line, $\kappa \rightarrow \kappa + i\epsilon_F$, and $\text{Disc}_\kappa f(\kappa) \equiv \lim_{\epsilon_F  \rightarrow 0^+} \left[ f(\kappa + i\epsilon_F)- f(\kappa - i\epsilon_F)\right] $ is the discontinuity across the branch cut. 

When the Euclidean region does not exist, and the branch cut extends along the entire real line, as with $\beta < 1$, one can no longer necessarily write the amplitude as the boundary value of a \textit{single} analytic function of external invariants \cite{Hannesdottir:2022bmo}. The Feynman-$i\epsilon_F$ prescription may no longer translate so simply into the kinematic invariants, and relations like \cref{eq::imdisc} may no longer hold. 

Thus, the takeaway is that even though the KLN theorem of course holds in the large mass-regulated theory, any analytic continuation of the final result from $\beta >1 $ to $\beta \rightarrow 0$ is blind to the dramatic change at $\beta =1$. As we have seen, this obstruction manifests in the calculation of the imaginary part of the box amplitude, which corresponds to the forward-scattering interference. It is therefore unsurprising that the pattern of cancellation for the large mass-regulator plays out so differently compared to the other regulators in \cref{tab::summaryoffpole,tab::summaryonpole}. 

For this reason, since it is unclear how to interpret crossing this special value of $\beta = 1$ that corresponds to the installation of a Euclidean region, we consider the large mass regulator a poor choice and have excluded it from our discussion in \cref{sec::inclusiveobservable} about constructing a physical observable.

One can also see this on simpler, more physical grounds: the asymptotic spectrum of a theory with $\beta > 1$ differs from that with $\beta < 1$ quite dramatically, since in the former, $\varphi$ is stable, while in the latter, it is unstable.

\section{Conclusion}
\label{sec::conclusion}
In this work, we studied both practical and conceptual subtleties of the KLN theorem \cite{Kinoshita:1962ur, Lee:1964is} as presented in \cite{Frye:2018xjj}, in which all the unitarity cuts (including disconnected/forward-scattering contributions) of a single Feynman diagram sum to zero. We were able to demonstrate the cancellation of $t$-channel divergences in an illustrative two-scalar model across several different regularization schemes,  and in doing so developed a prescription for handling issues including:
\begin{itemize}
    \item the non-commutativity of sending the Feynman $i\epsilon_F \rightarrow 0$ and analytically continuing the regulator $\epsilon$ in dimensional regularization or $m_0$ in mass regularization to zero, $m_0,\epsilon \rightarrow 0$;
    \item boundary contributions and non-trivial factors of $1/2$ arising from the distributional properties of the $\delta$-function;
    \item zero-energy emission diagrams as a necessary threshold contribution in certain regularization schemes.
\end{itemize}
The way that the cancellation played out revealed subtle regulator dependence in the KLN partial sums, ultimately rooted in the non-analyticity of unitarity cuts in external kinematics, masses and spacetime dimension. For our particular model, we were also able to trace some of the unusual threshold behavior back to the appearance of a singular hypersurface at the precise $p_i = p_f$ kinematic region being probed by the KLN theorem. 

In addition to demonstrating that all total cross-sections sum to zero in accordance with the KLN theorem, we also constructed a finite, fixed-order, inclusive $t$-channel cross-section more closely related to actual collider observables. This involved exploring subtleties with physical degeneracy and experimental indistinguishability when unstable external states are involved. However, our result here is clearly not the full story: there are certain values of physical kinematics for which our inclusive result goes negative. We leave a full exploration of the issue of negativity to future work, but do provide some speculation about what the next steps may look like.

 Moreover, given the relevance of $t$-channel divergences for early-universe cosmological calculations~\cite{Grzadkowski:2021kgi, Beneke:2014gla, Becker:2023vwd,Coy:2022unt, Garbrecht:2013gd,Garbrecht:2013bia}, another interesting extension to this work would be the inclusion of finite-temperature effects. Even though similar KLN-like fixed-order cancellations have been observed in many non-trivial examples ~\cite{Beneke:2014gla, Garbrecht:2013gd, Czarnecki:2011mr, Gabellini:1989yk, Grandou:1991qr, Baier:1989ub, Altherr:1988bg}, we are not aware of a general proof that holds in the finite-temperature setting. We intend to explore this in future work.

Ultimately, this work explores the KLN theorem as a fixed-order alternative to resummation for computing observable quantities in the presence of $t$-channel divergences. Being a fixed-order approach, it avoids well-known issues with gauge-invariance \cite{Kurihara:1994fz, Argyres:1995ym, Denner:1996gb, Carrington:2003ut} and double-counting \cite{Matak:2022qwc} that can arise from the mixing of perturbative orders, and furthermore applies even when $t$-channel mediators are stable, without needing to invoke regularizing thermal masses or widths \cite{Grzadkowski:2021kgi,Coy:2022unt,Iglicki:2022jjf}. On the other hand, we have shown that there are still many open questions about how to actually connect KLN cancellations to quantities which can be physically measured.

\section*{Acknowledgements}
The authors would like to thank Lachlan E. Tobin, Adam Lackner, Adrian Finke and Michael Klasen for helpful and enlightening discussions. M.B. acknowledges support from an Australian Government Research Training
Program Scholarship.
\appendix
\section{Kinematical Identities}
\label{app::kinematics}
For reference, let us quote here several useful kinematical identities involving the external invariants which will come up again and again in our calculations:
\begin{equation}
\label{eq::kinematical_identities}
\begin{aligned}
    p_1^2 &= p_3^2 = m^2 ,
     & 
     p_2^2 &=p_4^2=0,
    &
    s+ t+ u &= 2m^2,
    \\ 
    s &= (p_1+p_2)^2 = (p_3+p_4)^2,
    &  \qquad 
    p_1 \cdot p_2 &= \frac{s-m^2}{2},
    \\ 
    t &= (p_1-p_3)^2 = (p_2-p_4)^2,
    &  
    p_1 \cdot p_3 &= \frac{2m^2-t}{2},
    \\ 
    u &= (p_1-p_4)^2 = (p_2-p_3)^2,
    & p_2 \cdot p_3 &= \frac{m^2 -u}{2} = \frac{s+t-m^2}{2}.
\end{aligned}
\end{equation}
The majority of the calculations will be done with $t= 0$ (save for in \cref{app::trightarrow0minus}) in accordance with the forward-scattering restriction discussed in \cref{sec::cuttingrules} 

We also perform all our calculations in the centre-of-mass frame defined by $\vec{p_1} + \vec{p_2} = 0$, in which case
\begin{align}
\label{eq::comidentities}
    E_1 &= \frac{s+m^2}{2\sqrt{s}} , 
    & 
    \abs{\vec{p_1}} = \abs{\vec{p_2}} =E_2 &= \frac{s-m^2}{2\sqrt{s}}.
\end{align}
\section{Delta Function Identities}
\label{app::deltafunctions}
Here, we discuss some of the salient features of the Dirac $\delta$-function relevant to the computation of cuts.

\subsection{The \texorpdfstring{$i\epsilon_F$}{i epsilonF} Representation}
\label{app::iepsrep}
\begin{figure}[t!]
    \centering \includegraphics[width=0.5\linewidth]{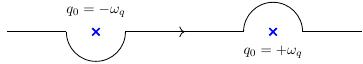}
    \caption{Integration contour $C$ in the complex energy domain. The particular deformation around the poles at $q_0 = \pm \omega_q$ is the one consistent with causality, which selects out the physical sheet. In order to evaluate this integral via the residue theorem, the contour is closed by a (limitingly) large semicircular arc in the lower/upper half planes, depending on the sign of $x_0$.}
    \label{fig:integrationcontour}
\end{figure}

To begin with, recall that we can write our propagators in position space as
\begin{align}
    iG(x) = \int_{C}\frac{dq_0}{2\pi} \int \frac{d^3 q}{(2\pi)^3} \frac{e^{-iqx}}{q^2-m^2},
\end{align}
where $C$ is the integration contour shown in \cref{fig:integrationcontour}, associated with causal time propagation. As the semi-circular contour deformations around the poles are taken smaller and smaller, $C$ can be mimicked by the Feynman $i\epsilon_F$ prescription,
\begin{align}
\label{eq::propdef}
    iG(x) = \lim_{\epsilon_F \rightarrow 0^+} \int \frac{d^4 q}{(2\pi)^4} \frac{e^{-iqx}}{q^2-m^2 + i\epsilon_F}, 
\end{align}
 whereby the propagator denominator is augmented with some small, positive imaginary part, and integration is over the real axis. Importantly, the limit may not exist before the integration (since the integrand would hit the pole), and must be taken \textit{after} the integral is performed.

Now, in order to make contact with $\delta$-functions, note that one can close the integration contour in the lower and upper half planes, depending on the sign of $x_0$, and write 
\begin{align}
    iG(x) = \theta(x_0) iG^+(x) + \theta(-x_0) iG^-(x) ,
\end{align}
where
\begin{align}
\label{eq::propplusminus}
    iG^\pm(x)=\int\frac{d^4q}{(2\pi)^4} 2\pi \theta(\pm q_0) \delta(q^2-m^2)e^{-iqx}.
\end{align}
As these $\delta$-functions arise from evaluating the residues at the poles of $iG$, they inherit the $i\epsilon_F$ representation of \cref{eq::propdef} and can be written using the notation of \cite{Bourjaily:2020wvq} in terms of $i\epsilon_F$ as
\begin{equation}
\label{eq::deltadef}
\begin{aligned}
    \delta^{\epsilon}(x) &\equiv -\frac{1}{2\pi i} \left[\frac{1}{x+i\epsilon_F} - \frac{1}{x-i\epsilon_F}\right] 
    \\ &=\frac{\epsilon_F}{\pi} \frac{1}{x^2+\epsilon_F^2},
\end{aligned}
\end{equation}
where, once again, $\epsilon_F \rightarrow 0^+$ only \textit{after} integration has been performed; the limit of \cref{eq::deltadef} does not exist before it has been interchanged with the integration.

These $\delta$ functions are precisely the ones that appear in the cutting rules, as per the position-space derivation following the largest-time equation in, say, \cite{tHooft:1973wag, Veltman:1994wz}.\footnote{A more recent (and stronger) result \cite{Bourjaily:2020wvq} shows that a similar relation also holds for individual diagrams in time-ordered perturbation theory, where the energy integrals are performed and all internal lines are on-shell.} With this, $\delta^\epsilon$ forms a \textit{representation} of $\delta(x)$, which allows us to define what happens in sticky situations involving $\delta$-functions at the boundary, derivatives of $\delta$-functions, products of $\delta$-functions, and so on. Let us now elucidate what to do in several scenarios relevant to our calculation.

\subsubsection{\texorpdfstring{$\delta$}{Delta}-Functions Localized at the Boundary}
\label{app::deltabdary}
Consider the situation where we have some $\delta$-function localized at the boundary of integration, schematically given by
\begin{align}
    I = \int_0^1 dx \delta(x) f(x) ,
\end{align}
where $f$ is a sufficiently integrable function which is non-singular at $x=0$. Substituting in our representation, we get
\begin{equation} 
\begin{aligned}
    I &= \lim_{\epsilon_F \rightarrow 0} \int_0^1 dx \frac{\epsilon_F}{\pi} \frac{1}{x^2 + \epsilon_F^2} f(x)
    \\
    &= \lim_{\epsilon_F \rightarrow 0} \int_0^{1/\epsilon_F} dx \frac{1}{\pi} \frac{f(\epsilon_Fx)}{x^2+1} ,
\end{aligned}
\end{equation}
where we have made the change of variables $x \rightarrow \epsilon_F x$ to get to the second line. Assuming $f$ is non-singular at $x=0$, we can interchange limit and integration and get 
\begin{align}
    I &= f(0) \int_0^\infty dx \frac{1}{\pi}\frac{1}{1+x^2}
    = \frac{1}{2} f(0) ,
\end{align}
which tells us that a single $\delta$-function localized at the boundary of integration results in an additional factor of $1/2$. Another way to see this is that since $\delta^\epsilon$ is an even function, we can simply expand our integration region $\int_0^1 dx \rightarrow\frac{1}{2} \int_{-1}^1 dx$, in which case the $\delta$-function is no-longer localized at the boundary. 

Note that when there is a product of $\delta$-functions localized at the boundary, the situation can get more complicated and does not necessarily amount to a $1/2^N$ factor; see for eg. Appendix G of \cite{Bourjaily:2020wvq} for an example where a factor of $1/N!$ arises from products of angular $\delta$-functions.

\subsubsection{Derivatives of \texorpdfstring{$\delta$}{delta}-Functions}
\label{app::deltaderiv}
The derivative of the $\delta$-function is defined by integration by parts. Accordingly, one must be careful to keep track of the boundary term, which is another $\delta$-function localized at the boundary of integration. 

In all the cases we will consider here, the derivative of the $\delta$-function lives under a multivariate integral. A schematic example is given by
\begin{align}
    R &= \int_{0}^1 dx \int _0^{1-x} dy f(x,y) \pdv{}{y} \delta(y-g(x)) , 
\end{align}
where $f, g$ are appropriately non-singular. 
Reinstating our representation, we have
\begin{equation} 
\begin{aligned}
    R &= \lim_{\epsilon_F \rightarrow 0} \int_0^1 dx \int_0^{1-x} f(x,y) \pdv{}{y} \left[\frac{\epsilon_F}{\pi} \frac{1}{(y-g(x))^2+\epsilon_F^2}\right]
    \\ &= \lim_{\epsilon_F \rightarrow 0}\left\{- \int_0^1 dx \int_0^{1-x} \pdv{f(x,y)}{y}\left[\frac{\epsilon_F}{\pi} \frac{1}{(y-g(x))^2 + \epsilon_F^2}\right] \right\}
    \\&+ \lim_{\epsilon_F \rightarrow 0}\left\{ \int_0^1 dx\left(  f(x,1-x) \left[\frac{\epsilon_F}{\pi} \frac{1}{(1-x-g(x))^2+\epsilon_F^2}\right]-f(x,0) \left[\frac{\epsilon_F}{\pi}\frac{1}{g(x)^2+\epsilon_F^2}\right]\right)\right\} 
    \\ &= -\int_0^1dx \int_0^{1-x} dy \pdv{f(x,y)}{y} \delta(y-g(x)) + \int_0^1 dx\left[f(x,1-x)\delta(1-x-g(x))-f(x,0)\delta(g(x))\right], 
\end{aligned}
\end{equation}
where terms like $\delta(1-x-g(x))$ and $\delta(g(x))$ can be defined via a change of variables in the standard way. If they lie on the boundary of the $x$ integration, then additional factors may be necessary as per the previous section.

There are two situations where derivatives of $\delta$-functions will arise in our calculation. First, as emphasized in \cite{Frye:2018xjj}, when there are cuts of on-shell propagators, one can use
\begin{equation}
\label{eq::derivativedeltaonshell}
\begin{aligned}
    \frac{\delta(x)}{x+i\epsilon_F} + c.c. &= \frac{\epsilon_F}{\pi} \frac{1}{x^2+\epsilon_F^2} \times \frac{2x}{x^2+\epsilon_F^2}
    \\ &= -\pdv{}{x}\left[\frac{\epsilon_F}{\pi} \frac{1}{x^2+\epsilon_F^2}\right]
    \\&= -\delta'(x)
\end{aligned}
\end{equation}
and second, when one is computing imaginary parts in Schwinger-space, one can use
\begin{align}
    \frac{1}{(x - i\epsilon_F)^n} - \frac{1}{(x+i\epsilon_F)^n} = 2i \frac{(-1)^{n+1}}{(n-1)!} \partial^{n-1}_{x} \delta(x), 
\end{align}
which can be seen by successively differentiating the representation of \cref{eq::deltadef} $n-1$ times.

Finally, let us remark that if the $x$ appearing in these identities is some function -- say, $x=x(a,b)$ -- of multiple integration variables $a,b$, then often one makes use of the chain rule $\partial_x = \left(\pdv{x}{a}\right)^{-1} \partial_a$ so that integration by parts can be used with an actual integration variable. Strictly speaking, for this equation to make sense, we need to be implicitly holding all the other variables -- in this case, $b$ -- constant, since the the full multivariate chain rule would also mandate additional $\partial_b$ contributions. But this is not a problem: since we introduced $\partial_x$ solely for bringing down powers of $1/(x\pm i\epsilon_F)$, we could have simply differentiated with respect to $a$ directly right from the beginning (holding $b$ constant), in which case    
\begin{align}
    \partial_a\left. \frac{1}{(x(a,b) \pm {i\epsilon_F})} \right\lvert_{b \text{ const}} = -\frac{1}{(x(a,b) \pm {i\epsilon_F})^2} \left(\left.\pdv{x}{a} \right\lvert_{b \text{ const}}\right) = \left(\partial_x \frac{1}{(x(a,b) \pm {i\epsilon_F})} \right)\left(\left.\pdv{x}{a} \right\lvert_{b \text{ const}}\right) ,
\end{align}
which can evidently be rearranged into the result we end up using. 

Thus, when we write $\partial_x \delta(x)$ with $x$ some multivariate function, it should be understood with some abuse of notation that we are keeping all variables constant, except for whichever one we choose to transfer the derivative to.

\subsubsection{The Product \texorpdfstring{$[\delta(q^2)]^2\theta(q_0)\theta(-q_0)$}{delta(q^2)^2 theta(q0)theta(-q0)}}
\label{app::delta2theta2}
This product $[\delta(q^2)]^2 \theta(q_0)\theta(-q_0)$ arises in the 4-body cut of \cref{sec::4body}, and seems rather gnarly; it is the square of a massless $\delta$-function, localized at single point $q_0 = \omega_q= 0$ in the phase space. Moreoever, noting that
\begin{align}
\label{eq::splittingpole}
    \delta(q^2) = \frac{1}{2\omega_q} \left[\delta(q_0-\omega_q) + \delta(q_0+\omega_q)\right]
\end{align}
it can be seen that at the very point where the integral is localized, the representation $\delta(q^2)$ is ill-defined, since $\omega_q = 0$!

Fortunately, we can make sense of this product by taking one step back and returning to the position-space representation \cref{eq::propplusminus} from which the cutting rule $\delta$-functions appearing in this product originated. Consider for a moment a single cut propagator position-space propagator given by \cref{eq::propplusminus}. When $m>0$, the energy $\omega_q \geq m$ is strictly positive, and so the $1/\omega_q$ denominator of \cref{eq::splittingpole} is well-defined. For the massless propagator, however, $\omega_q =0$ enters the phase space and the representation of \cref{eq::splittingpole} becomes problematic.

However, the $\omega_q$ in the denominator of \cref{eq::splittingpole} is not the only factor of $\omega_q$ present. Namely, explicitly writing out the measure $d^3q$ in spherical coordinates, one has (for, say, $G^+$) the massless propagator given by
\begin{align}
    iG^+(x) = \frac{1}{(2\pi)^3}\int_{0}^\infty d{q_0}\int d\Omega_3 \int_0^\infty d\omega_q\omega_q^2 \delta(q_0^2- \omega_q^2)  e^{-iqx},
\end{align}
where we have also redefined our ${q_0}$ integration range to $[0, \infty)$ to account for the step function range.

Written this way, it is clear that the correct distributional identity to use is
\begin{align}
    \omega_q^2 \delta(q^2) = \frac{\omega_q}{2}\left[\delta({q_0} -\omega_q) + \delta({q_0}+\omega_q)\right],
\end{align}
which also holds even at $\omega_q = 0$, where the right-hand side vanishes. Moreover, since $q_0 \in [0, \infty)$ due to the $\theta(q_0)$ step function, we can simply drop the $\delta(q_0 + \omega_q)$ contribution altogether, since the only place it can contribute is at $\omega_q = 0$, which is where the integrand vanishes. Thus, a somewhat better-behaved representation of our cut massless propagator is given by
\begin{align}
    iG^+(x) = \int_0^\infty d{q_0} \int \frac{d^3q}{(2\pi)^3} \frac{1}{2\omega_q} \delta({q_0} - \omega_q) e^{-iqx},
\end{align}
where we simply exclude the $\omega_q=0$ point\footnote{More precisely, we are excluding this point from the $\epsilon_F$-regulated expression, since obviously it is selecting out the precise value of $f$ at $\omega_q=0$ after the integration.} since it is regulated by the measure as argued above.

Now, this allows us to make sense of the product $[\delta(q^2]^2 \theta(q_0)\theta(-q_0)$ of two cut $\delta$-functions with opposite energies. Fourier-transforming the cut propagators into momentum space, we find (assuming $f$ is non-singular at $q_0 = \omega_q = 0$) that the expression $T$ in the first line should be replaced with the second line
\begin{equation} 
\begin{aligned}
    T &= \int d^4q [\delta(q^2)]^2 \theta(q_0)\theta(-q_0) f(q)
    \\ &\rightarrow f(0)\int_0^\infty d{q_1}_0 \int_{-\infty}^0 d{q_2}_0 \int d^3 q_1 \int d^3 q_2 \frac{1}{2\omega_1}\delta({q_1}_0-\omega_1) \frac{1}{2\omega_2} \delta({q_2}_0-\omega_2) \delta^4(q_1+q_2),
\end{aligned}
\end{equation}
where the $\delta^4(q_1+q_2)$ is a momentum-conserving $\delta$ function that comes from combining the exponential factors in the cut propagators and integrating over positions. Writing $\delta^4(q_1+q_2) = \delta({q_1}_0+{q_2}_0) \delta^3(q_1+q_2)$ and evaluating the $d^3q_2$ integral, which fixes $\omega_2 = \omega_1$, this becomes
\begin{align}
    T = f(0) \int_0^\infty d{q_1}_0 \int_{-\infty}^0 d{q_2}_0 \int d^3q_1 \frac{1}{4\omega_1^2}\delta({q_1}_0-\omega_1)\delta({q_2}_0-\omega_1) \delta({q_1}_0+{q_2}_0). 
\end{align}
Finally, we write the integral over 3-momentum using spherical coordinates and integrate over angles, $\int d^3q_1 = 4\pi \int_0^\infty \omega_1^2d\omega_1$, in which case the $\omega_1$ terms cancel. After evaluating the $\omega_1$ integral using one of the $\delta$-functions, we are left with 
\begin{align}
    T = \pi f(0)  \int_0^\infty d{q_1}_0 \int_{-\infty}^0 d{q_2}_0 \delta({q_1}_0 -{q_2}_0)\delta({q_1}_0+{q_2}_0).
\end{align}
Now, again instating our representation $\delta^\epsilon$, we see that as even functions, the boundaries can be extended to $\pm\infty$ for both integrals at the cost of an additional factor of $1/4 = 1/2 \times 1/2$. This gives
\begin{equation} 
\begin{aligned}
    T &= \frac{\pi}{4}f(0) \int_{-\infty}^\infty d{q_1}_0\delta({q_1}_0-{q_2}_0) \int_{-\infty}^\infty d{q_2}_0\delta({q_1}_0 + {q_2}_0)
    \\ &= \frac{\pi}{8} f(0)\int_{-\infty}^\infty d{q_1}_0 \delta({q_1}_0)
    \\ &= \frac{\pi}{8} f(0)
\end{aligned}
\end{equation}
where, to get to the second line, we used the fact that $\delta(2{q_1}_0) = \delta({q_1}_0)/2$.

Comparing both sides of this equation, this tells us then that for this representation, we have the identity
\begin{align}
\label{eq::illdefprod}
    [\delta(q^2)]^2 \theta(q_0) \theta(-q_0) = \frac{\pi}{8}\delta^4(q). 
\end{align}
This is a factor of $1/4$ off what we would have gotten had we failed to notice that both $\delta$-functions were localized at the $q_1^0 =q_2^0 = 0$ boundaries.

Now, let us also remark that in $d$-dimensions, the argument would run exactly the same, save for additional factors of $\omega^{-2\epsilon}$ appearing in the integration measure. Repeating our steps, then, we would arrive at the $d$-dimensional analogue
\begin{align}
    T_d = \frac{\pi}{8} f(0) \int_{-\infty}^\infty dq_0 \delta(q_0) q_0^{-2\epsilon} + \mathcal{O}(\epsilon) \text{ constant measure factors}.
\end{align}
This is essentially a question now about the order of limits, since, as per \cref{eq::deltadef}, there are factors of $\epsilon_F$ hiding inside the integration measure. Both orders of limits can be calculated
\begin{align}
   \lim_{\epsilon \rightarrow 0^-} \lim_{\epsilon_F \rightarrow 0^+} \int_{-\infty}^\infty dq_0 \frac{\epsilon_F}{\pi} \frac{q_0^{-2\epsilon}}{q_0^2 + \epsilon_F^2} &= 0
   \\ \lim_{\epsilon_F \rightarrow 0^+}\lim_{\epsilon \rightarrow 0^-}  \int_{-\infty}^\infty dq_0 \frac{\epsilon_F}{\pi} \frac{q_0^{-2\epsilon}}{q_0^2 + \epsilon_F^2} &= 1, 
\end{align}
which highlights a fundamental difference between the $d=4$ and $d>4$ result. Now, as per our argument in \cref{sec::practicaltakeaways}, we should send $\epsilon_F \rightarrow 0$ \textit{before} sending $\epsilon \rightarrow 0$, in which case
\begin{align}
     [\delta(q^2)]^2 \theta(q_0) \theta(-q_0) = 0 \quad \text{ for } d>4.
\end{align}

\subsection{The \texorpdfstring{$\delta_{\text{dim}}$}{delta_dim} Representation}
\label{app::deltaDim}
Finally, we now briefly comment on the "dimensionally-regulated" representation of the $\delta$-function given by
 \begin{align}
        \delta_{\text{dim}}(x) = \lim_{\epsilon \rightarrow 0^-} -\frac{1}{2} \epsilon \abs{x}^{-\epsilon -1}, 
    \end{align}
which comes up in calculations performed in $d = 4-2\epsilon$. As with the $i\epsilon_F$ representation, the limit appearing here should be understood as being performed \textit{after} some integration.

To see that this actually is a representation of a $\delta$-function, let us consider how it acts against a test function $f$. We have
\begin{align}
    I =\int_{-\Lambda}^\Lambda dx f(x) \delta_{\text{dim}}(x)
    \equiv \lim_{\epsilon \rightarrow 0^-} -\frac{1}{2} \int_{-\Lambda}^\Lambda dx \epsilon  \abs{x}^{-\epsilon -1} f(x). 
\end{align}
Here, $\Lambda > 0$ is some finite\footnote{Extending to $\Lambda \rightarrow \infty$ requires some additional technical qualifications, since the integral of $\delta_{\text{dim}}$ over infinite boundaries is divergent. See for eg. \cite{Strichartz:2003} for a more technical discussion on the theory of \textit{tempered distributions}. This is not a problem for us, however, since we only ever integrate $\delta_{\text{dim}}$ over compact intervals in this work.} boundary of integration. Let us also assume that $f$ is smooth and integrable over $[-\Lambda, \Lambda]$. 

We can  treat $f$ as an even function without loss of generality, because any function can be uniquely decomposed into a sum of odd and even functions, and clearly $I = f(0) = 0$  when $f$ is odd. We can also redefine $\epsilon \rightarrow - \bar{\epsilon}$, in which case
\begin{align}
    I = \lim_{\bar \epsilon \rightarrow 0^+} \int_0^{\Lambda} dx\, \bar \epsilon x^{-1 + \bar \epsilon} f(x) .
\end{align}
Integrating by parts, this can be written
\begin{align}
    I = \lim_{\bar \epsilon \rightarrow 0^+} \left\{ [x^{\bar \epsilon} f(x)]^{x=\Lambda}_{x=0} - \int_{0}^{\Lambda} dx\, x^{\bar \epsilon} f'(x) \right\}. 
\end{align}
First, for the boundary term, note that we only take $\bar \epsilon \rightarrow 0^+$ \textit{after} the integration, in which case the $x^{\bar\epsilon}$ contribution simply vanishes at $x=0$, because $0^{\bar \epsilon} = 0$ for any finite choice of $\bar \epsilon$. Thus, the boundary term only contributes $\lim_{\bar \epsilon \rightarrow 0^+} \Lambda ^{\bar \epsilon}f(\Lambda) = f(\Lambda)$.

For the remaining integral, since $f$ is finite and integrable over the interval, and $x^{\bar \epsilon} f(x)$ converges pointwise to $f(x)$ as $\bar \epsilon \rightarrow 0^+$ (save for at $x =0$, but this is measure-zero and non-singular, and can thus be ignored), we can simply set $\bar \epsilon = 0$ inside the integral to recover the limit by the dominated convergence theorem. Thus we arrive at
\begin{align}
    I = f(\Lambda) - \int_0^\Lambda dx f'(x) = f(0) ,  
\end{align}
which tells us that $\delta_{\text{dim}}$ selects out the value at $x=0$ as required.

Note finally that had we localized $\delta_{\text{dim}}$ at the boundary, we would have gotten an additional factor of $1/2$ 
\begin{align}
    \int_0^\Lambda dx f(x) \delta_{\text{dim}}(x) = \frac{1}{2} f(0)
\end{align}
due to the even-ness of our representation, since we could write
\begin{align}
    \int_0^\Lambda dx f(x) \delta_{\text{dim}}(x) = \frac{1}{2} \int_{-\Lambda}^\Lambda dx f(\abs{x}) \delta_{\text{dim}}(x)
\end{align}
and make use of our earlier manipulations.

\section{Feynman Parameterization for the Box Diagram}
\label{app::FeynparamFULL}
Here, we compute the Feynman-parameter representation for the box diagram shown in \cref{fig:box diagramp1neqp3}. Although it would be straightforward in this one-loop calculation to simply perform the Feynman parameterization and loop integral explicitly, we take this as an opportunity to demonstrate some efficient graph-theoretic technology well-suited for higher-loop calculations, where explicitly evaluating the loop integral becomes cumbersome.
\subsection{Graph-Theoretic Preliminaries}
\label{app::graphtheory}
Here we set up some of the graph-theoretic preliminaries and notation, following for eg. \cite{Weinzierl:2022eaz, Bogner:2010kv}, before going into our specific application. Consider a general $l$-loop scalar Feynman integral written in the form
\begin{align}
    I = \mu^{2l\epsilon}\int \prod_{r=1}^l \frac{d^d q_r}{i \pi ^{d/2}} \prod_{j=1}^n \frac{1}{(-k_j^2 + m_j^2-i\epsilon_F)^{\nu_j}} , 
\end{align}
where $q_r$ indexes of the $l$-loop momenta, and $k_j, m_j$ indexes over the momenta and masses respectively of the $n$ internal lines. Here, $\nu_j$ accounts for internal lines with repeated momenta, which in full-generality (in, for eg., analytic regularization) may be non-integer. The choice of sign for the propagators is chosen to be well-suited for the Wick-rotation one performs when evaluating the loop integral, but with integer $\nu_j$ as we have in our case, this causes no unusual behavior involving $(-1)^{\nu_j}$ to keep track of.

Transforming to Schwinger space and evaluating the loop momentum integrals under a Wick rotation, it can be shown after some manipulation that these integrals take the form
\begin{align}
\label{eq::feynparrep}
    I = \frac{\Gamma(\nu - ld/2)}{\prod_{j=1}^n \Gamma(\nu_j)} \mu^{2l\epsilon}\int_{x_j \geq 0} d^nx \delta(1-\sum_{i \in I}x_i) \left(\prod_{j=1}^n x_j^{\nu_j-1}\right) \frac{\mathcal{U}^{\nu - (l+1)d/2}}{(\mathcal{F}-i\epsilon_F)^{\nu - ld/2}} , 
\end{align}
where the $x_i$ are called Feynman parameters, $I$ is \textit{any} non-empty subset of Feynman parameters, $\nu = \sum_{j=1}^n \nu_j$, and $\mathcal{U}$ and $\mathcal{F}$ are respectively called the first and second Symanzik polynomials which we will define in a moment. The fact that one can choose \textit{any} non-empty subset $I$ and the integral evaluates to the same result is known as the Cheng-Wu theorem \cite{Cheng:1987ga}, and can be very useful in practice for choosing a convenient representation for calculations. In particular, we make use of it in \cref{app::trightarrow0minus}. 

Now, in graph-theoretic language, the Symanzik polynomials are defined as
\begin{equation}
\label{eq::symanzikpolynomials}
\begin{aligned}
    \mathcal{U} &= \sum_{T \in \mathcal{T}_1} \prod_{e_i \notin T} x_i ,
    \\ \mathcal{F} &=\sum_{(T_1, T_2) \in \mathcal{T_2}} \left(\prod_{e_i \notin (T_1, T_2)} x_i\right) (-s_{(T_1, T_2)}) + \mathcal{U}\sum_{i=1}^n x_i m_i^2 , 
\end{aligned}
\end{equation}
where $\mathcal{T}_1$ consists of all $1$-trees of the graph (where a $1$-tree is a connected subgraph that contains all vertices but has no loops), $\mathcal{T}_2$ consists of all $2$-trees of the graph (where a $2$-tree is a subgraph that contains all the vertices, has no loops, and contains exactly two connected components $(T_1, T_2)$), and $s_{(T_1, T_2)}$ is defined as
\begin{align}
    s_{(T_1, T_2)} =  \left(\sum_{e_j
    \notin (T_1, T_2)}q_j\right)^2. 
\end{align}
 $e_j \notin (T_1,T_2)$ denotes edges which are in neither of the two 2-trees $T_i$ and $q_j$ is the momentum flowing through the edge $e_j$ from $T_1$ to $T_2$. Here, we assume that internal lines are oriented to flow all from $T_1$ to $T_2$.
 
Alternatively, by momentum conservation, the invariant $s_{(T_1, T_2)}$ can be thought of equivalently as the inner product of all \textit{external} momenta $p_i \in P_{T_1}$ with all \textit{external} momenta $p_j \in P_{T_2}$, that is,
\begin{align}
   s_{(T_1, T_2)} =-\left(\sum_{p_i \in P_{T_1}} \eta_i p_i\right) \cdot \left(\sum_{p_j \in P_{T_2}} \eta_j p_j\right),
\end{align}
where $\eta = +1$ for incoming momenta and $\eta=-1$ for outgoing momenta, which accounts for orientations. Written this way, it is clear that such Feynman integrals are Lorentz-invariant functions of external kinematics.

\subsection{The Box Diagram}
\label{app::feynpars}

\begin{figure}[tbp!]\centering 

\includegraphics[width=0.4\linewidth]{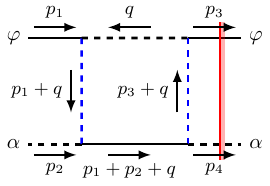}
\caption{The forward-scattering cut of the box diagram, to be summed with its complex conjugate. Here, we keep $p_3$ general, not enforcing $p_1 = p_3$. Everything to the right of the cut is to be complex-conjugated. The dashed lines in blue are given the mass $m_0$ in the LM and SM regulator schemes.}
\label{fig:box diagramp1neqp3}
\end{figure}
Now we consider our box diagram shown in \cref{fig:box diagramp1neqp3}, with
\begin{align}
\label{eq::tloopint}
    \begin{aligned}
    I_{\text{box}}(s, t; d, m_0) \equiv \mu^{2\epsilon}\int \frac{d^d q}{(2\pi)^d} \frac{1}{q^2+{i\epsilon_F}} \frac{1}{(p_1 + q)^2-m_0^2 + {i\epsilon_F}} & \frac{1}{(p_1+p_2+q)^2-m^2+{i\epsilon_F}}
    \\ & 
    \times \frac{1}{(p_3 + q)^2-m_0^2 + {i\epsilon_F}} + c.c.  .
\end{aligned}
\end{align}
Note that the expression appearing in \cref{sec::boxdiagram} sets $p_1 = p_3$ to begin with, but here we will consider the more general $p_1 \neq p_3$ case as well, so we can also use it in \cref{app::trightarrow0minus}.

\begin{figure}[tbp!]
    \centering \includegraphics[width=0.7\linewidth]{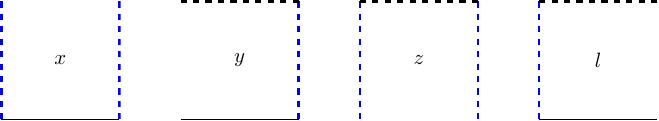}
    \caption{Set of $1$-trees for the box diagram, with the relevant Feynman parameter (associated with the deleted internal edge) displayed in the middle of each $1$-tree. External lines are omitted for visual clarity. }
    \label{fig::1trees}
\end{figure}

\begin{figure}[tbp!]
\captionsetup[subfigure]{labelformat=empty}
    \centering 
    \begin{subfigure}[]{0.7\linewidth}
\includegraphics[width=\linewidth]{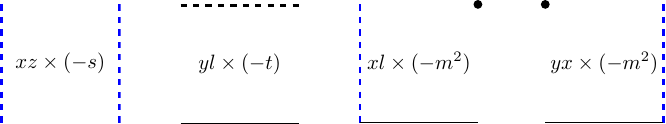}  
    \end{subfigure}
    \\ \vspace{3ex}\begin{subfigure}[]{0.32\linewidth}
\includegraphics[width=\linewidth]{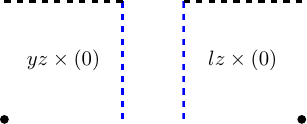}  
    \end{subfigure}
    \caption{Set of $2$-trees for the box diagram. The product of the relevant Feynman parameters (associated with the deleted internal edges) and external invariant coefficient $-s_{(T_1, T_2)}$ is displayed in the middle of each $2$-tree. The $2$-trees shown in the bottom line vanish, since they contain invariants $p_2^2 = 0$ and $p_4^2 = 0$. External lines are omitted for visual clarity, and isolated vertices are emphasized with a black dot.}
    \label{fig::2trees}
\end{figure}

With the $1$-trees shown in \cref{fig::1trees} and the $2$-trees shown in \cref{fig::2trees}, we can immediately read off
\begin{equation} 
\begin{aligned}
    \mathcal{U} &= x+y+z+l ,
    \\ \mathcal{F} &= -xzs -ylt-xlm^2-yxm^2 + (x+y+z+l) (m^2z + m_0^2(y+l))
\end{aligned}
\end{equation}
in accordance with \cref{eq::symanzikpolynomials}.
Then, with  $l = 1$, $d=4-2\epsilon$, and $\nu_j = 1$ for each internal line, we get 
\begin{align}
\label{eq::fullsymrep}
    I_{\text{box}} = \frac{i}{16\pi^2} \int_{x,y,z,l \geq 0} dxdydzdl \,\delta(1-\sum_{i \in I} x_i) \frac{(x+y+z+l)^{2\epsilon}}{(\mathcal{F} - i\epsilon_F)^{2+\epsilon}} + c.c.  ,
\end{align}
where we have multiplied \cref{eq::feynparrep} by $i\pi^{d/2}/(2\pi)^{d}$ to match the coefficient of \cref{eq::tloopint}, and set $\epsilon = 0$ in all prefactors out the front.  

Let us now consider the $t=0$ case of \cref{sec::boxdiagram} explicitly. In this case, it becomes convenient to choose the standard $I = \{x,y,z,l\}$ to fix our $\delta$-function. Moreover, with $t =0$, the two terms in the denominator with prefactors $y$ and $l$ contribute in the same way to $\Delta$, so the Feynman parameterization becomes a function of $(y+l)$ only. Thus, relabelling this as $y$, integrating out the extraneous coordinate, and then performing the $dz$ integral with our $\delta(1-x-y-z)$, we arrive at
\begin{align}
    I(s,0;d, m_0) = \frac{i}{16\pi^2}\int_0^1 dx \int_0^{1-x}dy \frac{ y }{(\Delta-{i\epsilon_F})^{2+\epsilon}} + c.c.
\end{align}
with
\begin{align}
    \Delta \equiv sx(-1+x+y)-m^2(-1+x+y+xy) +m_0^2y.
\end{align}

\mathversion{bold}
\section{Singular Hypersurfaces and \texorpdfstring{$t=0$}{t=0}}
\label{app::trightarrow0minus}
\mathversion{normal}
Now, let us investigate what's happening in the vicinity of $t = 0$, which is where the KLN theorem localizes our integral. For simplicity, let us focus on the region $s \in (m^2, 2m^2)$, and $t < 0$ sufficiently small in magnitude so as to avoid any other thresholds, and ask: as we increase $t$, at what value $t^*$ does $\sigma_{\text{box}}(s,t^*)$ jump discontinuously, signaling the presence of a new threshold?  

To compute this, we use the Feynman parameter representation of \cref{eq::fullsymrep}, without setting $t = 0$ as we do in \cref{sec::boxdiagram}. In this case, setting $s=\kappa m^2$, $m_0^2 = \beta m^2$, and $t = \tau m^2$ to define a dimensionless polynomial
\begin{align}
    \bar{\mathcal{F}} \equiv - xz \kappa -yl \tau -xl -yx +(x+y+z+l)(z+\beta(l+y)) , 
\end{align}
we have for the cross-section using a mass regulator $m_0$
\begin{align}
    \sigma =  \frac{g^4}{2(s-m^2)} \times\frac{1}{8\pi} \frac{1}{m^4} R_t
\end{align}
with 
\begin{equation} 
\label{eq:Rt}
\begin{aligned}
    R_t &= \frac{i}{2\pi} \int_{x,y,z,l \geq 0} dxdydzdl \delta(1-\sum_{i \in I} x_i) \frac{1}{(\bar{\mathcal{F}}-i\epsilon_F)^2} + c.c.
    \\ &= \frac{i}{2\pi}(-2\pi i)  \int_{x,y,z,l \geq 0} dxdydzdl \delta(1-\sum_{i \in I} x_i) \partial_{\bar{\mathcal{F}}} \delta(\bar{\mathcal{F}})
    \\ &= \int_{x,y,z,l \geq 0} dxdydzdl \delta(1-\sum_{i \in I} x_i) \frac{1}{\partial_{x_i}\bar{\mathcal{F}}} \partial_{x_i}\delta(\bar{\mathcal{F}}).
\end{aligned}
\end{equation}
Here, we have used some tricks from \cref{app::deltaderiv} to rewrite our expression in terms of the derivative of the $\delta$-function with respect to some Feynman parameter $x_i$. The exact nonempty subset $I$ we choose to fix our $\delta$-function in accordance with the Cheng-Wu theorem, and the exact variable $x_i$ we choose to, will now depend on whether we investigate the small or large mass regulator.
\subsection{Small Mass Regulator and No Regulator}
With the small mass and no regulator cases, it turns out to be convenient to fix $l=1$ (and rather inconvenient for the large mass regulator case, which we will comment on shortly), in which case, explicitly, our integral becomes
\begin{align}
    R_t = \int_0^\infty dz\int_0^\infty dy \int_0^\infty dx \frac{\eta_\pm\delta'(x-x_0)}{(\beta +\beta  y-y-\kappa  z+z-1)^2}
\end{align}
with
\begin{equation}
\begin{aligned}
    x_0 &\equiv \frac{-\beta -\beta  y^2-2 \beta  y+\tau  y-\beta  y z-y z-z^2-\beta  z-z}{\beta +\beta  y-y-\kappa  z+z-1},
    \\ \eta_\pm &\equiv \text{sgn{$\left(\beta +\beta  y-y-\kappa  z+z-1\right)$}}.
\end{aligned}
\end{equation}
We have chosen this integration order, because it allows us to integrate over a variable which is linear in the argument of the $\delta$-function, so that the roots are as simple as possible. This approach has proven useful throughout our calculations.

Integrating by parts, only a boundary term survives, since there is no $x$-dependence left in the integral. Now, in this representation, there are two boundaries of integration: $x= \infty$ and $x= 0$. For the former, note that any contribution with $y,z\rightarrow 0$ would not contribute anything, since the integrand is suppressed in that region. Thus, in order to get some contribution at $x=\infty$ with finite $y, z$, we need the denominator of $x_0$ to vanish. This occurs at
\begin{align}
    y_\infty = \frac{1-z-\beta + z\kappa}{\beta -1}, 
\end{align}
which is only contained in the integration region $y \geq 0$ (with $\kappa \in (1, 2]$) when $\beta \geq 1$, that is, in the large mass region. The fact that there is a contribution at the $x=\infty$ boundary is why this representation is not so convenient for the large mass, but works neatly for the small (and zero) mass.
Thus, integrating by parts, we are only left with the $x_0 = 0$ root
\begin{align}
     R_t = -\int_0^\infty dz\int_0^\infty dy \frac{\eta_\pm\delta(x_0)}{(\beta +\beta  y-y-\kappa  z+z-1)^2}.
\end{align}
where the minus sign comes from the fact that $x=0$ is the lower bound of integration.

To evaluate this remaining $\delta$-function, we can -- keeping in mind a subtlety which we will comment on in a moment -- solve for $x_0 = 0$ in terms of $y$ and evaluate it against the $dy$ integral, with the appropriate Jacobian factors. For the mass regulator $\beta > 0$, there are two roots
\begin{align}
    y_0^\pm \equiv \frac{-2 \beta +\tau -\beta  z-z\pm\sqrt{-4 \beta  \tau +\tau ^2+\beta ^2 z^2-2 \beta  z^2+z^2-2 \beta  \tau  z-2 \tau  z}}{2 \beta }, 
\end{align}
while  there is only a single root in the no regulator case,
\begin{align}
    y_0 \equiv \frac{-z-z^2}{z-\tau}.
\end{align}

The support of this $\delta$-function is what determines $t^*$. In the no regulator case, for any $t<0$, the only support is measure-zero at $z=0$, in which case the integrand (properly accounting for Jacobian factors with $\delta(y-y_0)$ takes the form $\abs{\tau}^{-1}$, which is non-singular for $\tau < 0$, and can thus be dropped. But, as we calculated in \cref{sec::boxdiagram}, at $t=0$, the no-regulator calculation was non-vanishing due to the emergence of a boundary contribution, in which case
\begin{align}
    \lim_{t \rightarrow 0^-}\sigma^{\text{NR}}(s,t) = 0 \neq \sigma^{\text{NR}}(s,0),
\end{align}
which tells us that $t^*_{\text{NR}} =0$.

On the other hand, with the mass regulator $\beta >0$ and $\kappa \in (1, 2)$, it can be straightforwardly checked that this $\delta$-function only switches on for $t^*_{\text{SM}}= 4 m_0^2$ ($\tau^*_\text{SM}=4\beta$).  As such, one has
\begin{align}
    \lim_{t \rightarrow 0^\pm}\sigma^{\text{SM}}(s,t) = 0 = \sigma^{\text{SM}}(s,0), 
\end{align}
which tells us that the $t = 0$ result can be identified with the $t \rightarrow 0^-$ approach from the physical region with the mass regulator in place.

Finally, as a sanity check, we might like to confirm that this representation can reproduce the $t = 0$ no-regulator result calculated in the main body under \cref{sec::boxnoreg}, which is where the subtlety alluded to earlier comes in. Namely, when writing $\delta(x_0)$ in terms of the roots of $y$, one must in principle \textit{also} check that there are no additional, independent roots in $z$ contained within the integration region, since the roots in $y$ may not necessarily be in one-to-one correspondence with the roots in $z$.\footnote{For instance, in the no regulator case, the equation is quadratic in $z$ but only linear in $y$.} In the $t < t^*$ case here, this works out and is not a problem, since both roots in $z$ are also excluded from the integration region. However, when both $\beta, \tau =0$ in the no-regulator calculation, one has
\begin{align}
     x_0\lvert_{\beta, \tau = 0} = \frac{z(1+y+z)}{1+y+z(\kappa-1)}
\end{align}
in which case there is a root at $z=0$ contained within the integration region, but no such roots for $y$. Accounting for this, computing the (trivial) Jacobian at $z=0$, and keeping track of the additional $1/2$ from boundary localization and the $\eta_\pm = -1$ factor, one gets
\begin{align}
    R_0 &=\int_0^\infty dy \int_0^\infty dz  \frac{ \delta(z)}{(-y-\kappa z + z-1)^2}
    =  \frac{1}{2} \int_0^\infty dy \frac{1}{(1+y)^2}
    =\frac{1}{2}, 
\end{align}
which exactly reproduces
\begin{align}
    \sigma^{\text{NR}}(s, t=0) = \frac{1}{32 \pi} \frac{g^4}{m^4}\frac{1}{s-m^2}.
\end{align}

\subsection{Large Mass Regulator}
For the large mass regulator, ie. $\beta > 1$, we choose the Feynman parameterization where $l= 1-x-y-z$, because we want to avoid boundary contributions at infinity. Explicitly, this integral becomes
\begin{align}
    R_t = \int_0^1 dx \int_0^{1-x}dy\int_0^{1-x-y} dz \frac{\eta_\pm \delta'(z-z_0)}{(1+x-\beta -x\kappa +y\tau)^2} , 
\end{align}
where
\begin{align}
    z_0 &\equiv \frac{\beta +x^2-\beta  x+\tau  x y-x+\tau  y^2-\tau  y}{\beta +\kappa  x-x-\tau  y-1} , 
    \\\eta_\pm &\equiv \text{sgn}(-\beta -\kappa  x+x+\tau  y+1) , 
\end{align}
 Integrating by parts, we are left only with pure boundary terms
 \begin{align}
     R_t = \int_0^1 dx \int_0^{1-x} dy \frac{\eta_{\pm}}{(1+x-\beta-x\kappa +y\tau)^2}\left[\delta(1-x-y-z_0)-\delta(z_0)\right]. 
 \end{align}
 It is this expression which determines $\tau^*_{\text{LM}}$. The $\delta(z_0)$ has no support for $\tau < \tau^*_{\text{LM}} = 4\beta$, and only switches on for $\tau \geq \tau^*_{\text{LM}}$. Thus, we have the same threshold as in the small-mass case. 
 
 Of course, as a sanity check, we should again confirm that we can reproduce the results of \cref{sec::boxmassreg} with this parameterization. The other $\delta$-function has support in the region $\tau < 4\beta$, and so we can write
\begin{multline}
    \delta(1-x-y-z_0) = \eta_{y_0} \delta(y_0)
    \\ \times \frac{\beta ^2-2 \beta +\tau +x^2 \left(\kappa ^2+\kappa  (\tau -2)+1\right)-x (-2 \beta  (\kappa -1)+\kappa  (\tau +2)+\tau -2)+1}{(\beta +(\kappa -1) x-1)^2}, 
\end{multline}
where $\eta_{y_0}$ enforces the positivity of the prefactor. 

We can now evaluate the $dy$ integral. Enforcing that $y_0 \in (0,1-x)$ sets bounds on the $x$ integral; namely
\begin{align}
    x \in (1/\kappa, 1), 
\end{align}
 where a measure zero contribution at $x=1$ has been dropped. In this region, $\eta_\pm = -1$ and $\eta_{y_0} = 1$. Thus, our integral becomes
\begin{align}
\label{eq::rt}
    R_t = -\int_{1/\kappa}^1 dx \frac{1}{(\beta +(\kappa -1) x-1)^2+\tau  (x-1) (\kappa  x-1)}.
\end{align}
Although this integral can be computed explicitly as some complicated  combination of logarithms, in the end we only wish to compute $\lim_{\tau \rightarrow 0^-} R_t$. Since the integrand converges pointwise to an integrable function as $\tau \rightarrow 0^-$ (because there are no zeroes in the denominator contained in the integration region for $\tau < 4\beta$), we can simply exchange the limit $\tau\to 0^-$ with the integral over $x$, and arrive at
\begin{align}
    \lim_{\tau \rightarrow 0^-} R_t &= - \int_{1/\kappa}^1 dx \frac{1}{(\beta +(\kappa -1) x-1)^2}
   = \frac{1-\kappa}{(\beta +\kappa -2) (\beta  \kappa -1)}
\end{align}
which, after restoring the appropriate factors, gives
\begin{align}
    \lim_{t \rightarrow 0^-} \sigma^{\text{LM}}(s,t) =  \frac{1}{16\pi} \frac{g^4}{(m^4-m_0^2s)} \frac{1}{(s-2m^2+m_0^2)}.
\end{align}
This is the same result as the one calculated with $t=0$ directly in \cref{sec::boxmassreg}, which is what one should expect, since the threshold is shifted away from $t=0$ to $t^*_{LM} = 4m_0^2 >0$. 

\section{Contributions to the Inclusive Cross-Section}
\label{app::contributionstoint}
Here, we compute all the individual contributions to our inclusive cross-section appearing in \cref{sec::fullincresult}, integrated in a window $[z^*(\kappa -\Lambda), z^*(\kappa + \Lambda)] \cap [-1,1]$ around the on-shell point $z^*$, where $0<\Lambda \ll 1$ is a fixed experimental resolution on $\kappa \equiv s/m^2$. Due to the constraint that $\cos \theta \in [-1,1]$, there are three relevant regions to consider: $\kappa < 2-\Lambda$, where the window around the on-shell point is excluded from the region of integration altogether, such that the contribution automatically vanishes (and will be omitted in the following for brevity); $\kappa \in [2-\Lambda, 2+\Lambda)$ where only the lower bound is included within $[-1,1]$; and $ \kappa \geq 2+\Lambda$, where both bounds appear inside $[-1,1]$. 

To cancel the $t$-channel divergence, the $t$-channel and $3$-body contributions are given by
\begin{align} [\sigma_t]_{ R^t_d}&= \frac{g^4}{\pi m^6}\begin{dcases}
         -\frac{(\kappa +\Lambda -2) (\kappa +\Lambda )}{16 (\kappa -2) \Lambda  (2 \kappa +\Lambda -2)}& 2-\Lambda \leq \kappa < 2+\Lambda
        \\ -\frac{(\kappa -1) \kappa }{4 \Lambda  (2 \kappa -\Lambda -2) (2 \kappa +\Lambda -2)}& \kappa \geq 2+\Lambda 
    \end{dcases}
   \\ [\sigma_3]_{ R^3_d}&= \frac{g^4}{\pi m^6}\begin{dcases}
        \frac{(\kappa +\Lambda -2) (\kappa +2 \Lambda )}{32 (\kappa -2) \Lambda  (\kappa +\Lambda -1)}& 2-\Lambda \leq \kappa < 2+\Lambda
        \\ \frac{\kappa ^2-\kappa -\Lambda ^2}{16 \Lambda  (\kappa -\Lambda -1) (\kappa +\Lambda -1)}& \kappa \geq 2+\Lambda 
    \end{dcases}
\end{align}
where we have omitted the squared $\delta$ function contribution which cancels between the two processes c.f. Eqs.~\eqref{eq::fulltchannelexppole}, \eqref{eq::squaredeltat}, \eqref{eq::sigma3SM}, and \eqref{eq::sigma3DR}. These contributions can be summed together and expanded in $\Lambda$ as 
\begin{align}
    \sigma_{\text{inc, t $+$ 3}}^{(g^4)} = \frac{g^4}{\pi m^6}  \begin{dcases}
       \frac{\kappa -2}{64 (\kappa -1)^2}+\frac{(4-\kappa ) \Lambda }{128 (\kappa -1)^3}+\mathcal{O}\left(\Lambda ^2\right)& 2-\Lambda \leq \kappa < 2+\Lambda
        \\ \frac{(4-\kappa) \Lambda }{64 (\kappa -1)^3}+\mathcal{O}\left(\Lambda ^2\right)& \kappa \geq 2+\Lambda 
    \end{dcases}
\end{align}
This is finite for all $\kappa$, but goes negative for $\kappa \gtrapprox 4$. However, individual diagrams are not physical, and this is not the full contribution; so, at $\mathcal{O}(g^4)$, 
what other processes could actually contribute in this region?\footnote{Note that are $\mathcal{O}(g^2)$ cuts of the $t$ and $s$ channel diagrams (without multiplication with their complex conjugates), but these do not have kinematic support in the relevant large $s$ region; the former contributes (divergently) only at $s = 2m^2$, and the latter only at $s = 0$.} 

Among the remaining cuts of the box diagram, note that both the forward-scattering interference of \cref{sec::boxdiagram} and the four-body diagram of \cref{sec::4body} only contribute at a single point in the phase space of external kinematics. Moreover, across all three NR, SM and DR regularization schemes, we have that $[\sigma_{\text{box}}]_{R^{\text{box}}} + [\sigma_4]_{R^4} = 0$ for any regions $R^{\text{box}}, R^4$ containing that point of support. As such, so long as both cross-sections are either included or excluded consistently, they will contribute nothing to the inclusive cross-section and can be ignored. 

Otherwise, there are also contributions not arising from cuts of the box diagram. Certainly, there is the finite $s$-channel scattering, in addition to the divergent $s-t$ interference of \cref{sec::stdiagraminterference} and (for finiteness of the $s-t$ divergence) the triple emission of \cref{sec::tripleemission}, wherein the cancellation and interpretation of the latter two are totally analogous to that of $\sigma_t$ and $\sigma_3$. In principle, one might also include the forward-scattering cut of the triangle diagram computed in \cref{sec::triangle} when $s \approx 2m^2$, but this only contributes in the forward-scattering region and is anyways subdominant, so we will exclude it for simplicity.\footnote{We also considered additional $\mathcal{O}(g^4)$ cuts of the $t$-channel diagram modified by on-shell self-energy corrections to the incoming and outgoing $\varphi$ legs. However, these diagrams are somewhat ill-defined, since they are involve on-shell propagators fixed with \textit{incoming} momenta, which one does not typically integrate over (although, it should be noted that a modified cross-section is proposed in \cite{Frye:2018xjj}, in which both initial and final state momenta are integrated over). Our estimate, which was computed by retaining the full $\epsilon_F$-dependence of both the propagators and $\delta$-functions c.f. \cref{app::iepsrep} was nonetheless subdominant for large $\kappa$, and is not included here. Alternative approaches to computing similar cuts are explored in \cite{Britto:2011cr, Bahl:2021rts}.} 

These additional contributions from the $s$-channel, $s$-$t$ interference, and triple emission can be computed straightforwardly and are given by 
\begin{align} [\sigma_s]_{ R^s_d}&= \frac{g^4}{\pi m^6}\begin{dcases}
       \frac{(\kappa +\Lambda -2) (\kappa +\Lambda )}{16 \kappa ^3 (\kappa +\Lambda -1)^2}& 2-\Lambda \leq \kappa < 2+\Lambda
        \\ \frac{(\kappa -1) \Lambda }{4 \kappa ^3 \left((\kappa -1)^2-\Lambda ^2\right)^2}& \kappa \geq 2+\Lambda  
    \end{dcases}
   \\ [\sigma_{st}]_{ R^{st}_d}&= \frac{g^4}{\pi m^6}\begin{dcases}
        \frac{\log \left(\frac{\Lambda ^2 (2 \kappa +\Lambda -2)^2}{(\kappa -2)^2 \kappa ^2 (\kappa +\Lambda -1)^4}\right)}{16 (\kappa -1)^2 \kappa }& 2-\Lambda \leq \kappa < 2+\Lambda
        \\ \frac{\log \left(\frac{(2 \kappa +\Lambda -2) (-\kappa +\Lambda +1)^2}{(2 \kappa -\Lambda -2) (\kappa +\Lambda -1)^2}\right)}{8\pi  (\kappa -1)^2 \kappa  } & \kappa \geq 2+\Lambda 
    \end{dcases}
    \\ [\sigma_{T}]_{ R^{st}_d}&= \frac{g^4}{\pi m^6}\begin{dcases}
        \frac{ \log \left(\frac{(\kappa -2)^2 \left(2 (\kappa -1) \Lambda +(\kappa -1) \kappa +\Lambda ^2\right)^2}{\Lambda ^2 (2 \kappa +\Lambda -2)^2}\right)}{16   (\kappa -1)^2 \kappa  }& 2-\Lambda \leq \kappa < 2+\Lambda
        \\ \frac{ \log \left(\frac{(-2 \kappa +\Lambda +2)^2 \left(2 (\kappa -1) \Lambda +(\kappa -1) \kappa +\Lambda ^2\right)^2}{(2 \kappa +\Lambda -2)^2 \left(\kappa ^2-\kappa  (2 \Lambda +1)+\Lambda  (\Lambda +2)\right)^2}\right)}{16   (\kappa -1)^2 \kappa  } & \kappa \geq 2+\Lambda 
    \end{dcases}
\end{align}
Observe a similar cancellation of divergences between $[\sigma_{st}]_{R^{st}_d}$ and $[\sigma_T]_{R^T_d}$. However, even with these contributions, the $\mathcal{O}(g^4)$ contribution will still remain negative for sufficiently large $\kappa$; summing all these processes together and expanding in $\Lambda$ gives 
\begin{align}
\sigma_{\text{inc}}^{(g^4)} = \begin{dcases}
        \begin{multlined}{\left( \frac{g^4 \left(2 \pi  (\kappa -2) \left(\kappa ^2+4\right)-16 \pi  \kappa  \log (\kappa -1)\right)}{128 \pi ^2 (\kappa -1)^2 \kappa ^2 m^6}\right) \\-  \left(\frac{g^4 \left(\kappa ^4-4 \kappa ^3+32 \kappa -16\right)}{128 \pi  (\kappa -1)^3 \kappa ^3 m^6}\right)\Lambda + \mathcal{O}(\Lambda^2)}\end{multlined} & 2-\Lambda \leq \kappa <  2 + \Lambda 
        \\ -\left(\frac{g^4 \left(\kappa ^4-4 \kappa ^3+32 \kappa -16\right)}{64 \pi  (\kappa -1)^3 \kappa ^3 m^6}\right) \Lambda + \mathcal{O}(\Lambda^2) & \kappa \geq 2+ \Lambda 
    \end{dcases}
\end{align}
which is indeed still negative as $\kappa \rightarrow \infty$.

\bibliographystyle{utphys}
\bibliography{refs}
\end{document}